\documentclass[fleqn,usenatbib]{mnras}

\usepackage{newtxtext,newtxmath}

\usepackage[T1]{fontenc}

\DeclareRobustCommand{\VAN}[3]{#2}
\let\VANthebibliography\thebibliography
\def\thebibliography{\DeclareRobustCommand{\VAN}[3]{##3}\VANthebibliography}

\usepackage{graphicx}	
\usepackage{amsmath}	
\usepackage{longtable}
\usepackage{makecell}

\newcommand{\mrm}[1]{\mathrm{#1}}	
\newcommand{\vect}[1]{\boldsymbol{#1}}
\newcommand{\ellsum}[1]{\ell_{\Sigma}} 
\newcommand{\kms}{\mathrm{km\,s^{-1}}} 
\newcommand{\dy}{\mathrm{d}}  
\newcommand{\Prot}{P_{\mathrm{rot}}} 

\newcommand{\Msun}[1]{$\mrm{M_{\odot}}$} 
\newcommand{\Rsun}[1]{$\mrm{R_{\odot}}$} 
\newcommand{\BV}{$\langle B_V \rangle$} 
\newcommand{\Btor}{$\langle B_{\mathrm{tor}} \rangle$} 
\newcommand{\Btormax}{$\sigma_{\langle B_\mathrm{tor} \rangle}$} 
\newcommand{\cc}[1]{#1} 

\title[Strong magnetic fields of slowly rotating M~dwarfs]{SPIRou reveals unusually strong magnetic fields of slowly rotating M~dwarfs}

\author[L. T. Lehmann et al.]{
L. T. Lehmann,$^{1}$\thanks{E-mail: lisa.lehmann@irap.omp.eu}
J.-F. Donati,$^{1}$
P. Fouqu{\'e},$^{1}$
C. Moutou,$^{1}$
S. Bellotti,$^{1,2}$
X. Delfosse,$^{3}$
P. Petit,$^{1}$
\newauthor
A. Carmona,$^{3}$
J. Morin,$^{4}$
A. A. Vidotto,$^{5}$
\,and the SLS consortium
\\
$^{1}$IRAP, Universit\'e de Toulouse, CNRS / UMR 5277, UPS-OMP, 14 Avenue E. Belin, Toulouse F-31400, France\\
$^{2}$Science Division, Directorate of Science, European Space Research and Technology Centre (ESA/ESTEC), Keplerlaan 1, 2201 AZ, Noordwijk, The Netherlands\\
$^{3}$Univ. Grenoble Alpes, CNRS, IPAG, 38000 Grenoble, France\\
$^{4}$Laboratoire Univers et Particules de Montpellier, Universit\'e de Montpellier, CNRS, F-34095 Montpellier, France\\
$^{5}$Leiden Observatory, Leiden University, PO Box 9513, 2300 RA Leiden, The Netherlands
}

\date{Accepted XXX. Received YYY; in original form ZZZ}

\pubyear{2022}

\begin{document}
\label{firstpage}
\pagerange{\pageref{firstpage}--\pageref{lastpage}}
\maketitle

\begin{abstract}
In this paper, we study six slowly rotating mid-to-late M~dwarfs (rotation period $\Prot \approx 40-190\,\dy$) by analysing spectropolarimetric data collected with SPIRou at the Canada-France-Hawaii Telescope as part of the SPIRou Legacy Survey from 2019 to 2022. 
From $\approx$100--200 Least-Squares-Deconvolved (LSD) profiles of circularly polarised spectra of each star, we confirm the stellar rotation periods of the six M~dwarfs and explore their large-scale magnetic field topology and its evolution with time using both the method based on Principal Component Analysis (PCA) proposed recently and Zeeman-Doppler Imaging. All M~dwarfs show large-scale field variations on the time-scale of their rotation periods, directly seen from the circularly polarised LSD profiles using the PCA method. We detect a magnetic polarity reversal for the fully-convective M~dwarf GJ~1151, and a possible inversion in progress for Gl~905. The four fully-convective M~dwarfs of our small sample (Gl~905, GJ~1289, GJ~1151, GJ~1286) show a larger amount of temporal variations (mainly in field strength and axisymmetry) than the two partly-convective ones (Gl~617B, Gl~408). Surprisingly, the six M~dwarfs show large-scale field strengths in the range between 20 to 200\,G similar to those of M~dwarfs rotating significantly faster. 
Our findings imply that the large-scale fields of very slowly rotating M~dwarfs are likely generated through dynamo processes operating in a different regime than those of the faster rotators that have been magnetically characterized so far. 
\end{abstract}




\section{Introduction}

M~dwarfs are known to host strong magnetic fields with large- and small-scale field strengths that may exceed 1\,kG \citep{Morin2010, Kochukhov2021}. Zeeman-Doppler-Imaging (ZDI, \citealt{DonatiBrown1997, Donati2006}) revealed different types of large-scale field topologies for M~dwarfs: the partly convective early M~dwarfs usually showing more complex, relatively weaker fields with non-axisymmetric poloidal fields and significant toroidal components \citep{Donati2008}. Mid M~dwarfs often display simpler and stronger, mainly poloidal and axisymmetric large-scale fields \citep{Morin2008} whereas the fully-convective late M~dwarfs of lowest masses may end up showing large-scale fields in either configuration, (\citealt{Morin2010}, see also the review by \citealt{Donati2009}).
 
Besides, magnetic activity (diagnosed by various proxies) increases for shorter rotation periods until it saturates, i.e. no longer increases with decreasing rotation periods (see e.g. \citealt{Saar1996, Wright2011}). In the unsaturated regime, both large- and small-scale fields, diagnosed by polarised and unpolarised Zeeman signatures on line profiles, increase with decreasing rotation periods \citep{Vidotto2014, See2015, Reiners2022}. 

The main parameter that describes magnetic fields and activity of most M~dwarfs is found to be the Rossby number $Ro$, equal to the rotation period divided by the convective turnover time (e.g.\ \citealt{Noyes1984, Wright2018} and references therein), with magnetic fields and activity increasing with decreasing $Ro$ until saturation occurs at $Ro \sim 0.1$ and below. Whereas large-scale fields of M~dwarfs featuring $Ro<1$ have been extensively studied, very little is known about those of very slow rotators with $Ro \sim 1$ or larger. 

In this paper, we explore large-scale fields of a small sample of very slowly rotating M dwarfs, whose fields and rotation periods were inaccessible to optical instruments. 
The six M~dwarfs were observed with the SpectroPolarimetre InfraRouge (SPIRou), the near-infrared spectropolarimeter and velocimeter recently mounted on the Canada-France-Hawaii Telescope (CFHT), in the framework of the SPIRou Legacy Survey (SLS, \citealt{Donati2020}). The SLS is a Large Programme carried out with SPIRou at CFHT from early 2019 to mid 2022, with a 310-night time allocation spread over this period. The two main goals of the SLS are (i) the search for habitable Earth-like planets around very-low-mass stars and (ii) the study of low-mass star and planet formation in the presence of magnetic fields. Its long timeframe (of 7 semesters) enables us to investigate the temporal variability in the time series of the monitored M dwarfs, and in particular to independently estimate rotation periods of up to a few hundred days and to study the temporal evolution of their large-scale magnetic fields, (\citealt{Fouque2023, Bellotti2023, Donati2023}, hereinafter D23). 

In Section~\ref{Sec:Obs} we will present the details of the observations and targets. To analyse the magnetic field properties and to determine the stellar rotation period, we use different methods explained in Section~\ref{Sec:Model}, before we present our results for each M~dwarf individually in Sec.~\ref{Sec:Gl905}-\ref{Sec:Gl408}. We conclude and discuss our results in Section~\ref{Sec:Conclusions}.

\section{SPIRou Observations}
\label{Sec:Obs}

   \begin{table*}
    \caption[]{The stellar characteristics of our sample are from \cite{Cristofari2022} including spectral type, effective temperature $T_{\mrm{eff}}$, stellar mass, stellar radius $R_\star$, metallicity [M/H], $\log g$. The rotation period $\Prot$ is copied from D23. For the Rossby number $Ro = \Prot / \tau$, we use $\Prot$ (8th column) and determine the convective turnover time $\tau$ via \citet[Eq.~6]{Wright2018}. In the last column, we give the projected equatorial velocity $v_e \sin i = \frac{2 \pi R_\star}{\Prot} \sin i $, determined from $R_\star$ (5th column), $\Prot$ (8th column) and an assumed inclination of $i = 60^{\circ}$ between the stellar rotation axis and the line-of-sight. }
    \label{tab:MdwarfPropoerties}
    \begin{tabular}{lccccccccc}
        \hline
        \noalign{\smallskip}
 star & spectral type &  $T_{\mrm{eff}}$ & Mass & Radius & [M/H] & $\log g$ & $\Prot$ & $Ro$ & $v_e \sin i$  \\
  & & [K] & [\Msun\ ]& [\Rsun\ ]& & & [d] & & $\mrm{[km\ s^{-1}]}$  \\
  		\hline
  		\noalign{\smallskip}
Gl 905 & M5.0V & $3069\pm31$ & $0.15\pm0.02$ & $ 0.165\pm0.004$ & $0.05 \pm 0.11$ & $4.78\pm0.08$ &$114.3\pm2.8$ & 0.88 & $0.06$  \\
GJ 1289 & M4.5V& $3238 \pm 32$ & $0.21 \pm 0.02$ & $ 0.233 \pm 0.005$ & $0.05 \pm 0.10$ & $5.00 \pm 0.07$ & $73.66\pm0.92$ & 0.67 & $0.14$ \\
GJ 1151 & M4.5V & $3178 \pm 31$ & $0.17 \pm 0.02$ & $0.193 \pm 0.004$ & $-0.04 \pm 0.10$ & $4.71 \pm 0.06$ & $175.6\pm4.9$ & 1.45 &$0.05$ \\
GJ 1286 & M5.5V & $2961 \pm 33$ & $0.12 \pm 0.02$ & $0.142 \pm 0.004$ & $-0.23 \pm 0.10$ & $4.55 \pm 0.12$ & $178\pm15$ & 1.25 &$0.03$ \\
Gl 617B & M3V & $3525 \pm 31$ & $0.45 \pm 0.02$ & $0.460 \pm 0.008$ & $0.20 \pm 0.10$ & $4.84 \pm 0.06$ & $40.4\pm3.0$ & 0.77 &$0.50$ \\
Gl 408 & M2.5V & $3487 \pm 31$ & $0.38 \pm 0.02$ & $0.390 \pm 0.007$ & $-0.09 \pm 0.10$ & $4.79 \pm 0.05$ & $171.0\pm8.4$ & 2.68 & $0.10$ \\
        \noalign{\smallskip}
        \hline
    \end{tabular}
  \end{table*}
  
We analyse here a total of 986 circularly polarised spectra collected with SPIRou. The spectra span a wavelength range of 0.95--2.5$\mu$m in the near-infrared with a resolving power of $R=70\,000$. Further details about SPIRou and its spectropolarimetric capabilities can be found in \cite{Donati2020}. To process the data, we used the new version of \textsc{Libre~ESpRIT}, i.e., the nominal reduction pipeline of ESPaDOnS at CFHT optimised for spectropolarimetry and specifically adapted for SPIRou \citep{Donati2020}. 

We applied Least-Squares~Deconvolution (LSD, \citealt{Donati1997b}) to all reduced unpolarised (Stokes~$I$) and circularly-polarised (Stokes~$V$) spectra using a M3 mask constructed from outputs of the VALD-3 database \citep{Ryabchikova2015} assuming \cc{a temperature $T_{\mrm{eff}}=3500$\,K,} a logarithmic surface gravity $\log g = 5$ and a solar metallicity $[\mrm{M/H}]$. \cc{Although our 6 stars do not have the exact same atmospheric properties (see Table~\ref{tab:MdwarfPropoerties}), we nonetheless used a single mask, whose impact on the LSD results is only marginal, especially in the near-infrared domain where the synthetic spectra only provide a rough match to observed ones (e.g., \citealt{Cristofari2022}). Besides, the mask we chose corresponds to the coolest atmospheric model available by default on the VALD-3 data base.} We have selected the atomic lines with a relative depth greater than 10\% and resulting in 575 lines for the mask. For further details see D23, Sec.~2. 
The ephemeris used to calculate the phase and rotation cycle in this paper are summarised in Tab.~\ref{tab:Ephemeris} for all targets of our sample.

The six M~dwarfs studied in this paper are part of the 43~star sample analysed by \cite{Fouque2023} and D23. These two papers aimed at determining, whenever possible, the rotation periods of the sample targets, by applying quasi-periodic Gaussain-Process-Regression (GPR) to times series of their longitudinal fields $B_\ell$, i.e., the line-of-sight-projected component of the vector magnetic field averaged over the visible stellar hemisphere. All six stars of our sample have well identified rotation periods according to D23, whereas \cite{Fouque2023}, using data reduced with the nominal SPIRou pipeline APERO (optimised for RV precision, \citealt{Cook2022}) was able to derive rotation periods for four of them (consistent with those of D23).   

The key stellar parameters of our sample are presented in Table~\ref{tab:MdwarfPropoerties} and are mostly extracted from \cite{Cristofari2022} who studied the fundamental parameters of these stars by analysing their SPIRou spectra.

\section{Model description}
\label{Sec:Model}

To analyse the magnetic properties of our six slowly rotating M~dwarfs, we use both the method based on Principal Component Analysis (PCA) recently proposed by \cite{Lehmann2022}, as well as Zeeman-Doppler Imaging (ZDI, \citealt{DonatiBrown1997, Donati2006}), applied to our set of LSD Stokes~$V$ profiles.

\subsection{PCA analysis of the LSD Stokes~$V$ profiles}
\label{SubSec:PCA}

\cite{Lehmann2022} proposed a method to retrieve key information about the large-scale stellar magnetic field directly from time series of Stokes~$V$ LSD profiles without the need of an elaborate model of the field topology or several stellar parameters (e.g., the projected equatorial velocity $v_e \sin i$ or the inclination of the stellar rotation axis $i$). The method provides information about the axisymmetry, the poloidal / toroidal fraction of the axisymmetric component, the field complexity and their evolution with time.

One first determines the mean profile of the whole Stokes~$V$ time series, which stores information about the axisymmetric component of the large-scale field. In contrast to \cite{Lehmann2022}, we use the weighted mean profile providing better results for time series such as ours, where all LSD profiles do not have the same SNR. The averaged Stokes $V$ LSD profile can be decomposed into its antisymmetric component (with respect to the line centre), which is related to the poloidal component of the axisymmetric large-scale field, and its symmetric component (with respect to the line centre), which is probing the toroidal component of the axisymmetric large-scale field (see e.g.\ Fig.~\ref{Fig:Gl905_PCA}a). 

To evaluate the non-axisymmetric field, we subtract the weighted mean profile (taken over all seasons) from the Stokes~$V$ time series removing the signal of axisymmetric field. The resulting mean-subtracted Stokes~$V$ profiles store now the information about the non-axisymmetric component of the large-scale field and are analysed using a weighted PCA \citep{Delchambre2015}
returning eigenvectors and coefficients. \cc{In the weighted PCA, the Stokes~$V$ profiles are weighted by the squares of their SNRs, taking into account the different noise levels. Thus, for the long time series analysed in this paper, with uneven SNR over the 7 semesters, the weighted PCA gives better results than a classical PCA where all profiles are treated equally.} The PCA coefficients, and in particular their fluctuation with time, can reveal the complexity of the large-scale field and its long-term temporal evolution. 

Given the long time range of the SLS data, we further split the Stokes~$V$ time series at successive observing seasons into 2--3 seasons per star. To evaluate the evolution of the axisymmetric field from season to season, we determine the weighted mean profiles per season and compare them to one another (see e.g.\ Fig.~\ref{Fig:Gl905_PCA}c left column). To study the evolution of the non-axisymmetric field, we compare the coefficients of the different seasons (see e.g.\ Fig.~\ref{Fig:Gl905_PCA}c middle and right column). We caution that the coefficients are derived from the weighted PCA of the mean-subtracted Stokes~$V$ time series using the weighted mean profile computed across all seasons (e.g.\ Fig.~\ref{Fig:Gl905_PCA}a) and not the weighted mean profile of each individual season (e.g.\ Fig.~\ref{Fig:Gl905_PCA}c left column). The usage of the weighted mean Stokes~$V$ profiles of each individual season would prevent a direct comparison of the different seasons. For example, it would centre the coefficients for each season, so that we will lose the information if the non-axisymmetric field becomes more or less positive / negative from one season to another, and also the amplitudes of the coefficients are no longer comparable. 
Further information about the PCA method can be found in \cite{Lehmann2022}.

In addition, \cite{Lehmann2022} showed that the sensitivity of the PCA method for toroidal fields decreases for low $v_e \sin i$. As all our stars have $v_e \sin i \leq 0.5\,\kms$, we are likely to miss large-scale toroidal fields. We provide a typical 1$\sigma$ error bar for the axisymmetric toroidal field for each star and each observing season of our sample.

\subsection{Gaussian Process modelling of the time series}
\label{SubSec:GP}

We analyse the temporal evolution of the M~dwarf's topology with the help of the PCA determined coefficients of the mean-substracted Stokes~$V$ time series. For our slowly rotating stars, most of the time only the first eigenvector and therefore only the first coefficient shows a signal. To directly compare the temporal evolution of the coefficients with the result from the longitudinal field $B_\ell$ (presented by D23) for the individual stars, we scale and translate the first coefficient, which we call $c_1$, using a linear model (scaling factor and offset), that minimises the distance between the first coefficients and $B_\ell$ taking into account the measurement errors on $B_\ell$.

We can re-determine the stellar rotation period $\Prot$ of our six M~dwarfs using a quasi-periodic (QP) GPR fit to $c_1$ allowing us a direct comparison with the QP GPR results of $B_\ell$ presented by D23. 

In contrast to D23, we use the python model presented by \cite{Martioli2022} based on \href{http://dfm.io/george/current/}{\textsc{george}} \citep{Ambikasaran2014}. Our adapted covariance function (or kernel) is given by
\begin{equation}
k(t_{ij}) = \alpha^2 \exp \left[- \frac{t^2_{ij}}{2l^2} - \frac{1}{2\beta^2} \sin^2 \left( \frac{\pi t_{ij}}{\Prot} \right) \right],
\end{equation}
where $t_{ij} = t_i-t_j$ is the time difference between the observations $i$ and $j$, $\alpha$ is the amplitude of the Gaussian Process (GP), $l$ is the decay time describing the typical time-scale on which the modulation pattern evolves, $\beta$ is the smoothing factor indicating the harmonic complexity of the QP modulation (lower values indicating higher harmonic complexity) and $\Prot$ is our new estimate of the stellar rotation period. 
The GP model parameters are fitted by maximising the following likelihood function $\mathcal{L}$ using the python package \textsc{scipy.optimize}:
\begin{equation}
\log \mathcal{L} = -\frac{1}{2} \left( N \log 2\pi + \log | \mathbf{K} + \mathbf{\Sigma} + \mathbf{S}| + \vect{y}^T (\mathbf{K} + \mathbf{\Sigma} + \mathbf{S})^{-1} \vect{y} \right),
\end{equation}
where $\mathbf{K}$ is the QP kernel covariance matrix, $\mathbf{\Sigma}$ the
diagonal variance matrix of $c_1$, $\mathbf{S}$ the diagonal matrix $\sigma^2 \mathbf{I}$ (with $\sigma$ an added amount of uncorrelated white noise \citep{Angus2018} and $\mathbf{I}$ the identity matrix), $N$ the number of observations and $\vect{y}$ corresponds to $c_1$.
The posterior distribution of the free parameters is sampled using a Bayesian Markov Chain Monte Carlo (MCMC) framework applying the package \textsc{emcee} \citep{emcee}. For the MCMC, we use 50 walkers, 200 burn-in samples and 1000 samples. Tab.~\ref{tab:GPFitParams_c1} provides a summary of the results for all $c_1$ GPR fits in this study. For three stars, the decay time $l$ was fixed as in D23 (see Tab.~\ref{tab:GPFitParams_c1}). Information about the assumed prior distributions and the posterior distributions for each parameter and each GPR fit can be found in appendix \ref{Sec:AddGPFigures}.

Furthermore, we applied the above GP model to the $B_\ell$ values of D23 (see appendix~\ref{App:GPComp}). This allows a direct comparison of the GP results for the same $B_\ell$ dataset with our GP routine and the GP routine used by D23, as well as a comparison of the GP results for $c_1$ and $B_\ell$ obtained by the same GP routine. In general, we find that $c_1$ has lower RMS, often shows lower $\sigma$ values and provides smaller errors for $\Prot$ when the topology is not highly axisymmetric.

\begin{table*}
    \caption[]{Summary of the best-fitting parameters of the QP GPR fits applied to $c_1$ for the six M~dwarfs in our sample, where rms is the root-mean-square of the residuals and $\chi^2_r$ the reduced chi-square value of the GPR fit. Fixed parameters are shown in italics. A comparison with the results of the GPR fit to the $B_\ell$ data, and with those of D23, is given in Table~\ref{tab:GPFitParams}.}
    \label{tab:GPFitParams_c1}
    \setlength{\extrarowheight}{.3em}
    \begin{tabular}{lccccccc}
        \hline
        \noalign{\smallskip}
 star &  rotation period & decay time & smoothing factor & amplitude & white noise & rms & $\chi^2_{\rm r}$  \\
 & $P_{\rm rot}$ [d] & $l$ [d] & $\beta$ & $\alpha$ [G] & $\sigma$ [G] & [G] & \\
        \noalign{\smallskip}
        \hline
        \noalign{\smallskip}
Gl~905 &  $111.7^{+3.0}_{-3.2}$ & $133^{+18}_{-22}$ & $0.50^{+0.09}_{-0.07}$ & $12.9^{+3.1}_{-2.1}$ & $0.6^{+0.5}_{-0.4}$ & 3.8 & 0.79 \\
        \noalign{\smallskip}
GJ~1289 &  $75.62^{+0.85}_{-0.79}$ & $129^{+26}_{-24}$ & $0.40\pm 0.05$ & $41.9^{+7.9}_{-6.0}$ &  $1.2^{+1.1}_{-0.8}$ & 6.9 & 0.72 \\
        \noalign{\smallskip}
GJ~1151  & $175.8^{+3.2}_{-3.4}$ & \textit{300} & $0.40^{+0.10}_{-0.09}$ & $12.7^{+3.2}_{-2.4}$ &  $3.3\pm0.7$ & 5.8 & 0.99 \\
        \noalign{\smallskip}
GJ~1286  & $186.8^{+9.5}_{-5.8}$ & \textit{300} & $0.23^{+0.05}_{-0.04}$ & $18.0^{+4.3}_{-3.1}$  & $2.7^{+1.5}_{-1.7}$ & 7.5 & 1.02 \\
        \noalign{\smallskip}
Gl~617B  & $37.8^{+8.5}_{-2.6}$ & $35^{+8}_{-4}$ & $0.47^{+0.09}_{-0.05}$ & $5.9^{+1.2}_{-0.8}$ & $0.7^{+0.6}_{-0.5}$ & 2.2 & 0.66 \\
        \noalign{\smallskip}
Gl~408 & $175^{+12}_{-14}$ & \textit{200} & $0.18^{+0.10}_{-0.05}$ & $4.2^{+0.9}_{-0.8}$ & $2.2^{+0.5}_{-0.6}$ & 3.8 & 1.19 \\
          \noalign{\smallskip}
        \hline
    \end{tabular}
  \end{table*}

\subsection{Zeeman-Doppler-Imaging}

We determined the large-scale vector magnetic field at the surface of the six M~dwarfs, for each season, using ZDI. ZDI iteratively builds up the large-scale magnetic field and compares the synthetic Stokes profiles corresponding to the current magnetic map, assuming solid body rotation, with the observed Stokes profiles until it converges on the requested reduced chi-square value $\chi^2_r$ between the observed and synthetic data.  The problem being ill-posed, i.e., with an infinite number of solutions featuring the requested agreement to the data, ZDI chooses the one with maximum entropy, (i.e., minimum information in our case, \citealt{Skilling1984}). The surface magnetic field is described with a spherical harmonics expansion given in \cite{Donati2006}, where the $\alpha_{\ell,m}$ and $\beta_{\ell,m}$ coefficients of the poloidal component are modified as indicated in \citet[Eq.~B1]{Lehmann2022}. To compute synthetic Stokes profiles, the stellar surface is decomposed into a grid of 1000 cells. For each cell the local Stokes~$V$ and $I$ profiles are determined using Unno-Rachkovsky's analytical solution to the equations of the polarised radiative transfer in a plane-parallel Milne-Eddington atmosphere \citep{Landi2004}. The Stokes profiles are integrated over the visible hemisphere for each observing phase applying a mean wavelength of 1700\,nm and a Land\'{e} factor of 1.2.

For the slowly rotating M~dwarfs of our sample, we see no obvious variations in the Stokes~$I$ LSD profiles beyond those attributable to radial velocity variations, so that we only use the Stokes~$V$ profiles for determining the magnetic field map via ZDI. Nevertheless, we make sure that the synthetic Stokes~$I$ profiles computed with ZDI agree well with the averaged observed Stokes~$I$ profile, especially in terms of width and depth. We assumed a fraction $f_V$ of each grid cell, which actually contributes to the Stokes~$V$ profile. This fraction $f_V$ is called filling factor of the large-scale field and is set equally to all cells, see also \cite{Morin2010, Kochukhov2021}. For each star, we set $f_V = 0.1$ motivated by the results of \cite{Klein2021ProxCen} for the slow rotator Proxima~Centauri and the results of \cite{Moutou2017} for the SPIRou sample. We confirm the choice of $f_V = 0.1$ by finding lower $\chi^2_r$ values with $f_V = 0.1$ compared to $f_V=1$ for each season of the different M~dwarfs. The filling factor for the Stokes~$I$ profiles is set to $f_I = 1.0$ in consistency with the literature, \citep{Morin2010, Kochukhov2021}.
The $v_e \sin i$ of our sample is $\leq 0.5\,\kms$ (see Tab.~\ref{tab:MdwarfPropoerties}) and prevents us from reliably determining the inclination of the stellar rotation axis for the M~dwarfs, so that we set the inclination to $60^\circ$ for all M~dwarfs. This is motivated by the steep modulation patterns seen for $B_\ell$ and $c_1$ for most targets, that can not be obtained for pole-on viewed stars. Another reason is that higher values of $i$ are intrinsically more likely than smaller ones. We restrict the spherical harmonics of the ZDI reconstructions to $\ell = 7$, as we see little magnetic energy stored in $\ell \sim 5-7$.

\section{Gl~905}
\label{Sec:Gl905}

The first star in our sample is the M5.5V dwarf Gl~905 (HH~And, Ross~248) with a mass of $0.15\pm0.02\,$\Msun\ , \cc{\citep{Cristofari2022}}. For our analysis we use 219 Stokes~$IV$ LSD profiles observed with SPIRou between 2019 Apr and 2022 June and split the data in three seasons (2019 Apr -- Dec, 2020 May -- 2021 Jan, 2021 June -- 2022 Jan) for the per-season analysis. The 15 profiles collected in 2022 June at the beginning of a new season, only covering 9\% of a rotation cycle, were left out of the per-season analysis.

\subsection{PCA analysis of Gl~905}

\begin{figure}
	\raggedright \textbf{a.} \hspace{2.7cm} \textbf{b.} \\
	\centering
	\includegraphics[width=0.34\columnwidth, trim={0 0 0 0}, clip]{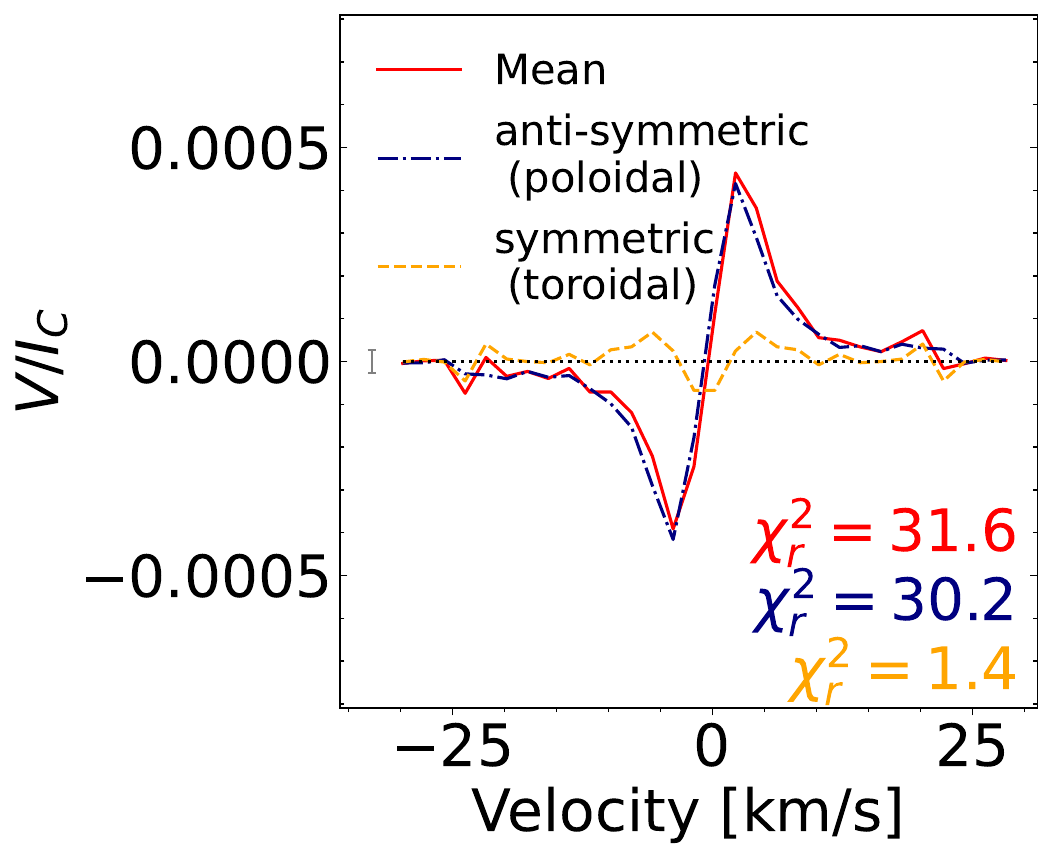}
	\includegraphics[width=0.64\columnwidth, trim={0 400 430 0}, clip]{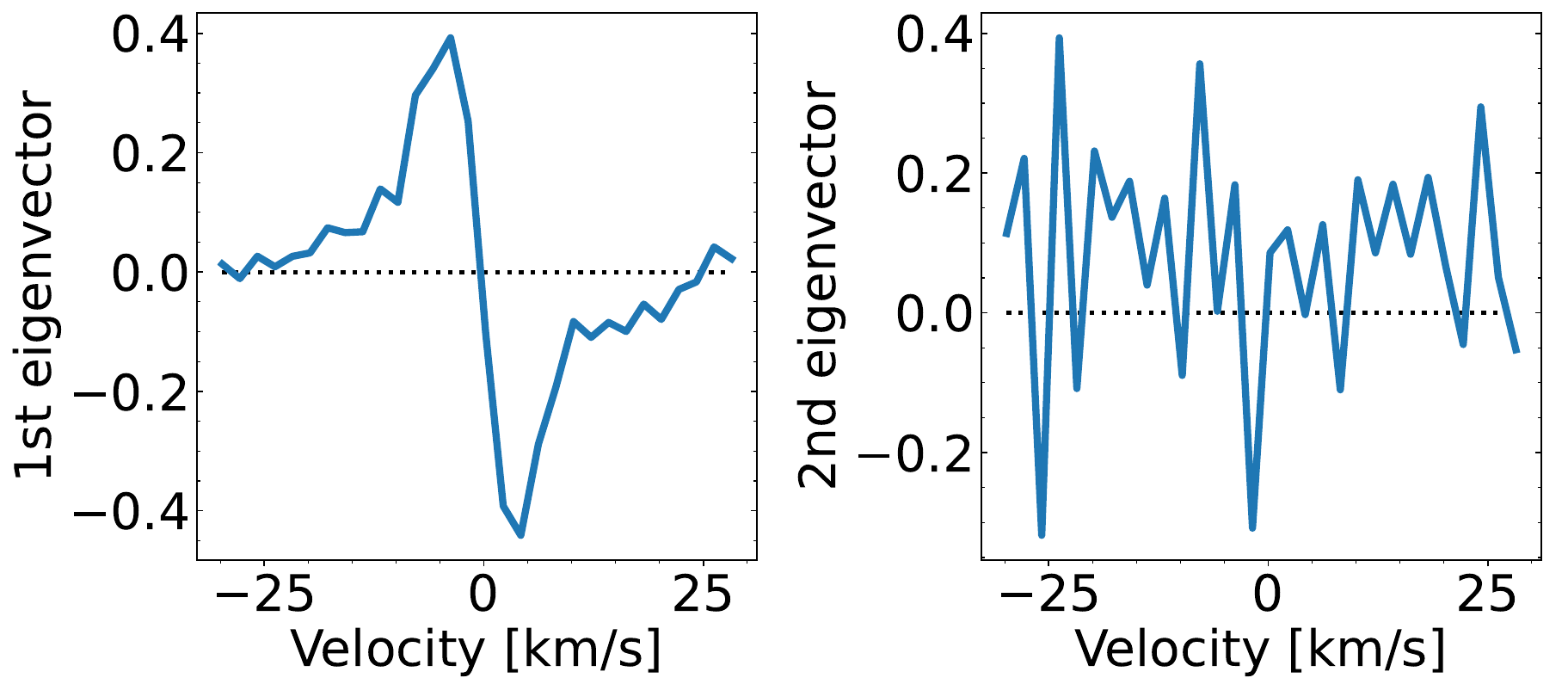}\\ 
	\rule{7cm}{0.3mm}\\
	\raggedright \textbf{c.} \\
	\centering
	\includegraphics[width=0.35\columnwidth, trim={0 0 0 0}, clip]{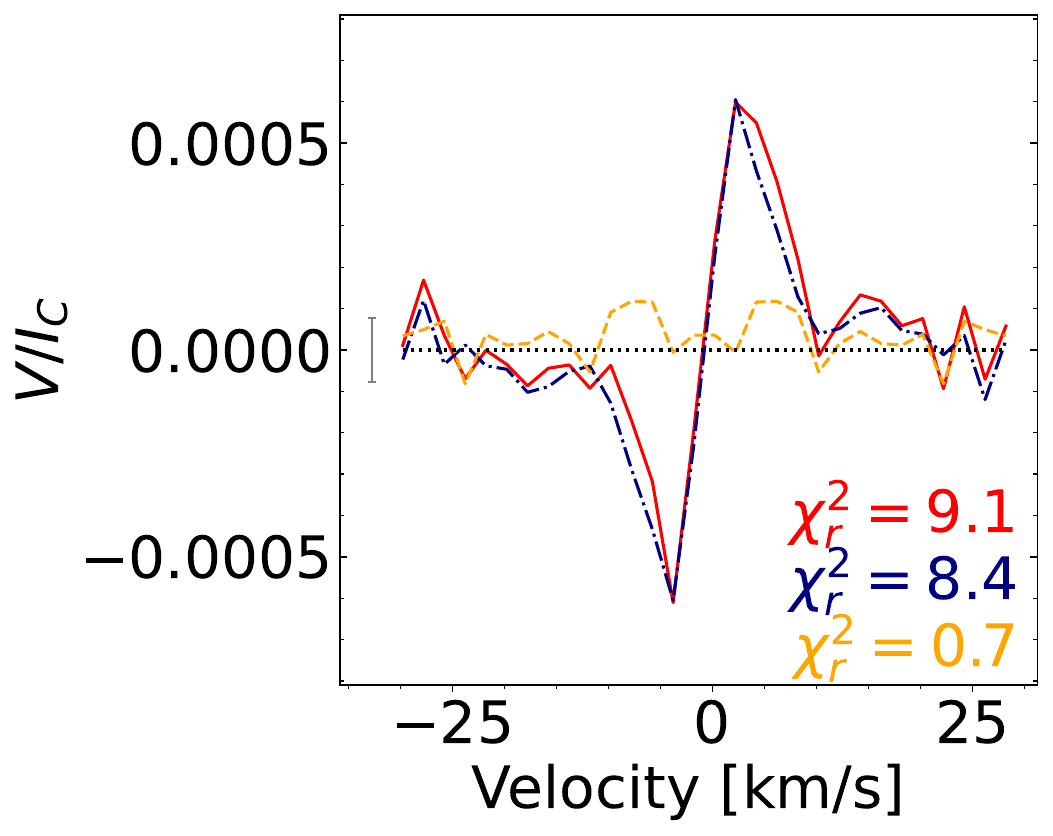}
	\includegraphics[width=0.63\columnwidth, trim={30 400 445 0}, clip]{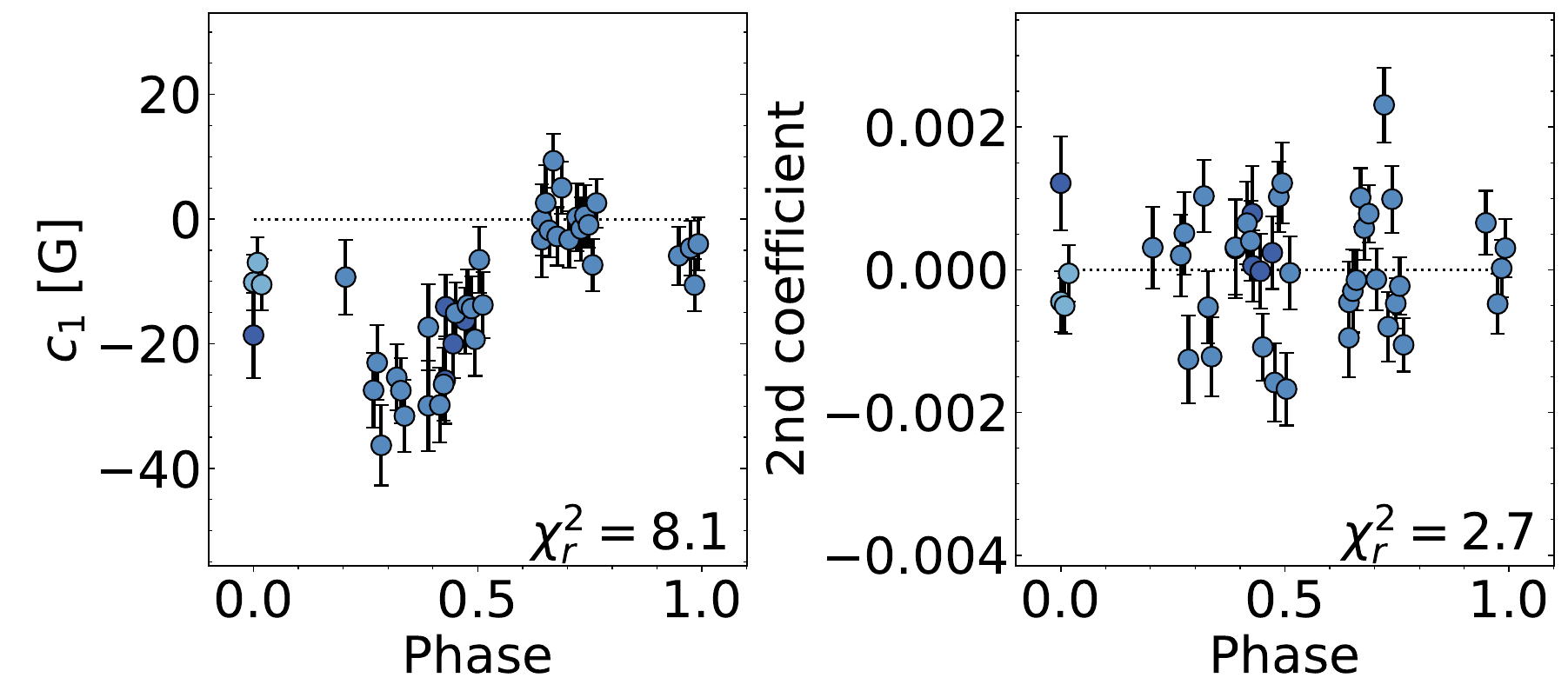}\\
		\includegraphics[width=0.35\columnwidth, trim={0 0 0 0}, clip]{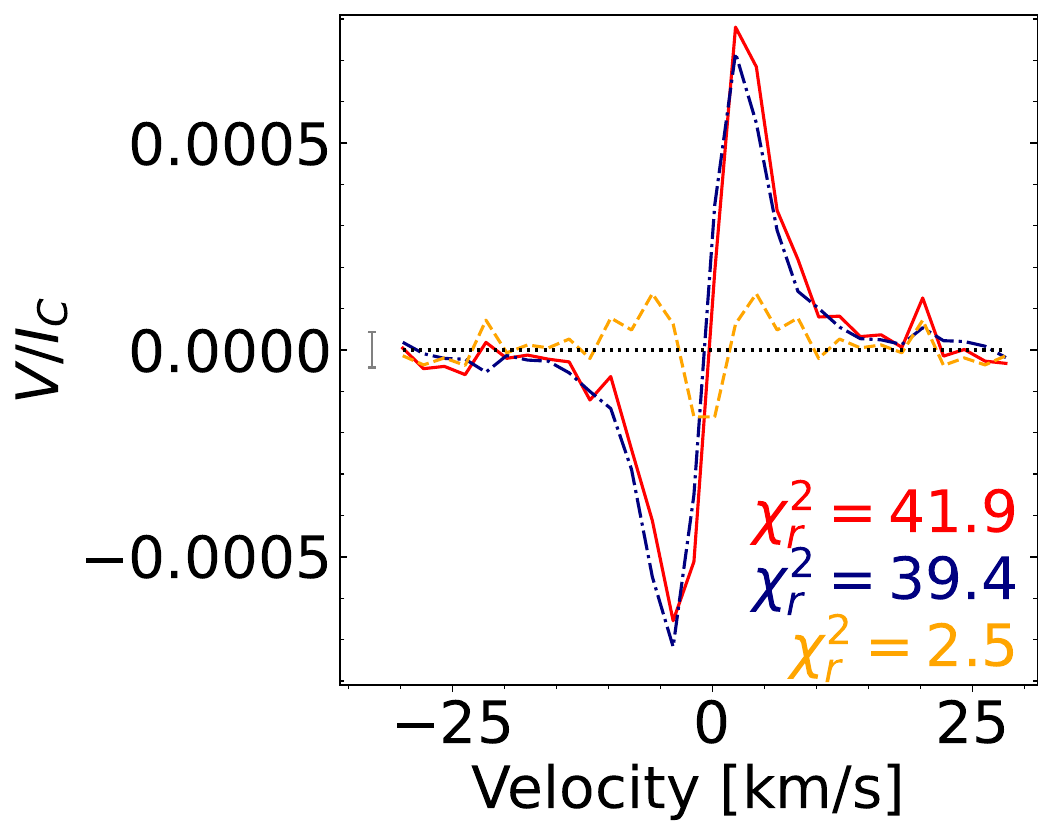}
	\includegraphics[width=0.63\columnwidth, trim={30 400 445 0}, clip]{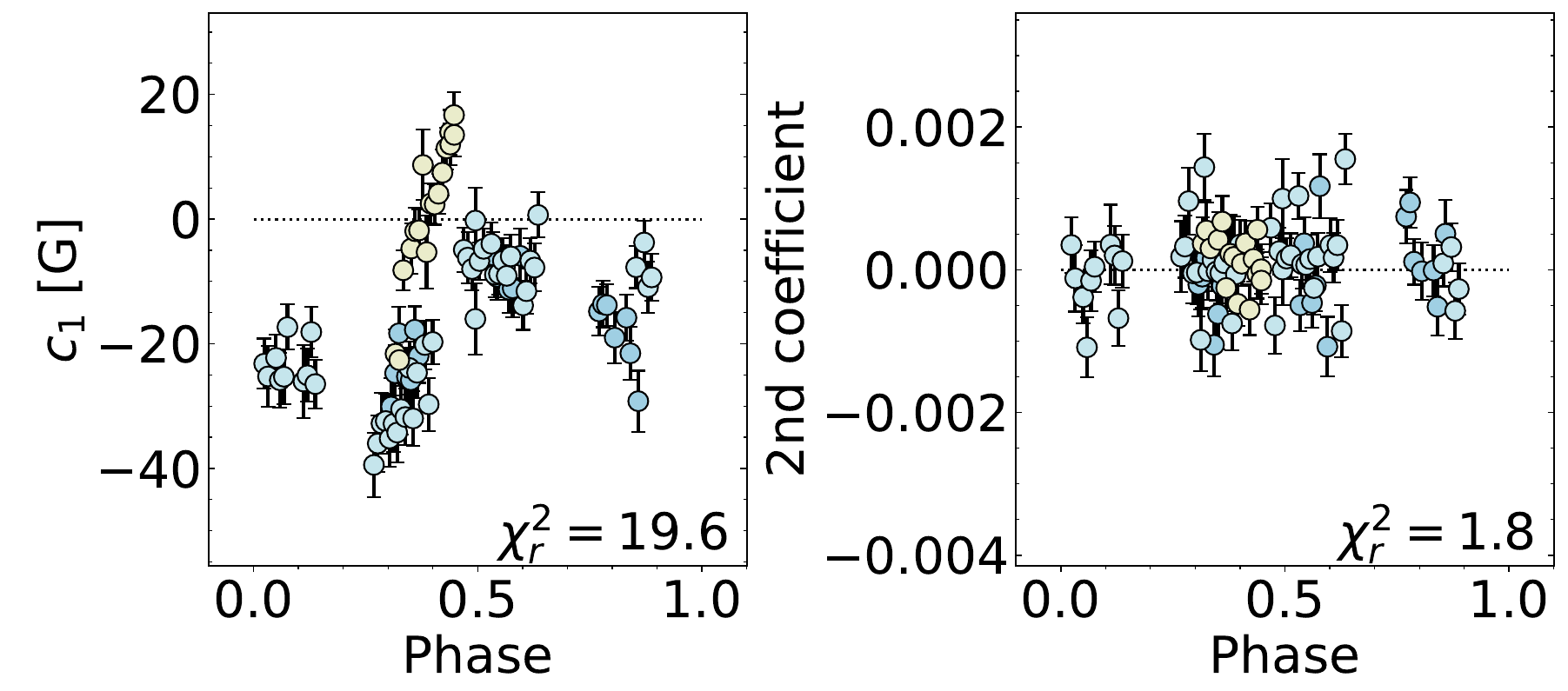}\\
		\includegraphics[width=0.35\columnwidth, trim={0 0 0 0}, clip]{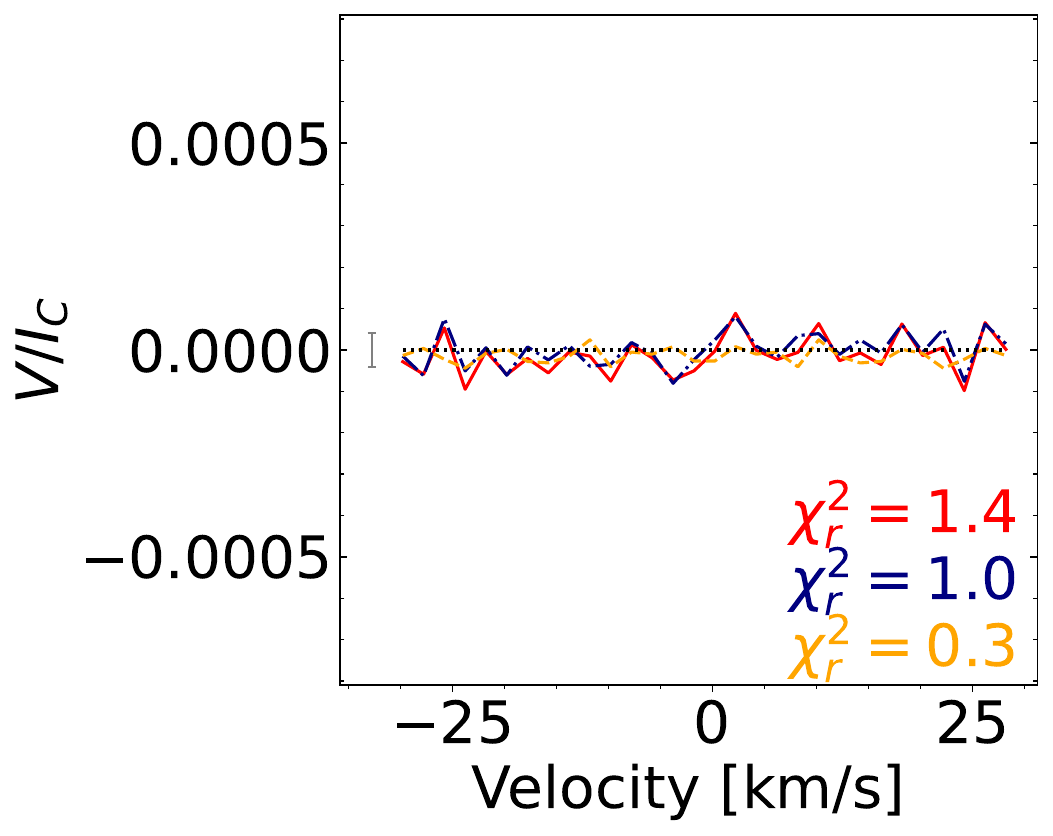}
	\includegraphics[width=0.63\columnwidth, trim={30 400 445 0}, clip]{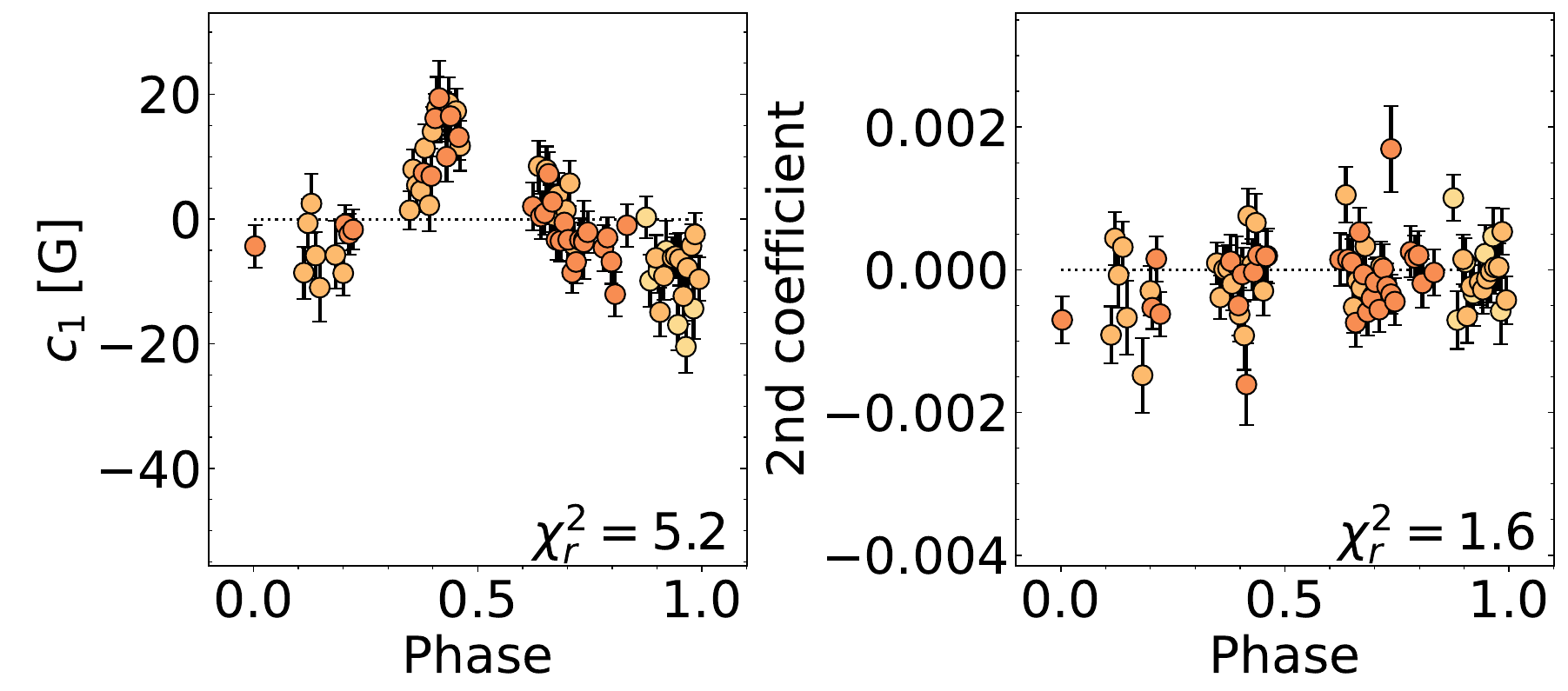}\\
    \caption{The PCA analysis for Gl~905. \textbf{a.} The mean profile (red) for all observations and its decomposition in the antisymmetric (blue dashed) and symmetric (yellow dotted) components (with respect to the line centre) related to the poloidal and toroidal axisymmetric field, respectively. This mean profile is used to determine the mean-subtracted Stokes~$V$ profiles to which we apply PCA, yielding the eigenvectors and coefficients shown in panels b and c. \textbf{b.} The first two eigenvectors of the mean-subtracted Stokes~$V$ profiles. \textbf{c.} The mean profile (left column), $c_1$ (the scaled and translated first PCA coefficient introduced in Sec.~\ref{SubSec:GP}, middle column) and the coefficients of the second eigenvector (right column) for each season (one season per row). The mean profiles of the individual seasons are plotted in the same format as above. The coefficients are colour-coded by rotation cycle.} 
    \label{Fig:Gl905_PCA}
\end{figure}

First, we investigate the large-scale field topology using PCA \citep{Lehmann2022}. The weighted mean profile of all Stokes~$V$ profiles is antisymmetric with respect to the line centre indicating a poloidal axisymmetric large-scale field (see Fig.~\ref{Fig:Gl905_PCA}a). The symmetric component of Gl~905's mean profile exceeds the noise level ($\chi^2_r = 1.4$) but is likely due to an uneven phase coverage in the season 2020/21 (see  Sec.~\ref{Subsec:Gl905ZDI}). 
In Fig.~\ref{Fig:Gl905_PCA}b, we show the first two eigenvectors of the mean-subtracted Stokes~$V$ profiles allowing the analysis of the non-axisymmetric field. Only the first eigenvector shows an antisymmetric signal with respect to the line centre. All other eigenvectors display noise. 

\begin{figure}
	%
	%
	\centering
	\includegraphics[width=\columnwidth, trim={0 0 0 0}, clip]{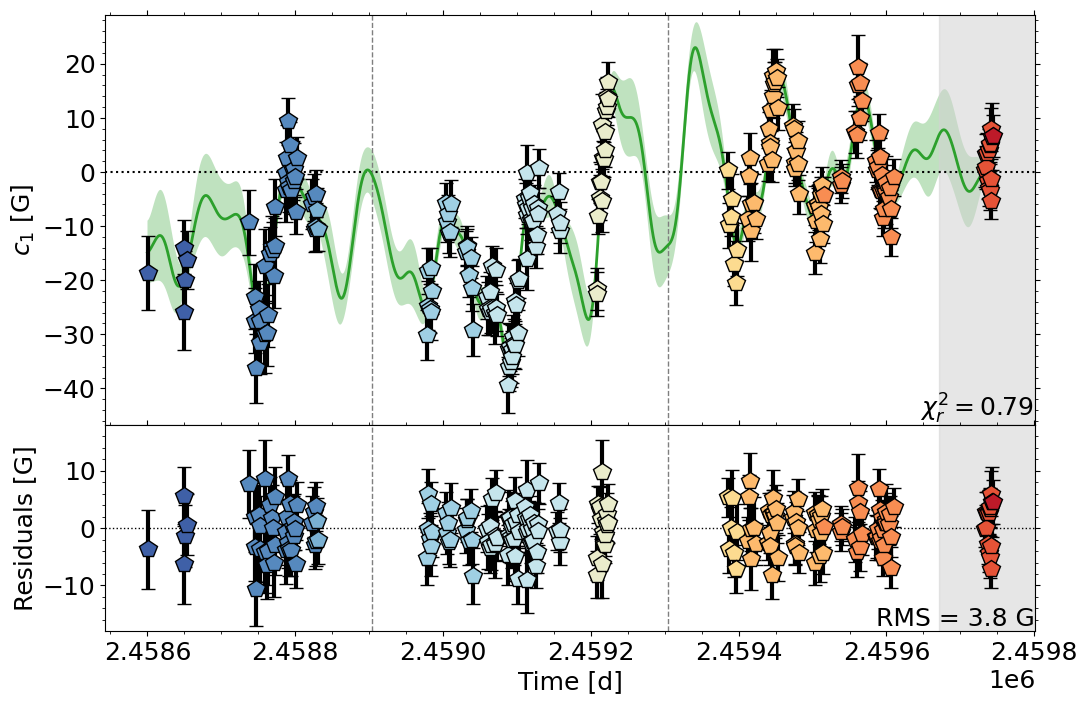}\\ 
	\includegraphics[width=\columnwidth, trim={0 0 0 0}, clip]{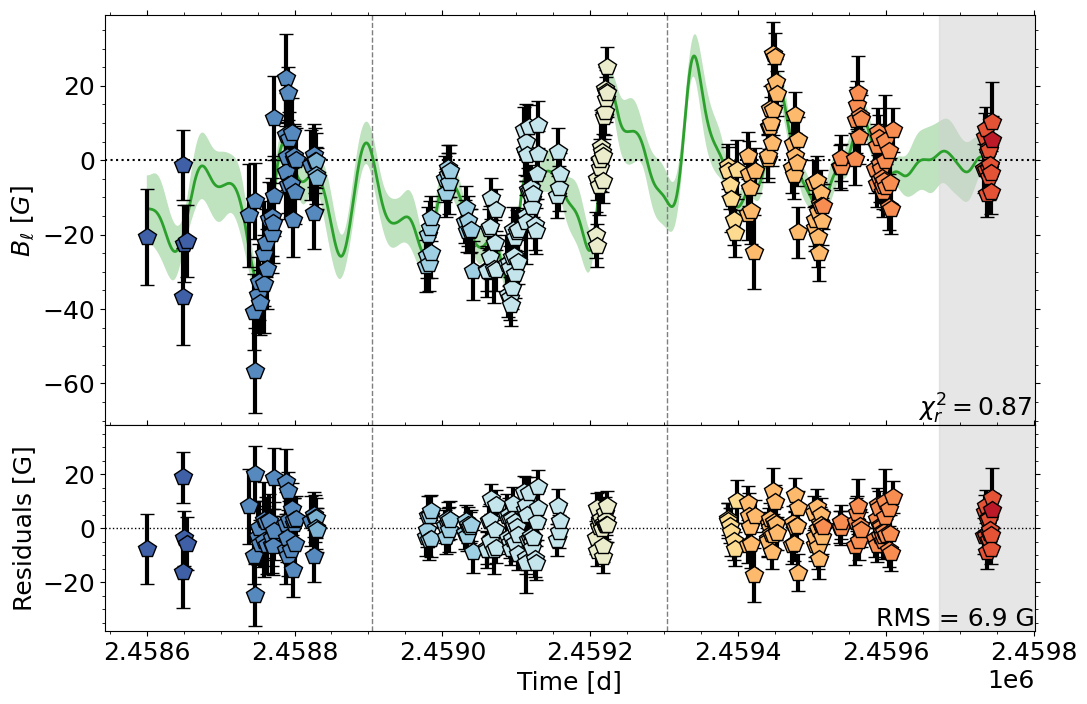}\\ 
    \caption{Temporal variations of $c_1$ (top) and longitudinal field $B_\ell$ (bottom) for Gl~905. We show the QP GPR fit and its 1$\sigma$ area as green shaded region in the top panel and the residuals in the bottom panel for both variables. The plot symbols are colour-coded by rotation cycle. The vertical grey lines separate the analysed seasons. The grey shaded region indicates a season for which not enough data were available for a reliable PCA and ZDI analysis.}
    \label{Fig:Gl905_CoeffvsTime}
\end{figure}

In Fig.~\ref{Fig:Gl905_CoeffvsTime} top, we plot the temporal evolution of $c_1$, which appears very similar to the temporal evolution of $B_\ell$ determined with our GP model (see Fig.~\ref{Fig:Gl905_CoeffvsTime} bottom) and to D23's results (see Fig.~A12 middle in D23). When only one eigenvector is significant, as it is the case here for Gl~905, the $c_1$ curve mimics that of $B_\ell$ \citep{Lehmann2022}.

Our QP GPR model of $c_1$ finds a rotation period of $\Prot = 111.7^{+3.0}_{-3.2}\,\dy$ and decay time of $l = 133^{+18}_{-22}\,\dy$ very similar to the values derived by D23 ($\Prot = 114.3\pm2.8\,\dy$ and $l = 129^{+25}_{-21}\,\dy$) and \cite{Fouque2023} ($\Prot = 109.5^{+4.9}_{-5.4}\,\dy$ and $l = 149^{+26}_{-25}\,\dy$) and also consistent with our GP fit of D23's $B_\ell$ values ($\Prot = 114.4^{+3.5}_{-2.4}\,\dy$ and $l = 130^{+25}_{-32}\,\dy$), see Tab.~\ref{tab:GPFitParams_c1} and \ref{tab:GPFitParams}. 
For consistency, we use the rotation periods found by D23 to determine the rotation phase (see Tab.~\ref{tab:Ephemeris}) and to model the ZDI maps for all six stars in our sample (see Tab.~\ref{tab:MdwarfPropoerties}).

In Fig.~\ref{Fig:Gl905_PCA}c, we plot the mean profiles (left column) and the phase-folded coefficient curves colour-coded by rotation phase (middle and right columns) for the three seasons (one season per row).
They exhibit large changes in the large-scale field topology from season to season, allowing us to draw first conclusions about the field evolution of Gl~905. We recall that the coefficients for all three seasons are computed from the mean-subtracted Stokes~$V$ profiles, using the weighted mean derived from the full data set, and not from the profiles of each season. The same applies to the five M~dwarfs discussed in Secs.~\ref{Sec:GJ1289}-\ref{Sec:Gl408}.  

The mean profiles of the first two seasons (2019 and 2020/21) are antisymmetric with respect to the line centre and indicate a mostly poloidal axisymmetric field although the symmetric component is larger for 2020/21 (see Fig.~\ref{Fig:Gl905_PCA}c). This may reflect an increasing toroidal field but is more likely due to the uneven phase coverage of this season (with more than 75\% of the observations concentrating between phase 0.3 and 0.75). 

For the first season 2019, $c_1$ features a roughly sinusoidal behaviour indicating a mainly dipolar configuration. For 2020/21, $c_1$ appears more complex than for 2019 implying that the field becomes more complex, too. For the last season 2021/22, the topology changes more drastically: the mean profile is close to zero indicating a much lower axisymmetric component than before. The phase at which $c_1$ reaches its maximum is shifted, with $c_1$ being more positive now, while it is mainly negative before. Considering the sign of the mean profile and the eigenvector, this suggest that the main polarity of the large-scale field is evolving from a predominantly negative polarity to a positive polarity.
We can conclude from the PCA analysis, that the large-scale field topology becomes more complex from 2019 to 2020/21 before it becomes mostly non-axisymmetric and possibly initiates a polarity reversal.

\subsection{ZDI reconstructions of Gl~905} 
\label{Subsec:Gl905ZDI}

\begin{figure}
	%
	%
\centering
\begin{minipage}{0.32\columnwidth}
\centering
\includegraphics[height=0.85\columnwidth, angle=270, trim={140 0 0 29}, clip]{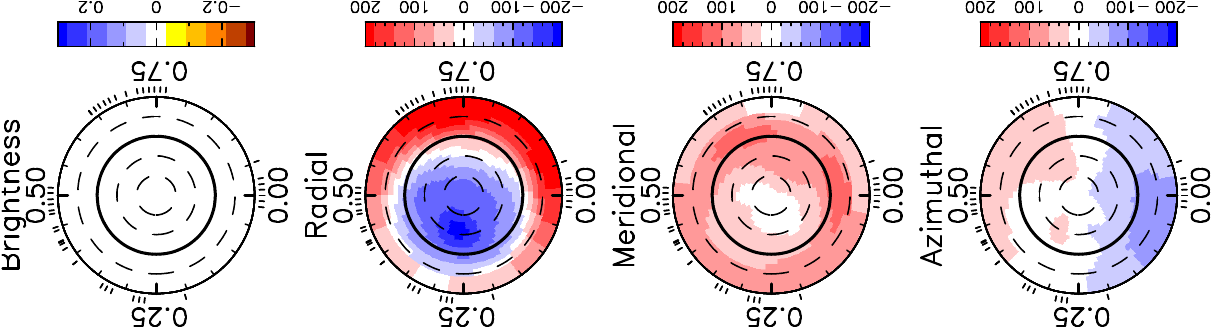} 
\end{minipage}
\begin{minipage}{0.32\columnwidth}
\centering
\includegraphics[height=0.85\columnwidth, angle=270, trim={140 0 0 29}, clip]{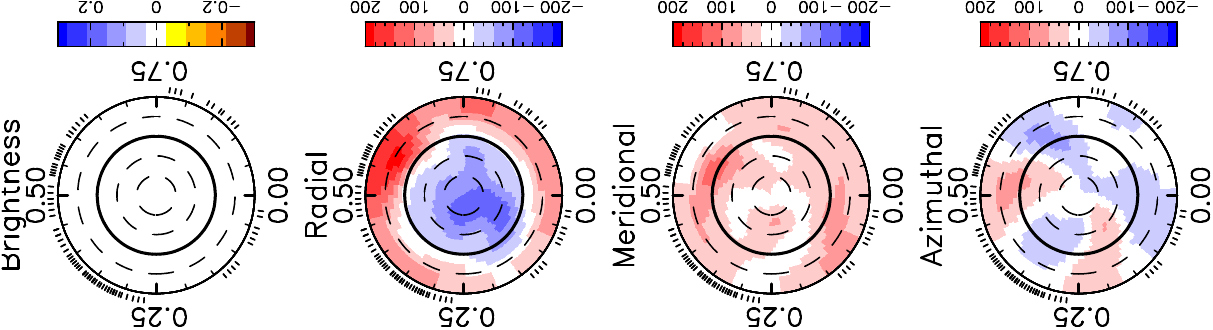} 
\end{minipage}
\begin{minipage}{0.32\columnwidth}
\centering
\includegraphics[height=0.85\columnwidth, angle=270, trim={140 0 0 29}, clip]{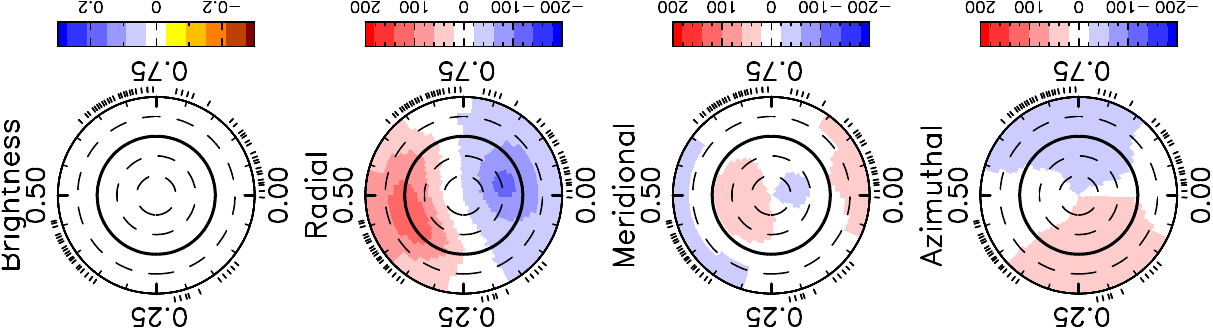} 
\end{minipage}
\includegraphics[width=0.3\columnwidth, angle=180, trim={460 130 2 0}, clip]{Figures/Gl905_ZDIMap_JFDLSD_epo3_v23.pdf}

\vspace*{2mm}
\includegraphics[width=0.95\columnwidth, clip]{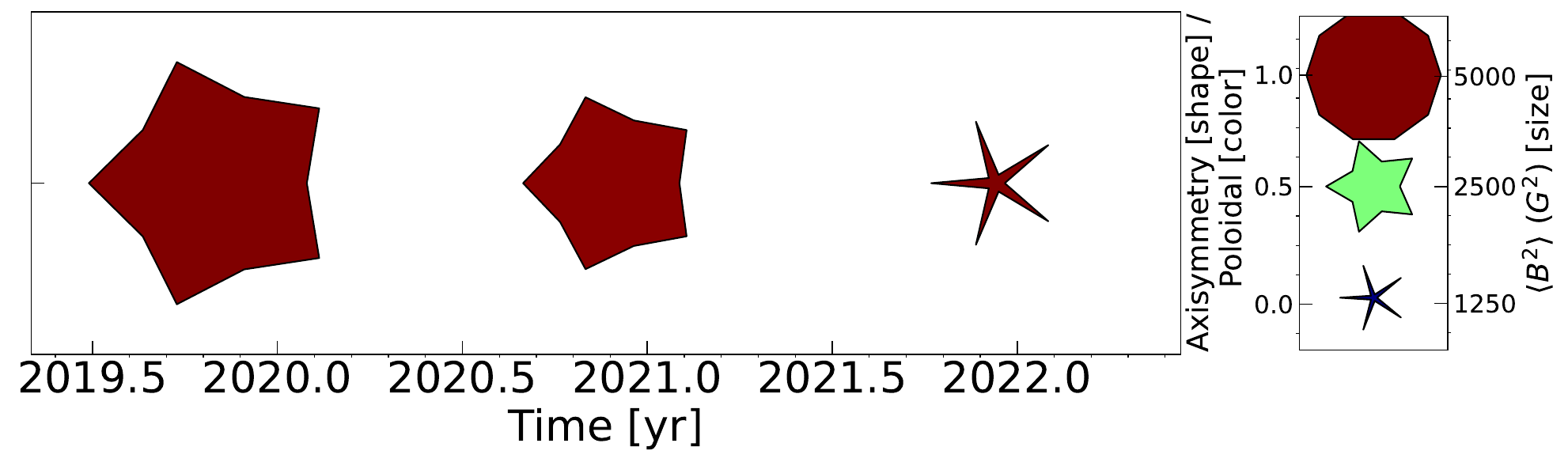}
    \caption{The magnetic field maps of Gl~905 shown in a flattened polar view for the radial (top row), meridional (middle row) and azimuthal component (bottom row). In each plot, the visible north pole is in the centre, the thick line depicts the equator and the dashed line the latitudes in $30^\circ$ step. The ticks outside the plot illustrate the observing phases. The different seasons are shown next to each other (one season per column). The colour bar below the third row is used for all maps and indicates the magnetic field strength in G. The bottom panel summarises the main characteristics of the large-scale field of Gl~905 and its evolution  with time. For each season, the symbol size indicates the magnetic energy, the symbol shape the fractional energy in the axisymmetric component and the symbol colour the fractional energy stored into the poloidal component of the field (see legend to the right). }
    \label{Fig:Gl905_ZDIMaps}
\end{figure}

We conclude our analysis by deriving vector magnetic field maps for Gl~905 using ZDI, for each of the three main observing seasons. The maps are shown in Fig.~\ref{Fig:Gl905_ZDIMaps} and their magnetic properties are summarised in Table~\ref{tab:MagProp_Gl905}. We were able to fit all three ZDI maps down to $\chi_r^2 \approx 1.0$ assuming $P_{\mrm{rot}} = 114.3\,\mrm{d}$, $v_e \sin i = 0.06\,\kms$, $i = 60^\circ$, $f_V = 0.1$. 

The ZDI maps confirm the conclusions we derived from the PCA analysis. The topology gets indeed more complex from 2019 to 2020/21 and the degree of axisymmetry decreases from around 70\% to 4\% for the last season 2021/22. The surface mean magnetic field decreases from $128\,\mrm{G}$ to $64\,\mrm{G}$. Most prominent is the hint of an ongoing polarity reversal from negative to positive radial field taking place in the last season.

To test whether the symmetric component of the mean profile for season 2020/21 indeed results from an uneven phase coverage, we simulate 24 evenly phased Stokes~$V$ LSD profiles from the 2020/21 ZDI map (see Fig.~\ref{Fig:Gl905_ZDIMaps} middle column) and determine the corresponding mean profile and its symmetric and antisymmetric components (see Fig.~\ref{Fig:Gl905_EvenPhase}). The symmetric component disappears with even phase sampling, confirming that the mean LSD profile provides no observational hint for a large-scale axisymmetric toroidal field at the surface of Gl~905.

The reconstructed surface averaged toroidal field \Btor\ is lower than 10\,G for each season. To derive a 1$\sigma$ error bar on the simplest possible large-scale axisymmetric toroidal field (described with spherical harmonics coefficients $\ell=1$ and $m=0$) at the stellar surface, we proceed in the following way: (1) artificially add an axisymmetric toroidal field of strength \Btor\ to the reconstructed ZDI map, (2) simulate the corresponding Stokes~$V$ LSD profiles with the phase coverage and SNR of the actual observations and (3) compute the new $\chi^2$ with respect to the observed LSD profiles and raise \Btor\ until $\chi^2$ is increased by 1 with respect to the optimal ZDI fit. We find 1$\sigma$ error bars ranging from 180\,G in 2019 down to 55\,G in 2020/21.

\begin{table}
    \caption[]{Magnetic properties of Gl~905 extracted from the ZDI maps per season: the start and end month of the observations used for the ZDI maps, the surface averaged unsigned magnetic field \BV [G], the surface average unsigned dipole magnetic field $\langle B_\mrm{dip} \rangle$[G], the typical 1$\sigma$ error bar on the ZDI reconstructed surface averaged toroidal field \Btormax , the fractional energy of the poloidal field, the fractional energy of the axisymmetric field (only $m=0$ modes), the fractional energy of the dipole component, the tilt angle of the dipole ($\ell=1$) with respect to the negative pole, the phase at which the dipole field faces the observer, the reduced $\chi^2$ values for the Stokes~$V$ profiles ($\chi^2_{r,V}$; corresponding to a $\textbf{B} = 0\,$G fit), for the ZDI fit of the Stokes~$V$ profiles ($\chi^2_{r,V,\mrm{ZDI}}$)  and for the Null profiles ($\chi^2_{r,N}$) and the number of observations per season (nb. obs). }
    \label{tab:MagProp_Gl905}
    \begin{center}
    \begin{tabular}{lccc}
        \hline
        \noalign{\smallskip}
      season & \bf{2019} & \bf{2020/21} & \bf{2021/22} \\
      start & 2019 Apr & 2020 May & 2021 June \\
      end & 2019 Dec & 2021 Jan & 2022 Jan \\
        \noalign{\smallskip}
        \hline
        \noalign{\smallskip}
\BV [G] & 128 & 89 & 64 \\
$\langle B_\mrm{dip} \rangle$[G] & 124 & 80 & 64 \\ 
\Btormax [G] & 179 & 55 & 84 \\ 
$f_{\mrm{pol}}$ & 1.0 & 0.99 & 1.0 \\ 
 $f_{\mrm{axi}}$ & 0.68 & 0.7 & 0.04 \\
 $f_{\mrm{dip}}$ & 0.93 & 0.79 & 0.94 \\
        \noalign{\smallskip}
        \hline
        \noalign{\smallskip}
dipole tilt angle &  $33^\circ$ & $22^\circ$ & $83^\circ$ \\ 
pointing phase & 0.29 & 0.12 & 0.96 \\ 
        \noalign{\smallskip}
        \hline
        \noalign{\smallskip}
 $\chi^2_{r,V}$ & 1.94 & 3.81 & 1.47\\ 
$\chi^2_{r,V,\mrm{ZDI}}$ & 1.04 & 1.12 & 1.02 \\ 
 $\chi^2_{r,N}$ & 1.22 & 1.10 & 1.04 \\  
 nb. obs & 43 & 84 & 77 \\ 
        \hline
    \end{tabular}
    \end{center}
\end{table}

\section{GJ~1289}
\label{Sec:GJ1289}

The fully convective M~dwarf GJ~1289 ($M = 0.21\pm0.02\,$\Msun\ , \cc{\citealt{Cristofari2022}}) is the next star in our sample. SPIRou observed GJ~1289 from 2019 Sept until 2022 June providing a time series of 204 LSD profiles split into three seasons (2019 June -- Dec, 2020 May -- 2021 Jan, 2021 June -- 2022 Jan) for the per-season analysis.  As for Gl~905, the 14 profiles collected in 2022 June at the beginning of a new season, only covering 16\% of a rotation cycle, were left out of the per-season analysis.

\subsection{PCA analysis of GJ~1289}

\begin{figure}
	\raggedright \textbf{a.} \hspace{2.7cm} \textbf{b.} \\
	\centering
	\includegraphics[width=0.345\columnwidth, trim={0 0 0 0}, clip]{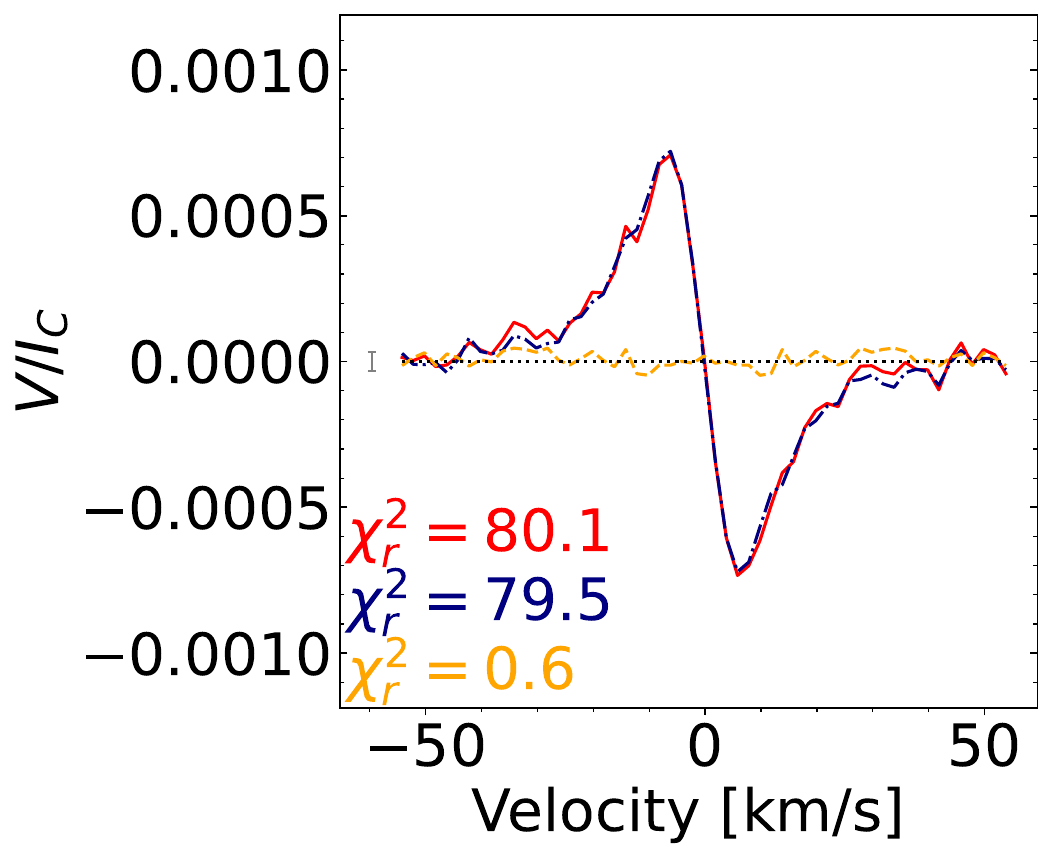}
	\includegraphics[width=0.64\columnwidth, trim={0 400 440 0}, clip]{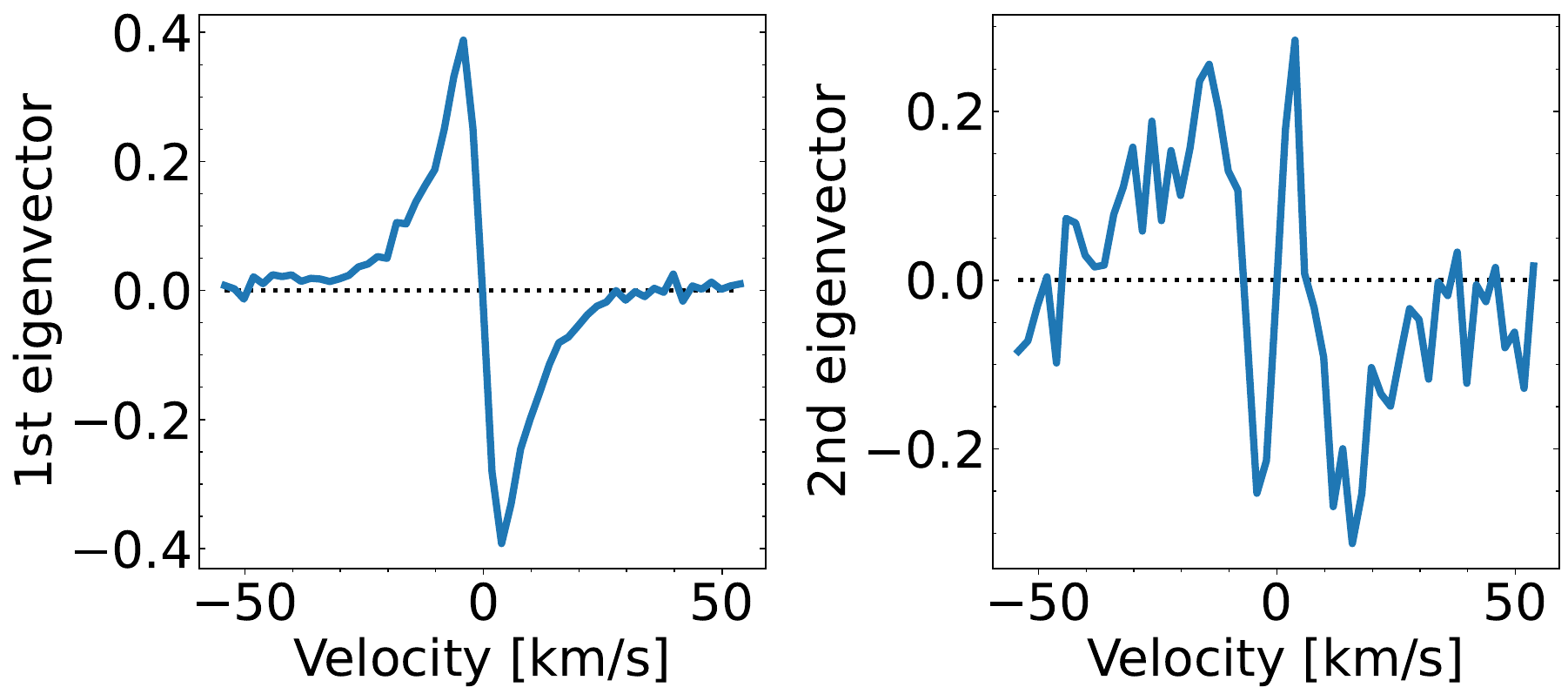}\\ 
	\rule{7cm}{0.3mm}\\
	\raggedright \textbf{c.} \\
	\centering
	\includegraphics[width=0.35\columnwidth, trim={0 0 0 0}, clip]{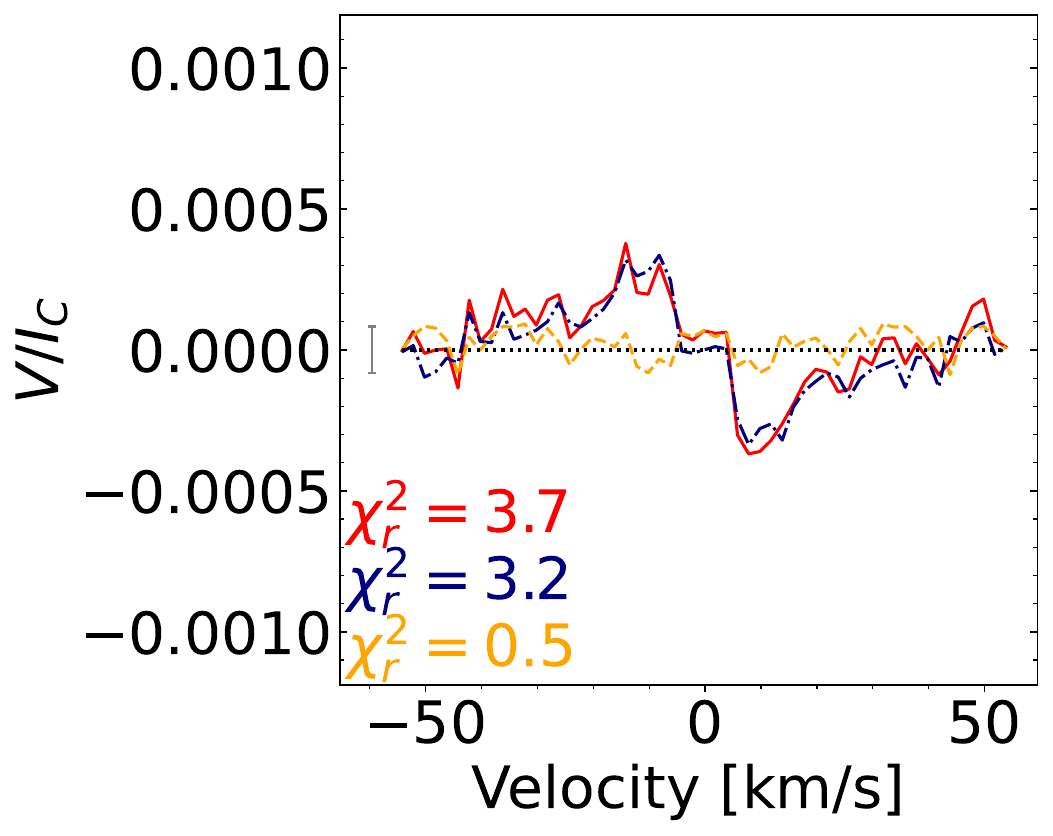}
	\includegraphics[width=0.63\columnwidth, trim={30 400 445 0}, clip]{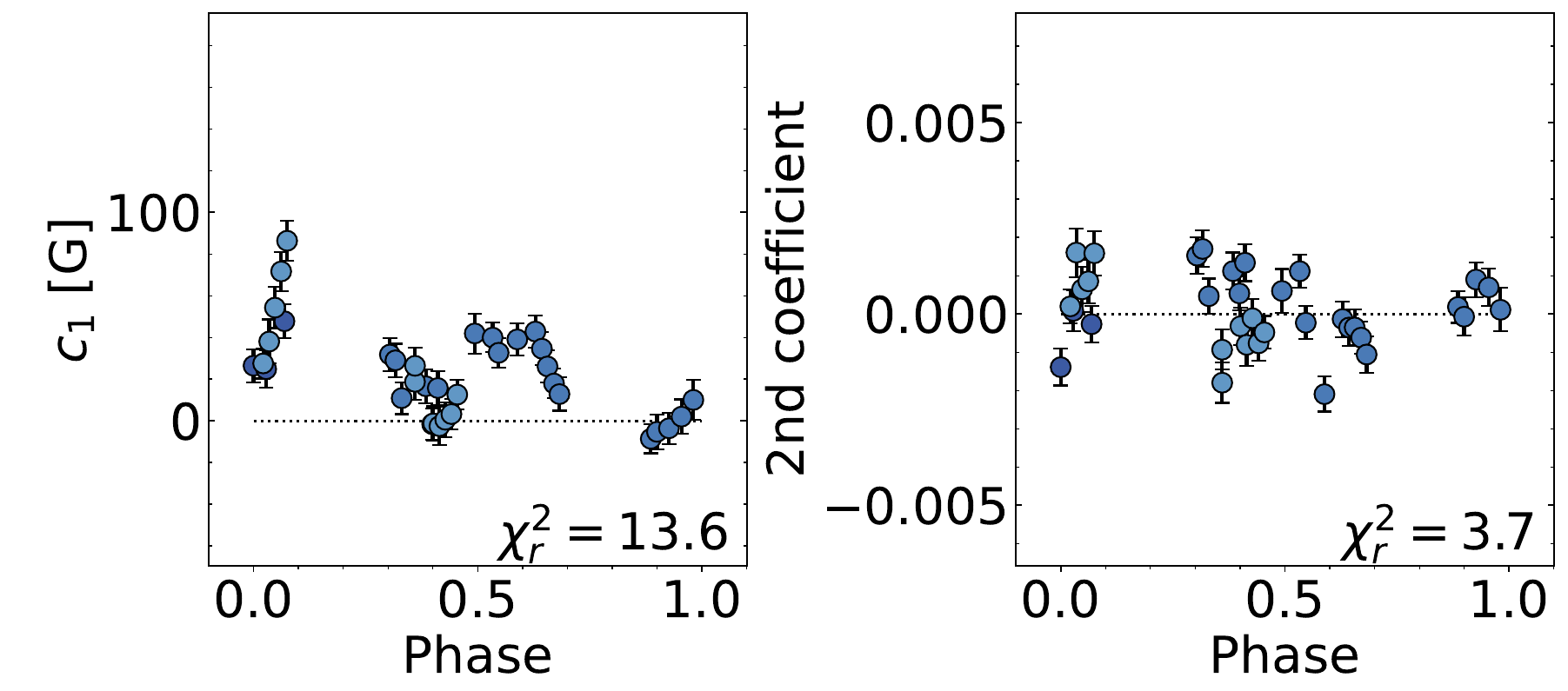}\\
		\includegraphics[width=0.35\columnwidth, trim={0 0 0 0}, clip]{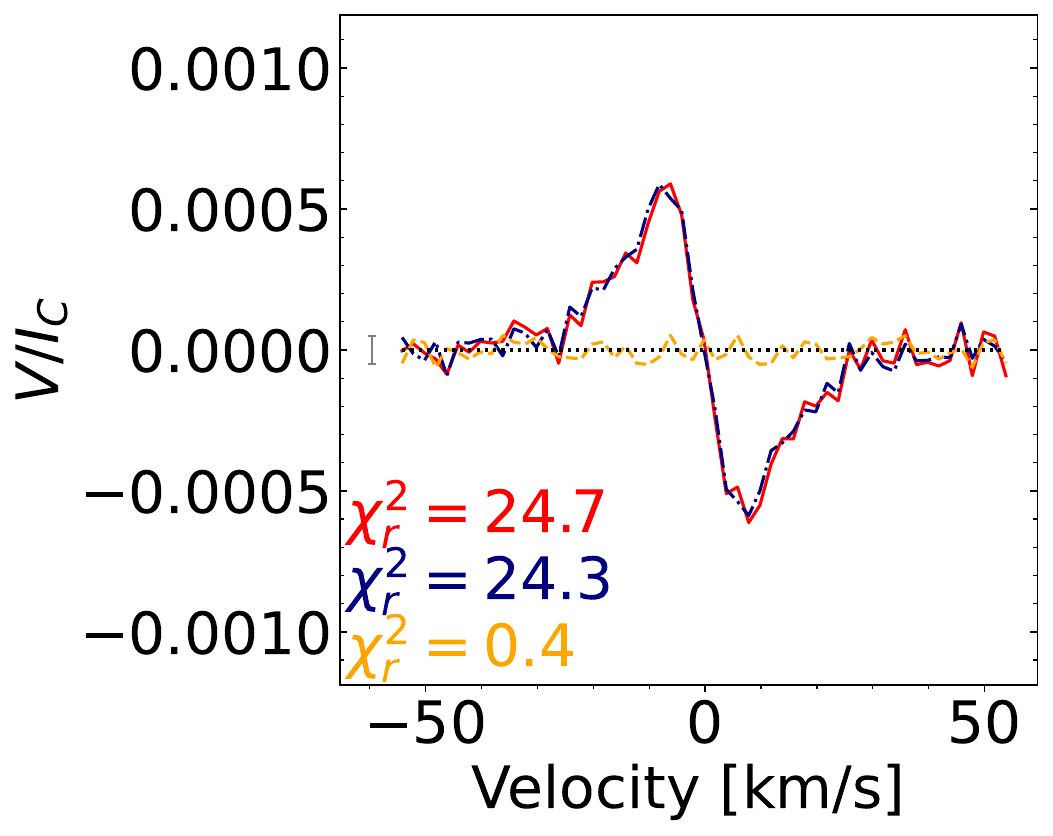}
	\includegraphics[width=0.63\columnwidth, trim={30 400 445 0}, clip]{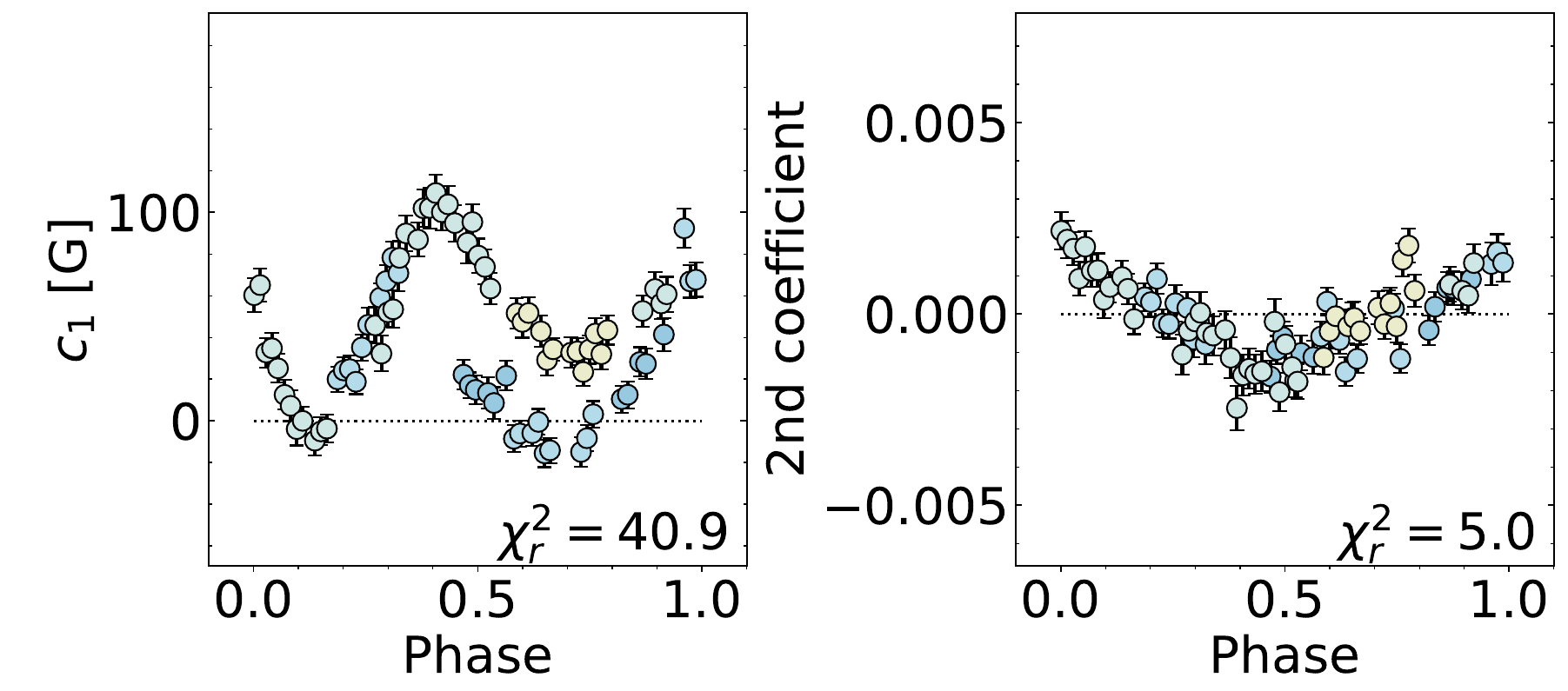}\\
		\includegraphics[width=0.35\columnwidth, trim={0 0 0 0}, clip]{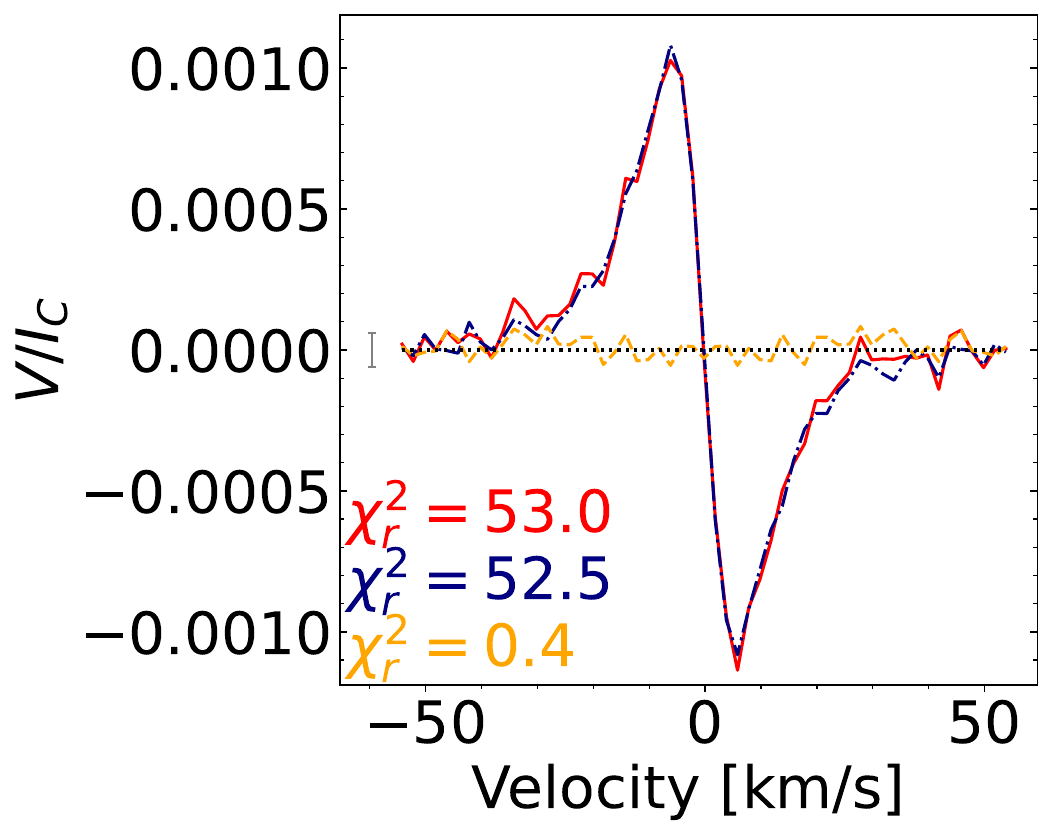}
	\includegraphics[width=0.63\columnwidth, trim={30 400 445 0}, clip]{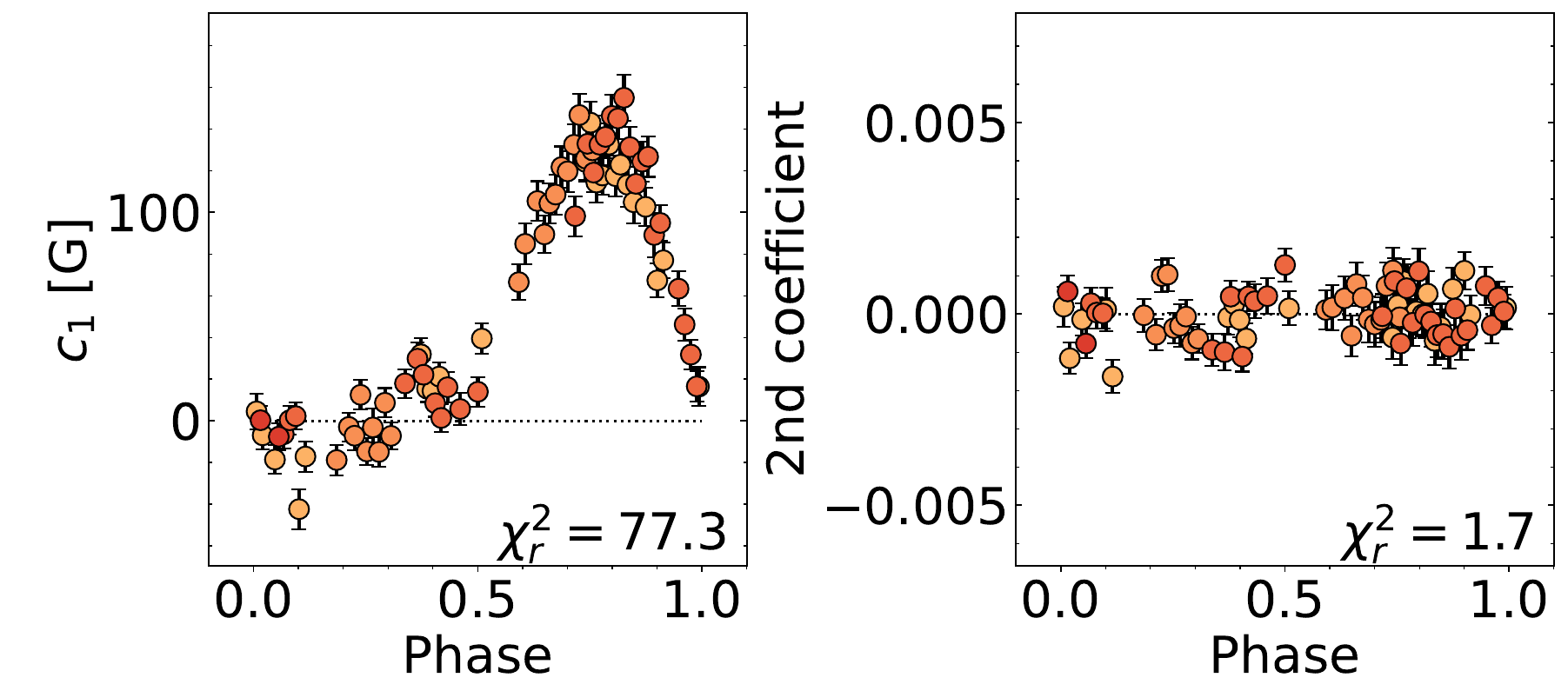}\\
    \caption{Same as Fig.~\ref{Fig:Gl905_PCA} for GJ~1289.}
    \label{Fig:GJ1289_PCA}
\end{figure}

The mean profile is perfectly antisymmetric with respect to the line centre indicating a dominant axisymmetric poloidal component (see Fig.~\ref{Fig:GJ1289_PCA}a). Both the first and second eigenvectors are found to be antisymmetric with respect to the line centre, which is a strong hint of a non-axisymmetric poloidal component (see Fig.~\ref{Fig:GJ1289_PCA}b). All further eigenvectors trace noise. 

\begin{figure}
	%
	%
	\centering
	\includegraphics[width=\columnwidth, trim={0 0 0 0}, clip]{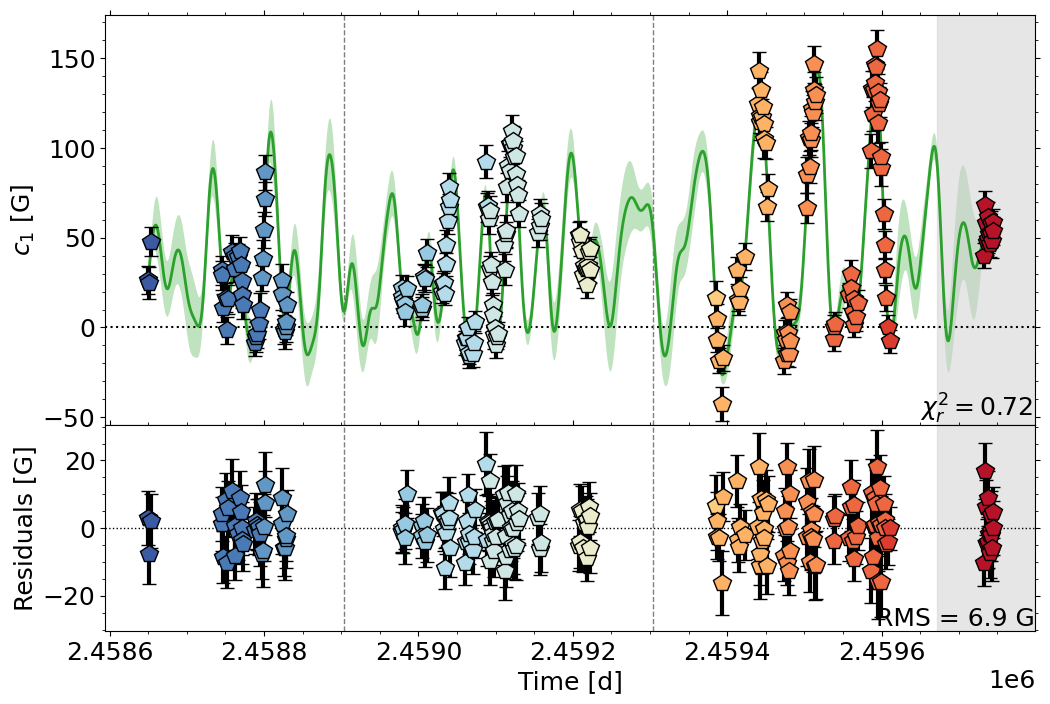}\\ 
		\includegraphics[width=\columnwidth, trim={0 0 0 0}, clip]{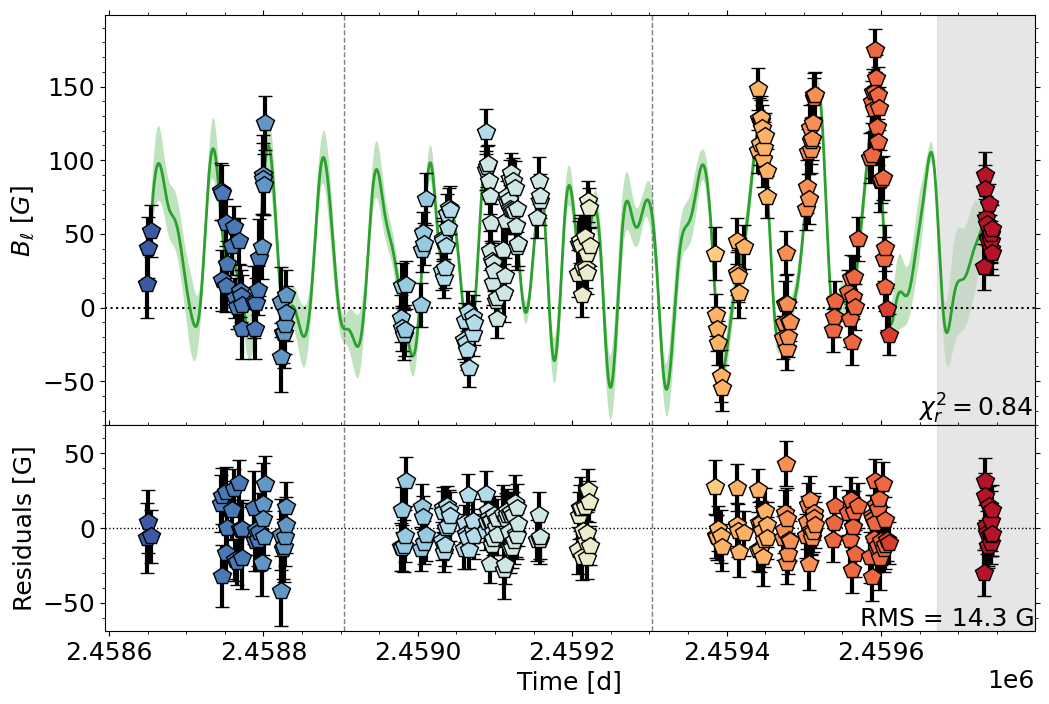}\\ 
    \caption{Same as Fig.~\ref{Fig:Gl905_CoeffvsTime} for GJ~1289.}
    \label{Fig:GJ1289_CoeffvsTime}
\end{figure}

The QP GPR fit applied to $c_1$ (see Fig.~\ref{Fig:GJ1289_CoeffvsTime} top) finds $\Prot = 75.62^{+0.85}_{-0.79}\,\dy$ with a decay time of $l = 129^{+26}_{-24}\,\dy$ fitting all 5 parameters with a $\chi^2_r = 0.72$ (see Tab.~\ref{tab:GPFitParams_c1}). 
The rotation period and decay time agree with the values found from the GPR fits of $B_\ell$ found by D23 ($\Prot = 73.66\pm0.92\,\dy$ and $l = 152^{+32}_{-27}\,\dy$) and \cite{Fouque2023} ($\Prot = 74.0^{+1.5}_{-1.3}\,\dy$ and $l = 142^{+33}_{-26}\,\dy$) and are also consistent with our GP fit of the $B_\ell$ values of D23 ($\Prot = 73.67^{+1.01}_{-0.91}\,\dy$ and $l = 152^{+30}_{-28}\,\dy$, see Fig.~\ref{Fig:GJ1289_CoeffvsTime} bottom).

In Fig.~\ref{Fig:GJ1289_PCA}c, we show the mean profile and the phase-folded coefficients split by season. 
 Comparing the mean profiles of the three seasons, the axisymmetric component grows in amplitude and stays always poloidal. As the amplitude of the coefficients increases as well, the magnetic field becomes in general stronger. 

The phase-folded coefficient curves indicate a rapidly evolving and complex large-scale field, as we see variations from one rotation cycle to the next (see season 2020/21) and trends that are more complex than sine waves (e.g.\ for 2019 and 2020/21). Furthermore, season 2020/21 stands out, with the second eigenvector contributing significantly to the Stokes~$V$ signal before disappearing again for the last season 2021/22, for which $c_1$ shows a simpler trend.

\subsection{ZDI reconstructions of GJ~1289}

\begin{figure}
\centering
\begin{minipage}{0.32\columnwidth}
\centering
\includegraphics[height=0.85\columnwidth, angle=270, trim={140 0 0 29}, clip]{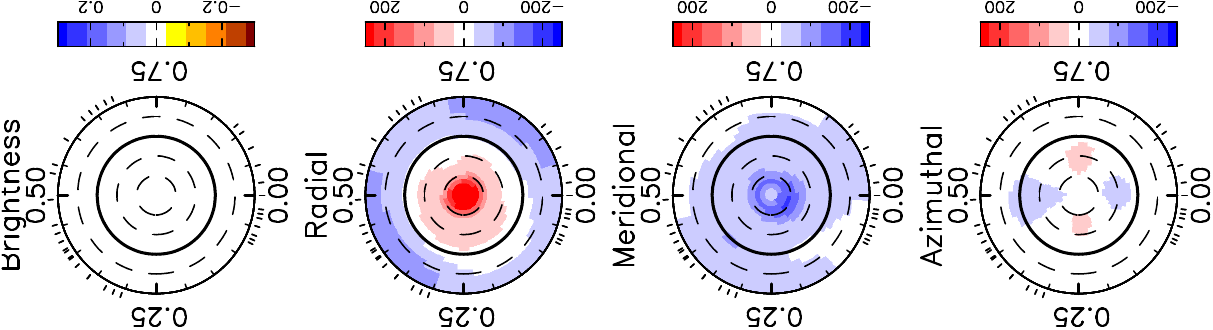} 
\end{minipage}
\begin{minipage}{0.32\columnwidth}
\centering
\includegraphics[height=0.85\columnwidth, angle=270, trim={140 0 0 29}, clip]{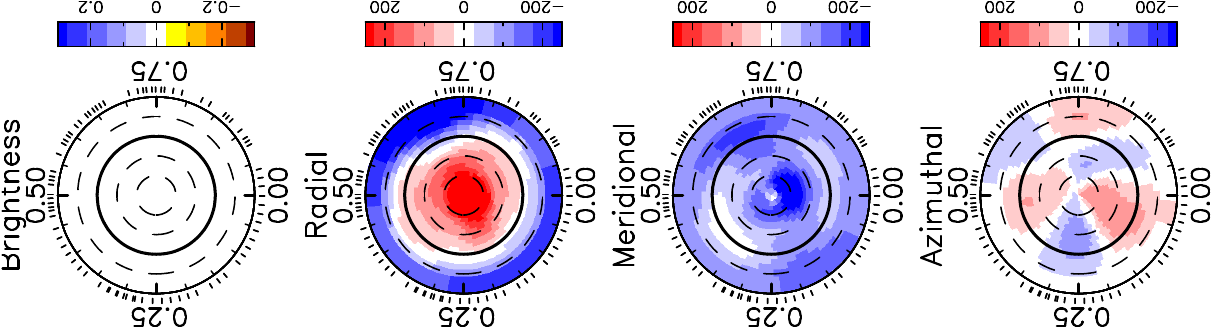} 
\end{minipage}
\begin{minipage}{0.32\columnwidth}
\centering
\includegraphics[height=0.85\columnwidth, angle=270, trim={140 0 0 29}, clip]{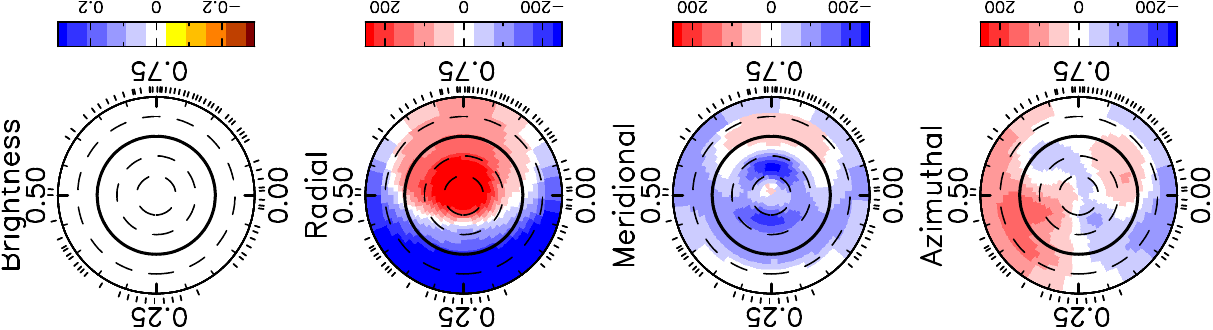} 
\end{minipage}
\includegraphics[width=0.3\columnwidth, angle=180, trim={460 130 2 0}, clip]{Figures/GJ1289_ZDIMap_JFDLSD_epo3_v51.pdf}
\vspace*{2mm}
\includegraphics[width=0.95\columnwidth, clip]{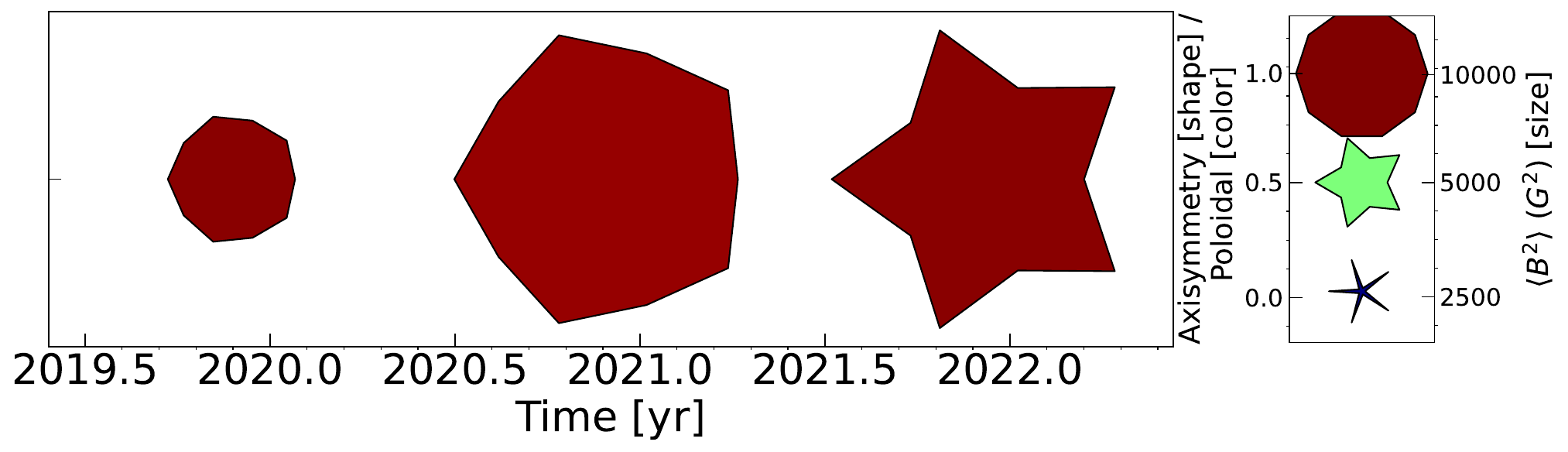}
    \caption{Same as Fig.~\ref{Fig:Gl905_ZDIMaps} for GJ~1289.}
    \label{Fig:GJ1289_ZDIMaps}
\end{figure}

We were able to fit the Stokes~$V$ profiles for all seasons down to $\chi^2_r \approx 1.0$ assuming $\Prot = 73.66\,\mrm{d},\ v_e \sin i = 0.14\,\kms,\ i = 60^\circ,\ f_V = 0.1$.
In the first season, the data set only includes about half of the observations of those from the two other seasons and the achieved $\chi^2_{r,V} = 1.46$ is several times lower than the $\chi^2_{r,V}$ of the following seasons (see Tab.~\ref{tab:MagProp_GJ1289}). In the first season, ZDI reveals a weak marginally complex field topology. In 2020/21, the surface averaged field \BV\ becomes twice as large due to a growing axisymmetric poloidal dipole and 
ZDI reconstructs a more complex azimuthal field, featuring a quadrupolar non-axisymmetric azimuthal structure (see Fig.~\ref{Fig:GJ1289_ZDIMaps}). 
In the last season 2021/22, the dipole tilts more strongly to $39^{\circ}$ and dominates the field topology (see Tab.~\ref{tab:MagProp_GJ1289}).
The toroidal field of the ZDI maps varies between $7-25$\,G, which is again lower than the typical 1$\sigma$ error bar that we derive ranging from 40 to 100\,G. 

\begin{table}
%
%
    \caption[]{Same as Table~\ref{tab:MagProp_Gl905} for GJ~1289. The tilt angle now refers to the positive pole.}
    \label{tab:MagProp_GJ1289}
    \begin{center}
    \begin{tabular}{lccc}
        \hline
        \noalign{\smallskip}
      season & \bf{2019} & \bf{2020/21} & \bf{2021/22} \\
      start & 2019 June & 2020 May & 2021 June \\
      end & 2019 Dec & 2021 Jan & 2022 Jan \\
        \noalign{\smallskip}
        \hline
        \noalign{\smallskip}
\BV [G] & 83 & 199 & 214 \\
$\langle B_\mrm{dip} \rangle$[G]  & 79 & 194 & 200 \\
\Btormax [G] & 42 & 57 & 100 \\
$f_{\mrm{pol}}$ & 0.99 & 0.98 & 0.99 \\
 $f_{\mrm{axi}}$ & 0.93 & 0.86 & 0.57 \\
 $f_{\mrm{dip}}$ & 0.67 & 0.79 & 0.82 \\
        \noalign{\smallskip}
        \hline
        \noalign{\smallskip}
dipole tilt angle &  $11^\circ$ & $8^\circ$ & $39^\circ$ \\
pointing phase & 0.19 & 0.18 & 0.74 \\
        \noalign{\smallskip}
        \hline
        \noalign{\smallskip}
 $\chi^2_{r,V}$ & 1.46 & 3.60 & 8.34\\
 $\chi^2_{r,V,\mrm{ZDI}}$ & 0.98 & 1.04 & 0.98 \\
 $\chi^2_{r,N}$ & 0.95 & 0.94 & 0.97 \\
 nb. obs & 35 & 80 & 75 \\
        \hline
    \end{tabular}
    \end{center}
\end{table}

\section{GJ~1151}

Our next star, GJ~1151, is also a fully-convective M~dwarf ($M = 0.17\pm0.02\,$\Msun\ , \cc{\citealt{Cristofari2022}}) and was observed between 2019 Dec and 2022 June with SPIRou providing us 158 LSD profiles (seasons: 2019 Dec -- 2020 July, 2020 Dec -- 2021 July, 2021 Dec -- 2022 June).

\subsection{PCA analysis of GJ~1151}

\begin{figure}
	\raggedright \textbf{a.} \hspace{2.7cm} \textbf{b.} \\
	\centering
	\includegraphics[width=0.355\columnwidth, trim={0 0 0 0}, clip]{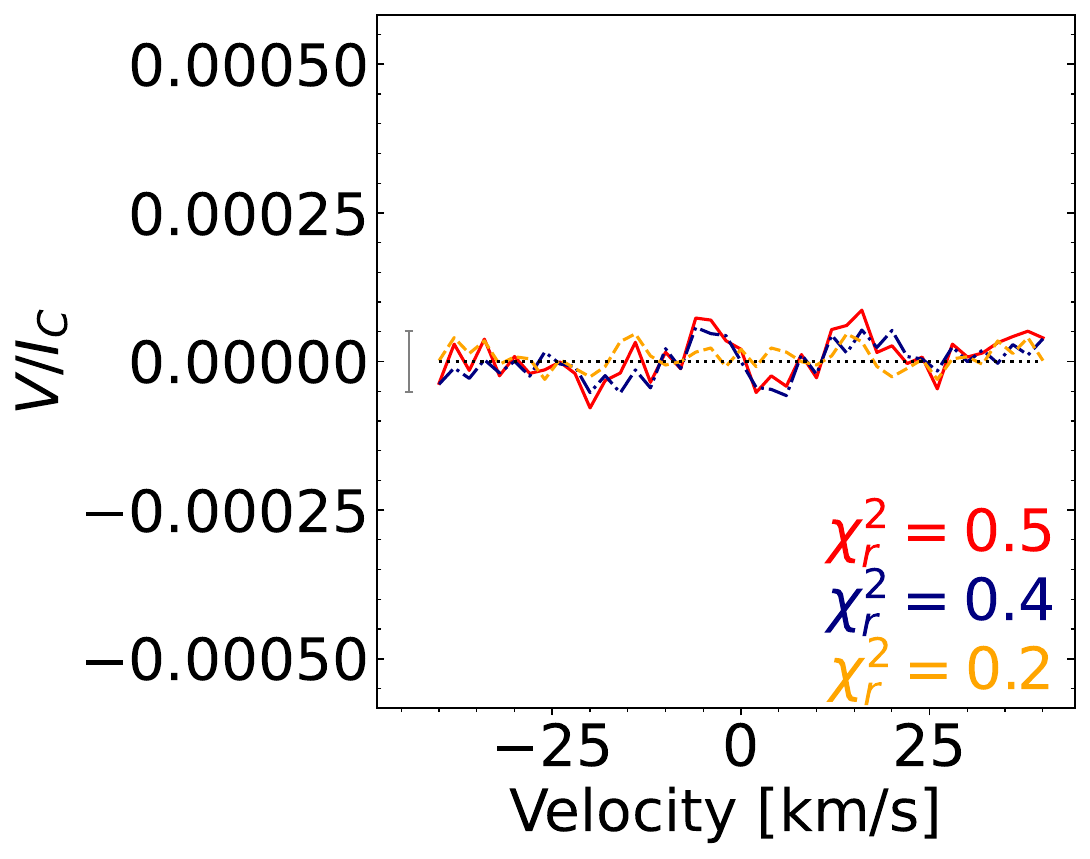}
	\includegraphics[width=0.63\columnwidth, trim={0 400 443 0}, clip]{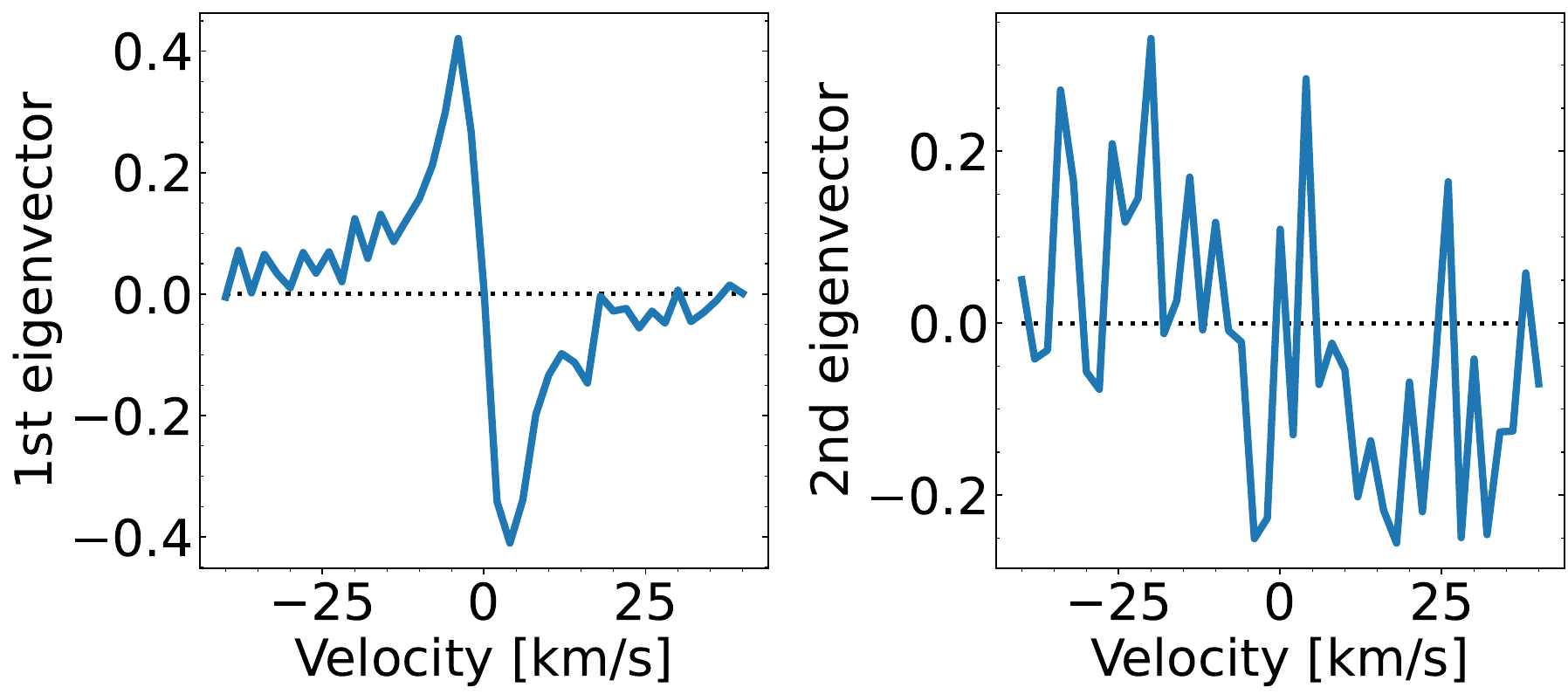}\\ 
	\rule{7cm}{0.3mm}\\
	\raggedright \textbf{c.} \\
	\centering
	\includegraphics[width=0.36\columnwidth, trim={0 0 0 0}, clip]{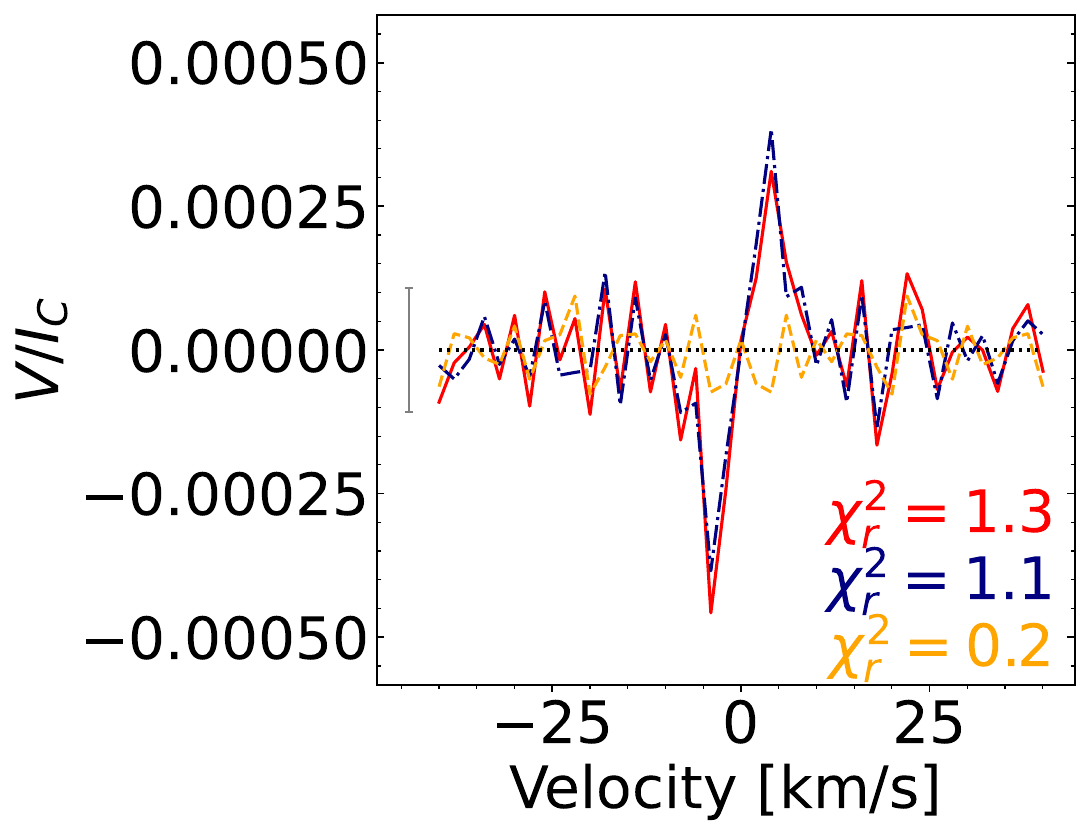}
	\includegraphics[width=0.63\columnwidth, trim={30 400 445 0}, clip]{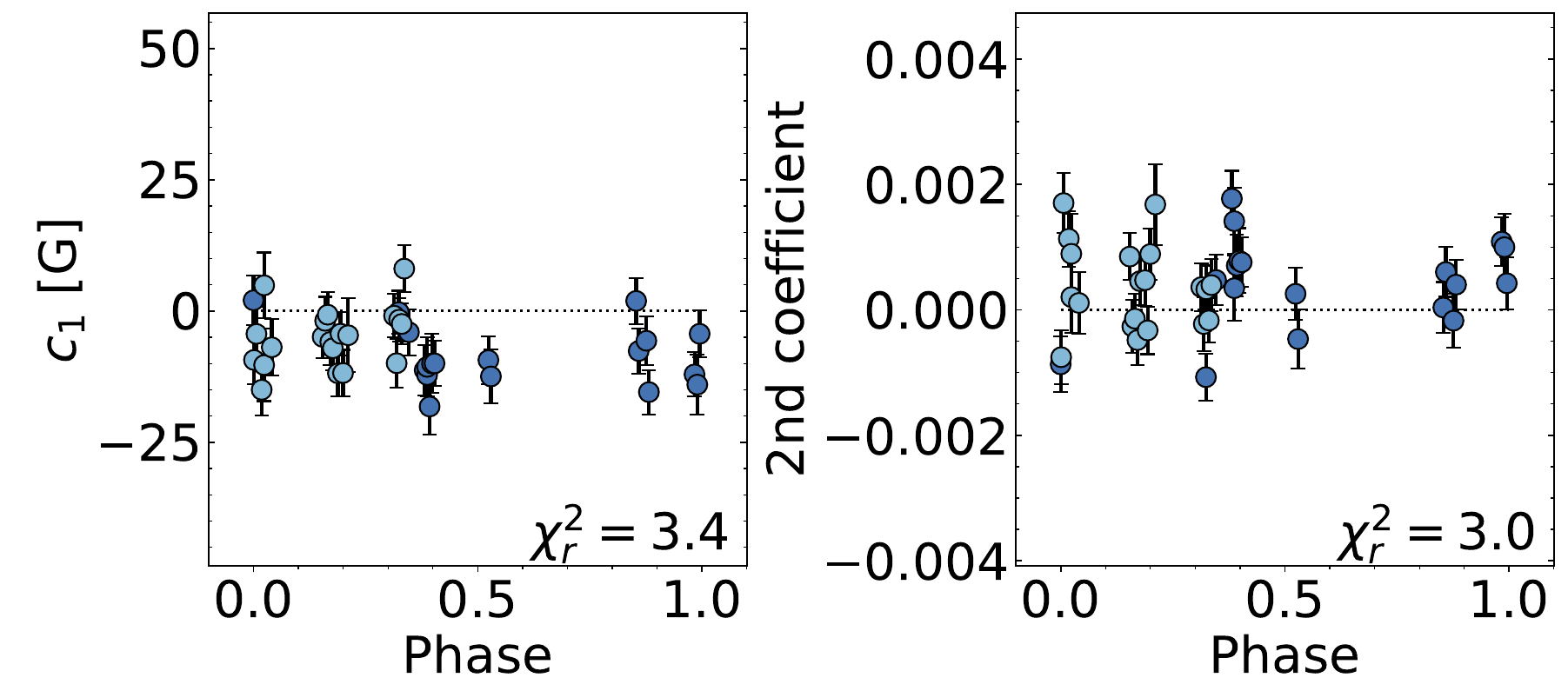}\\
		\includegraphics[width=0.36\columnwidth, trim={0 0 0 0}, clip]{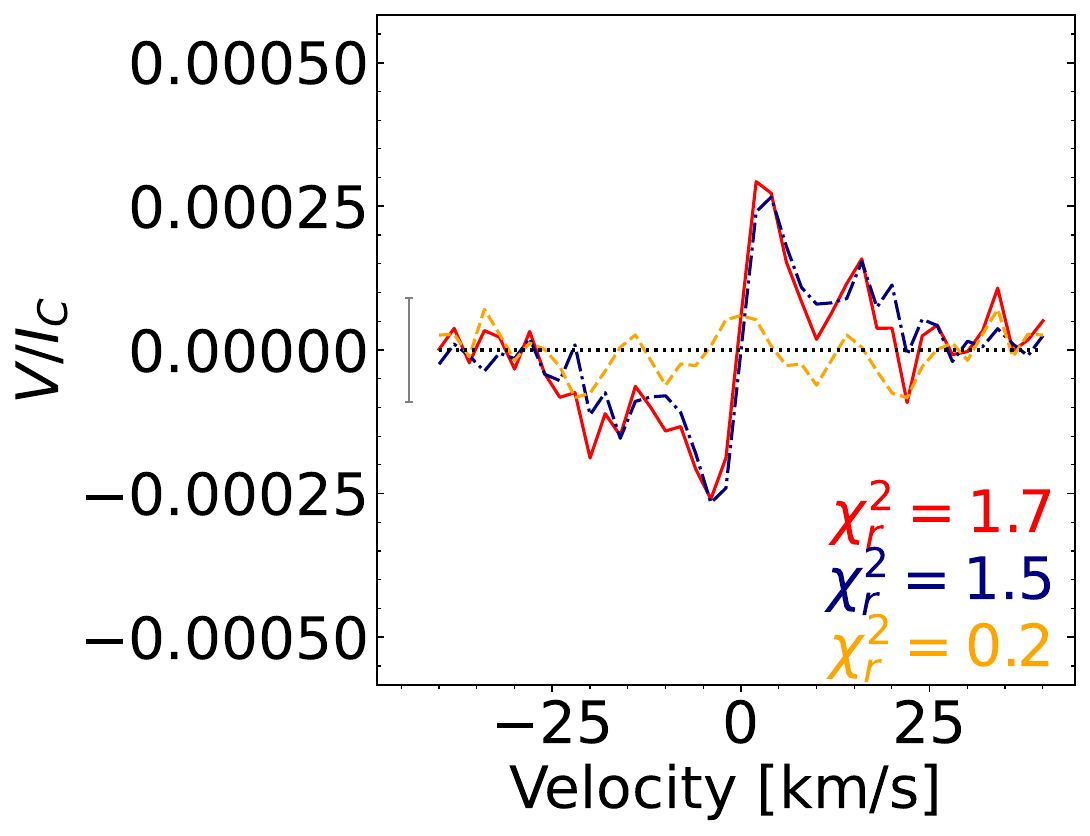}
	\includegraphics[width=0.63\columnwidth, trim={30 400 445 0}, clip]{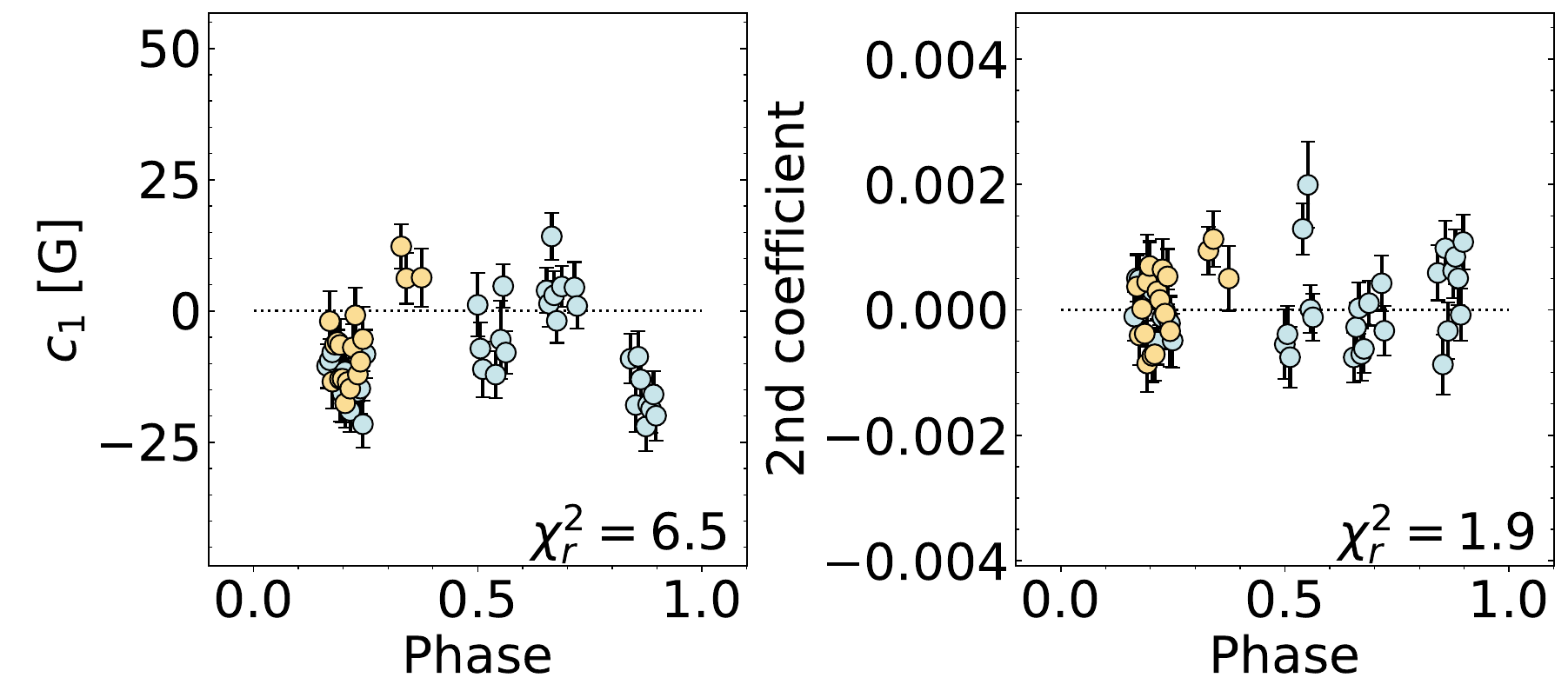}\\
		\includegraphics[width=0.36\columnwidth, trim={0 0 0 0}, clip]{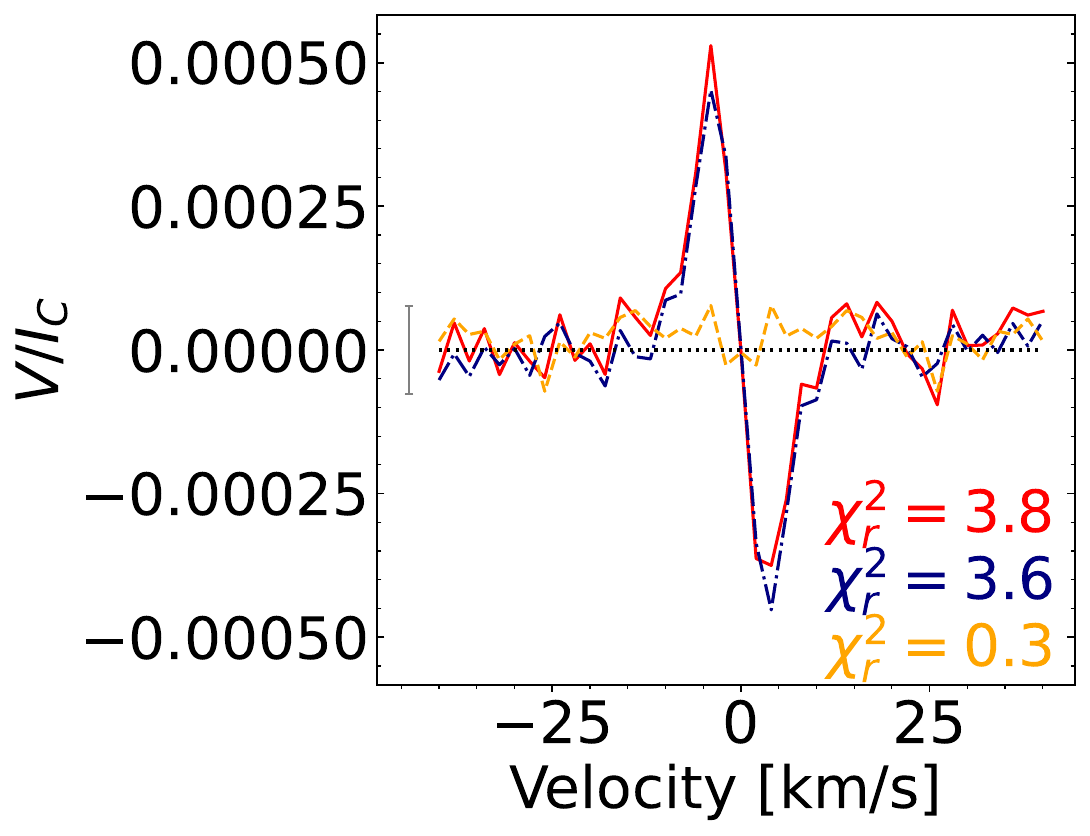}
	\includegraphics[width=0.63\columnwidth, trim={30 400 445 0}, clip]{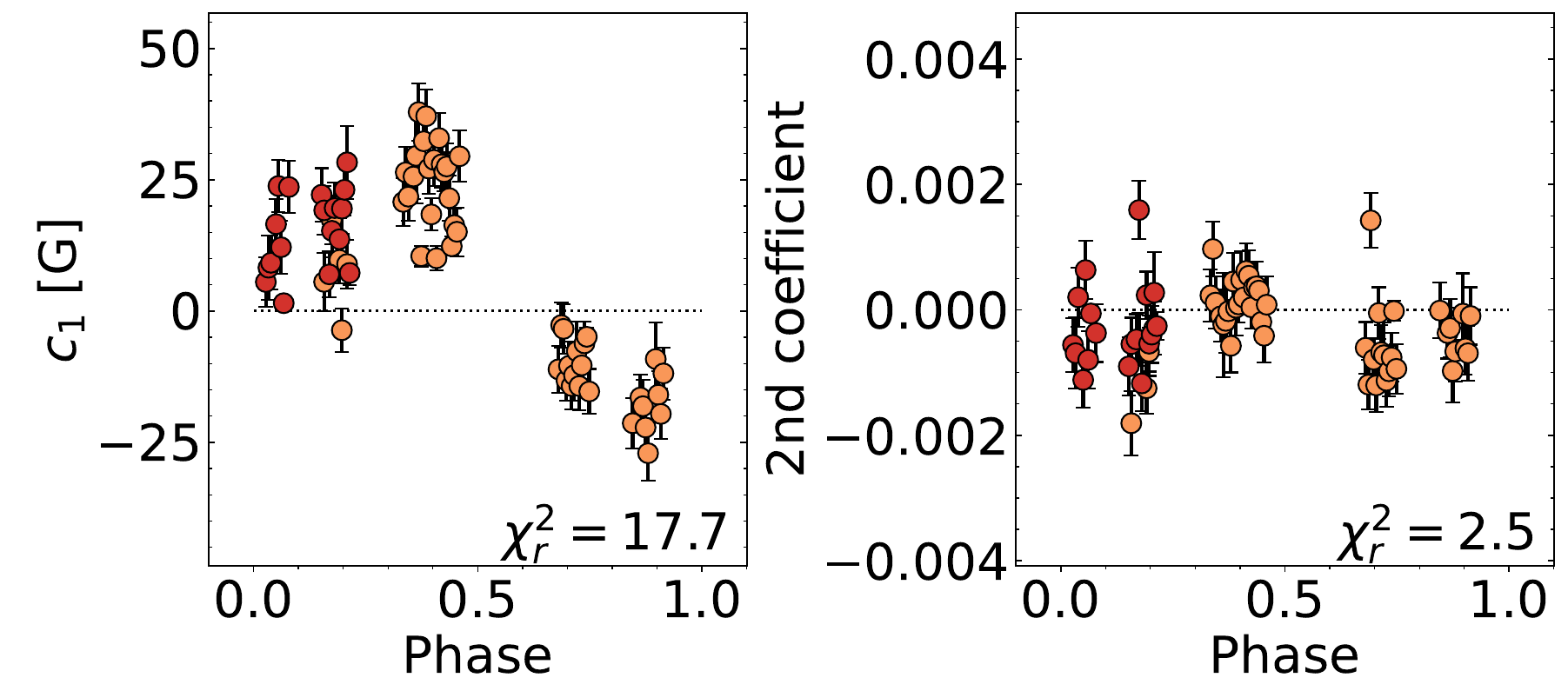}\\
    \caption{Same as Fig.~\ref{Fig:Gl905_PCA} for GJ~1151.}
    \label{Fig:GJ1151_PCA}
\end{figure}

The mean profile is close to zero indicating a strongly non-axisymmetric topology (see Fig.~\ref{Fig:GJ1151_PCA}a). Only the first eigenvector of the mean-subtracted Stokes~$V$ profiles significantly differs from the noise and features an antisymmetric signal with respect to the line centre (see Fig.~\ref{Fig:GJ1151_PCA}b). 

\begin{figure}
	%
	%
	\centering
	\includegraphics[width=\columnwidth, trim={0 0 0 0}, clip]{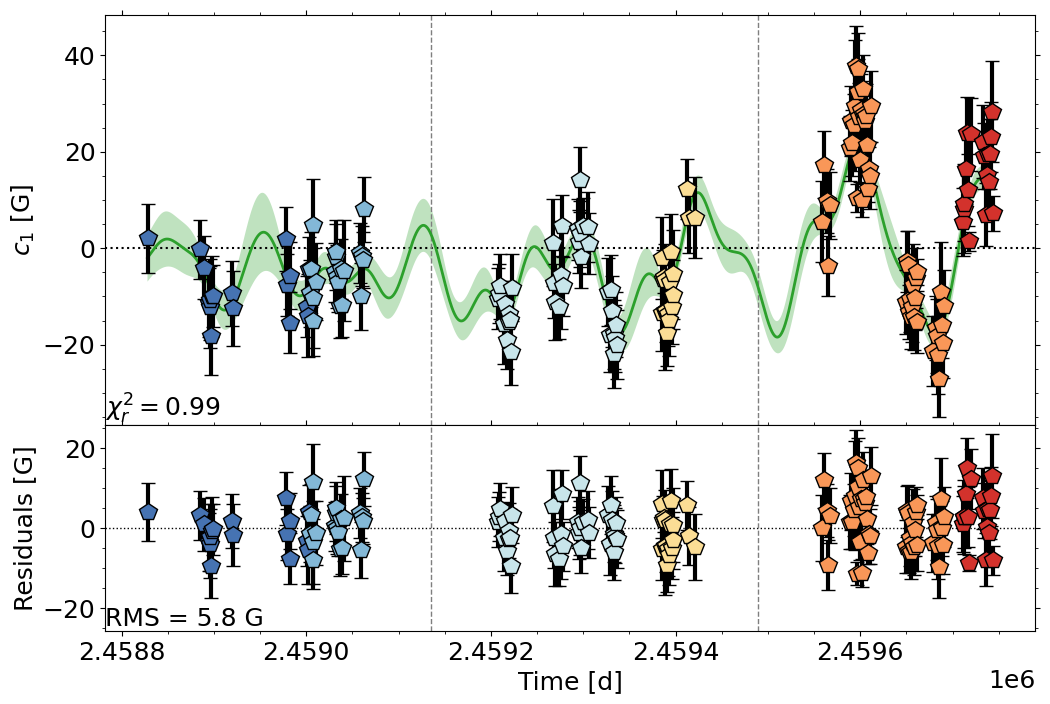}\\ 
		\includegraphics[width=\columnwidth, trim={0 0 0 0}, clip]{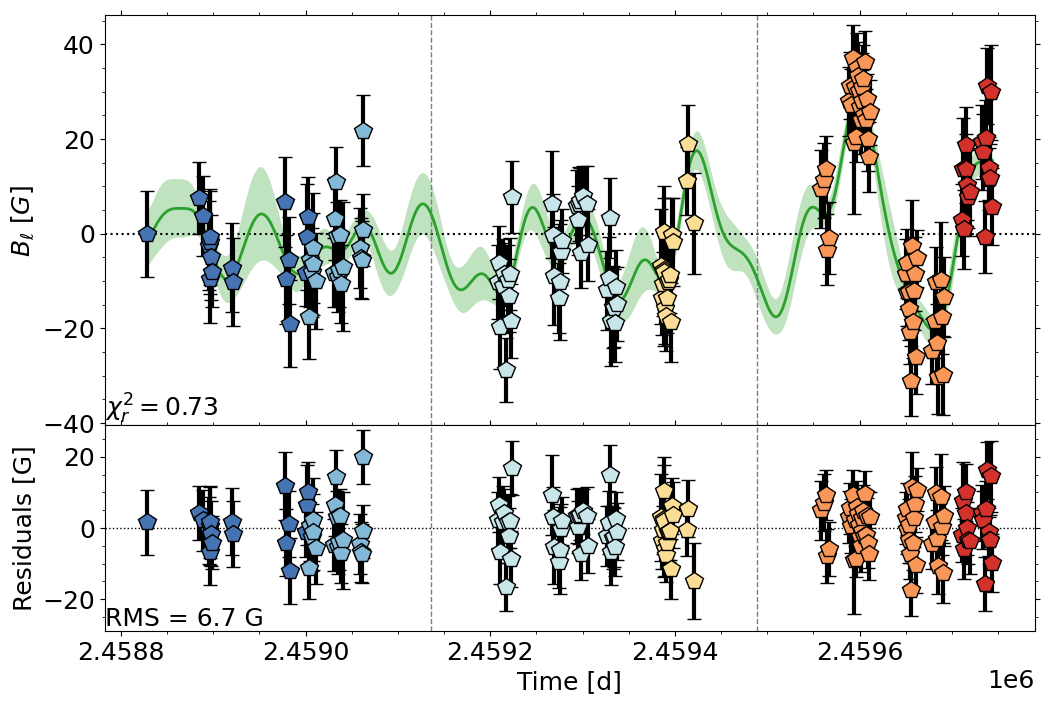}\\ 
    \caption{Same as Fig.~\ref{Fig:Gl905_CoeffvsTime} for GJ~1151.}
    \label{Fig:GJ1151_CoeffvsTime}
\end{figure}

The QP GPR model fits $c_1$ down to a $\chi^2_r = 0.99$ (see Fig.~\ref{Fig:GJ1151_CoeffvsTime} top). We fix the decay time to 300\,d similar to D23 and find a $\Prot = 175.8^{+3.2}_{-3.4}\,\dy$ similar to the results of D23 ($\Prot = 175.6\pm4.9\,\dy$) and our own GP fit of $B_\ell$  ($\Prot = 176.1^{+3.6}_{-4.1}\,\dy$), see also Fig.~\ref{Fig:GJ1151_CoeffvsTime} bottom). 
Our rotation period is a bit higher than the one found by \cite{Fouque2023} ($\Prot = 158\pm12\,\dy$) but compatible at $1\sigma$.

The mean profile of the first two seasons is antisymmetric with respect to the line centre (axisymmetric poloidal field) and is relatively weak (see Fig.~\ref{Fig:GJ1151_PCA}c). The coefficient $c_1$ shows no obvious trend with phase for the first season 2019/20 and just start to display a weak variation with phase for 2020/21. The low amplitude of the mean profiles and coefficients indicate that the magnetic field must be very weak during the first two seasons. For the last season 2021/22, the amplitude of the mean profile is twice as high as before and also $c_1$ shows a higher amplitude ($\chi^2_r = 7.9$), indicating that the magnetic field increases significantly for 2021/22. We also notice that the sign of the mean profile (and therefore the projected main polarity of the large-scale magnetic field) changed from negative to positive for the last season, hence why the mean profile over the whole time series is close to zero (see Fig.~\ref{Fig:GJ1151_PCA}a).

\subsection{ZDI reconstructions of GJ~1151}

We could fit the Stokes~$V$ profiles down to $\chi^2_r \approx 1.0$ assuming $\Prot = 175.6\,\dy $, $f_V = 0.1, v_e \sin i = 0.05\,\kms $ and inclination of $i = 60^\circ$.

For the first two seasons, the Stokes~$V$ profiles are weak (see Fig.~\ref{Fig:GJ1151_StVFit}) and so is the reconstructed field, with a dominant negative polarity in the upper hemisphere that is consistent with the corresponding mean profiles.  
We see a small increase of \BV\ in the second season of 2020/21 explaining the higher amplitude of $c_1$ seen in the PCA analysis (see Fig.~\ref{Fig:GJ1151_PCA}c). For the last season 2021/22, the ZDI map shows a strongly tilted dipole (tilt angle = $55^\circ$), that flipped polarity, and the surface averaged field is more than twice as high as before (\BV\ = 63\,G). To the best of our knowledge this is the first polarity reversal seen in the vector magnetic field map of an M~dwarf. 

The reconstructed toroidal field \Btor\ is lower than 8\,G for all seasons of GJ~1151 and we find that the 1$\sigma$ error bar on the toroidal field ranges between 370 and 450\,G.

\begin{figure}
\centering
\begin{minipage}{0.32\columnwidth}
\centering
\includegraphics[height=0.85\columnwidth, angle=270, trim={140 0 0 29}, clip]{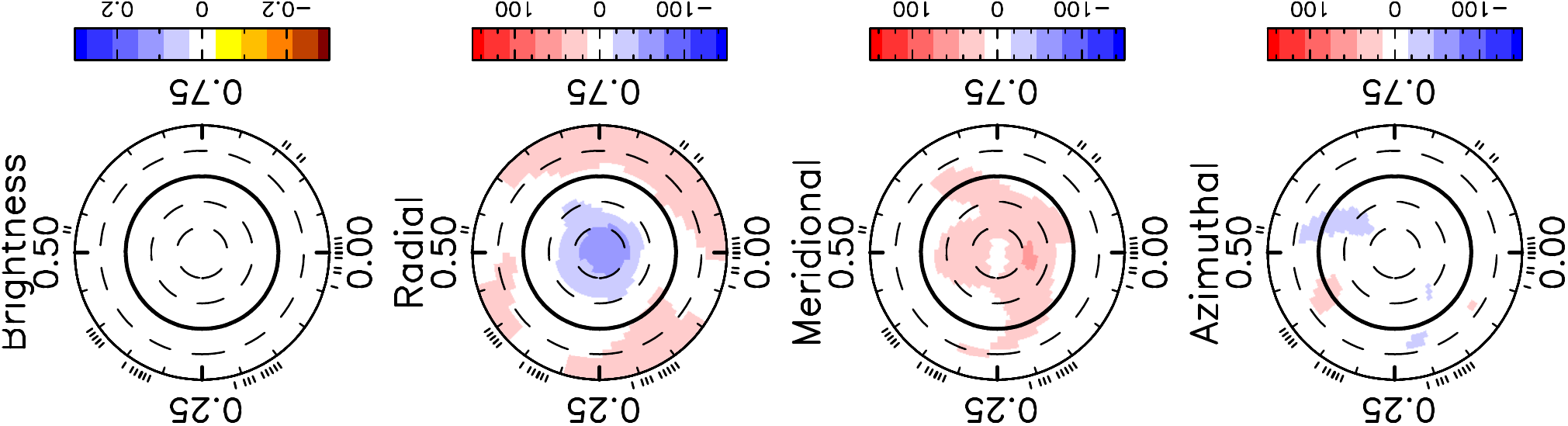} 
\end{minipage}
\begin{minipage}{0.32\columnwidth}
\centering
\includegraphics[height=0.85\columnwidth, angle=270, trim={140 0 0 29}, clip]{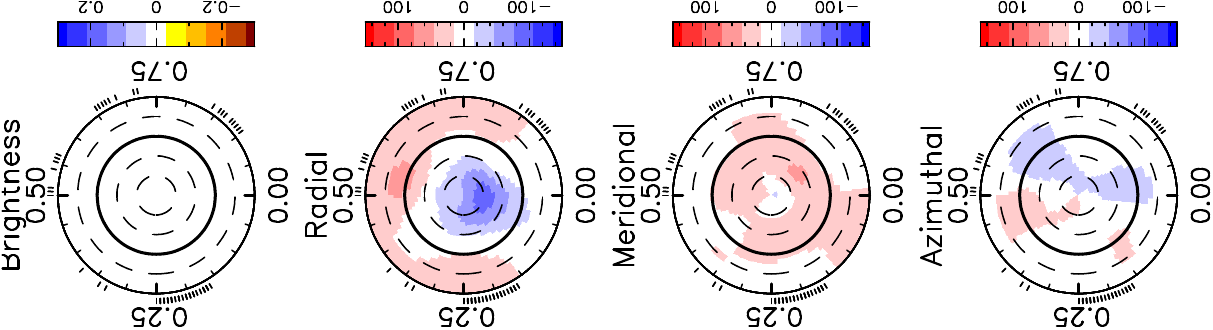} 
\end{minipage}
\begin{minipage}{0.32\columnwidth}
\centering
\includegraphics[height=0.85\columnwidth, angle=270, trim={140 0 0 29}, clip]{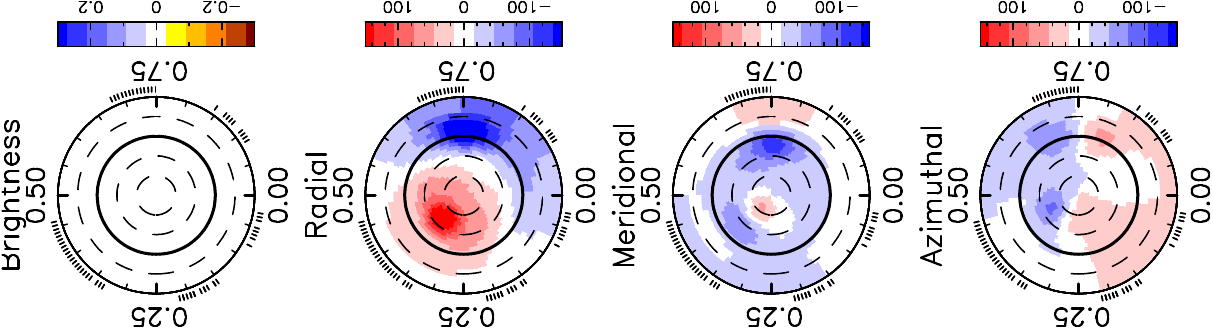} 
\end{minipage}
\includegraphics[width=0.3\columnwidth, angle=180, trim={460 130 2 0}, clip]{Figures/GJ1151_ZDIMap_JFDLSD_epo3_v30.pdf}
\vspace*{2mm}
\includegraphics[width=0.95\columnwidth, clip]{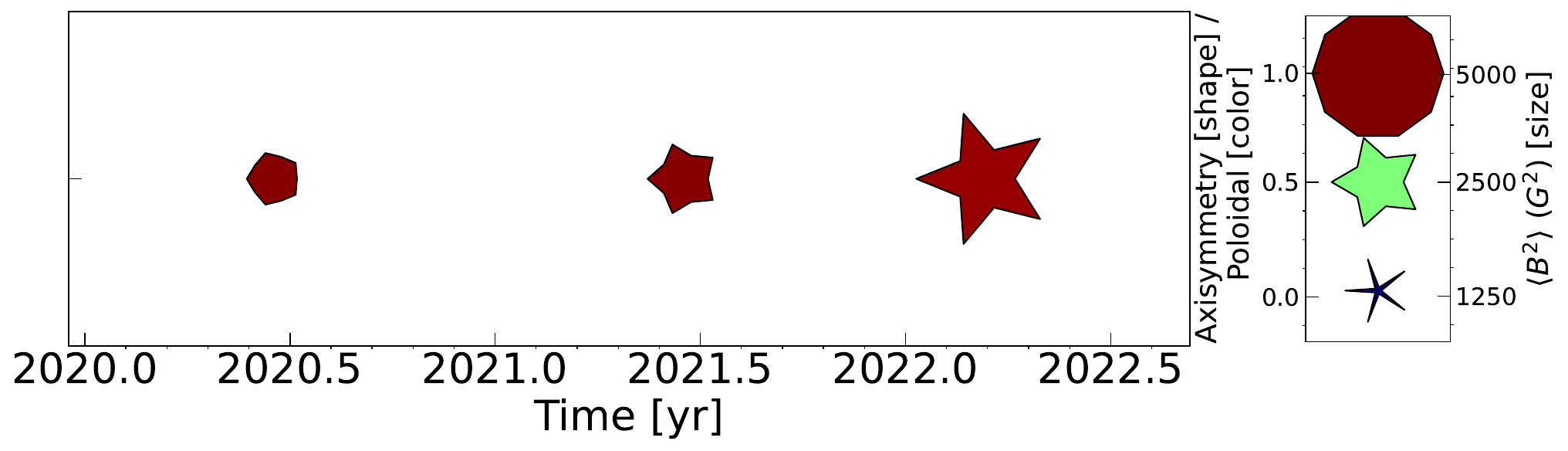}
    \caption{Same as Fig.~\ref{Fig:Gl905_ZDIMaps} for GJ~1151. Note, that the magnetic field switches polarity.}
    \label{Fig:GJ1151_ZDIMaps}
\end{figure}

\begin{table}
%
%
    \caption[]{Same as Table~\ref{tab:MagProp_Gl905} for GJ~1151.}
    \label{tab:MagProp_GJ1151}
    \begin{center}
    \begin{tabular}{lccc}
        \hline
        \noalign{\smallskip}
      season & \bf{2019/20} & \bf{2020/21} & \bf{2021/22} \\
      start & 2019 Dec & 2020 Dec & 2021 Dec \\
      end & 2020 July & 2021 July & 2022 June \\
        \noalign{\smallskip}
        \hline
        \noalign{\smallskip}
\BV [G] & 26 & 35 & 63 \\
$\langle B_\mrm{dip} \rangle$[G] & 23 & 32 & 62 \\
\Btormax [G] & $448$ & $378$ & $372$\\
$f_{\mrm{pol}}$ & 0.99 & 0.99 & 0.98 \\
 $f_{\mrm{axi}}$ & 0.84 & 0.64 & 0.38 \\
 $f_{\mrm{dip}}$ & 0.64 & 0.72 & 0.75 \\
        \noalign{\smallskip}
        \hline
        \noalign{\smallskip}
dipole tilt angle &  $4^\circ$ & $32^\circ$ & $-55^\circ$ \\
pointing phase & 0.44 & 0.05 & 0.33 \\
        \noalign{\smallskip}
        \hline
        \noalign{\smallskip}
 $\chi^2_{r,V}$ & 1.16 & 1.25 & 2.57\\
 $\chi^2_{r,V,\mrm{ZDI}}$ & 1.00 & 0.87 & 0.91 \\
 $\chi^2_{r,N}$ & 1.00 & 1.01 & 0.88\\
 nb. obs & 38 & 53 & 67 \\
        \hline
    \end{tabular}
    \end{center}
\end{table}

\section{GJ~1286}

GJ~1286 (LHS~546) is the lowest mass M~dwarf in our sample ($M = 0.12\pm0.02\,$\Msun\ , \cc{\citealt{Cristofari2022}}). We analyse here 104 observations, which we split into two seasons (2020 June -- Dec, 2021 Aug -- Dec) for the per-season analysis.  As the first and last season (2019 Sep -- Dec and 2022 June) do not contain enough observations, we once more left 21 LSD profiles out of the per-season analysis.

\subsection{PCA analysis of GJ~1286}

\begin{figure}
	\raggedright \textbf{a.} \hspace{2.7cm} \textbf{b.} \\
	\centering
	\includegraphics[width=0.35\columnwidth, trim={0 0 0 0}, clip]{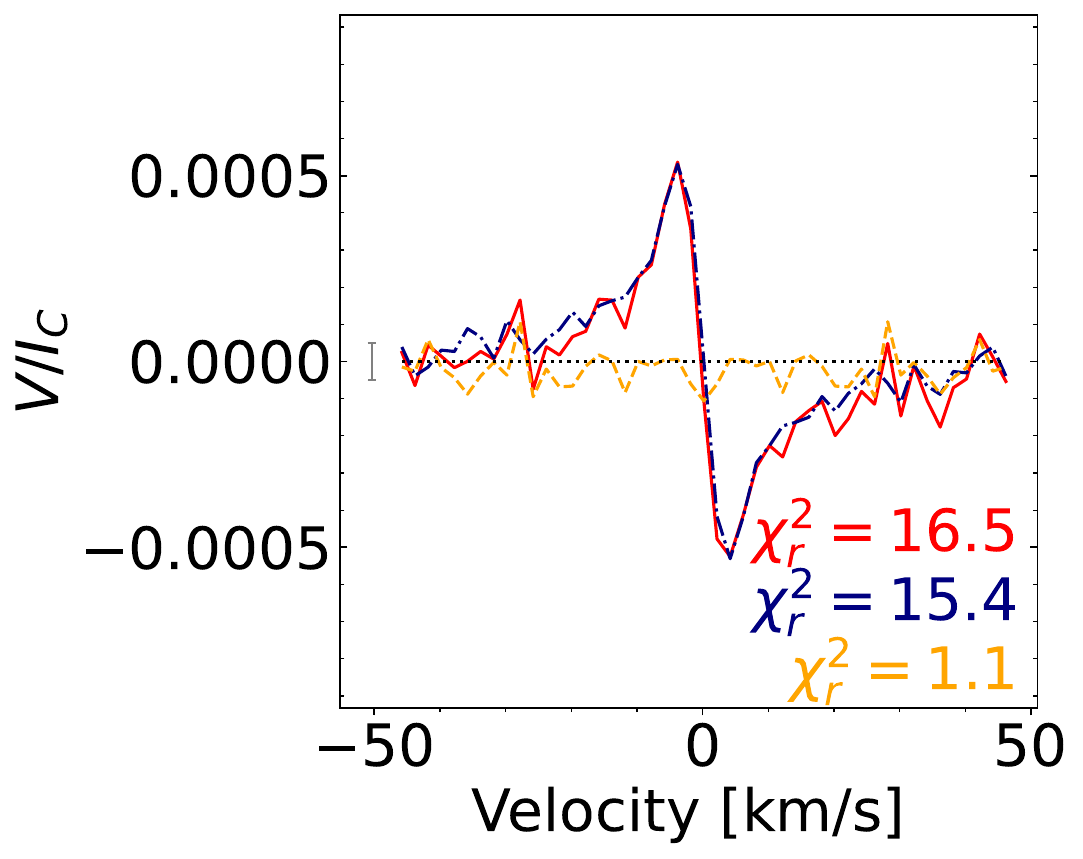}
	\includegraphics[width=0.64\columnwidth, trim={0 400 445 0}, clip]{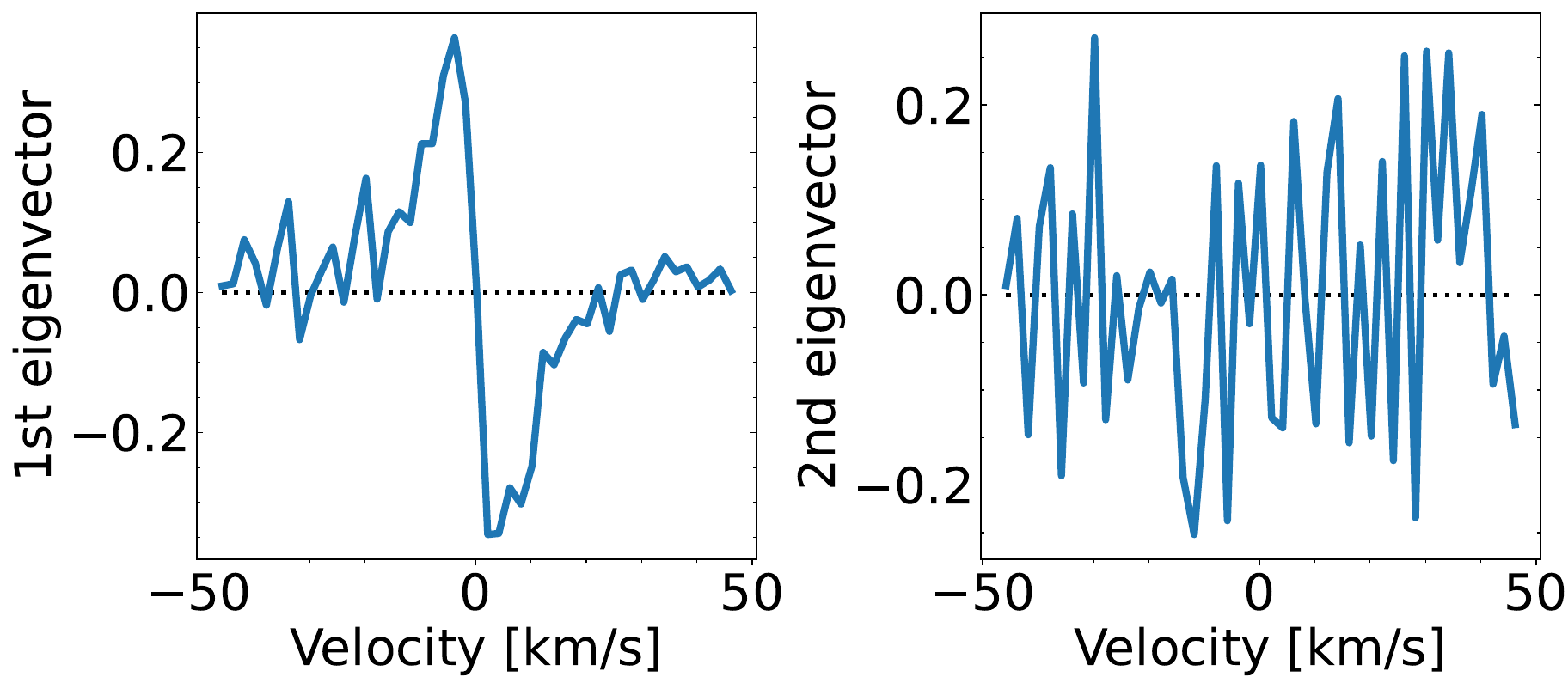}\\ 
	\rule{7cm}{0.3mm}\\
	\raggedright \textbf{c.} \\
	\centering
	\includegraphics[width=0.35\columnwidth, trim={0 0 0 0}, clip]{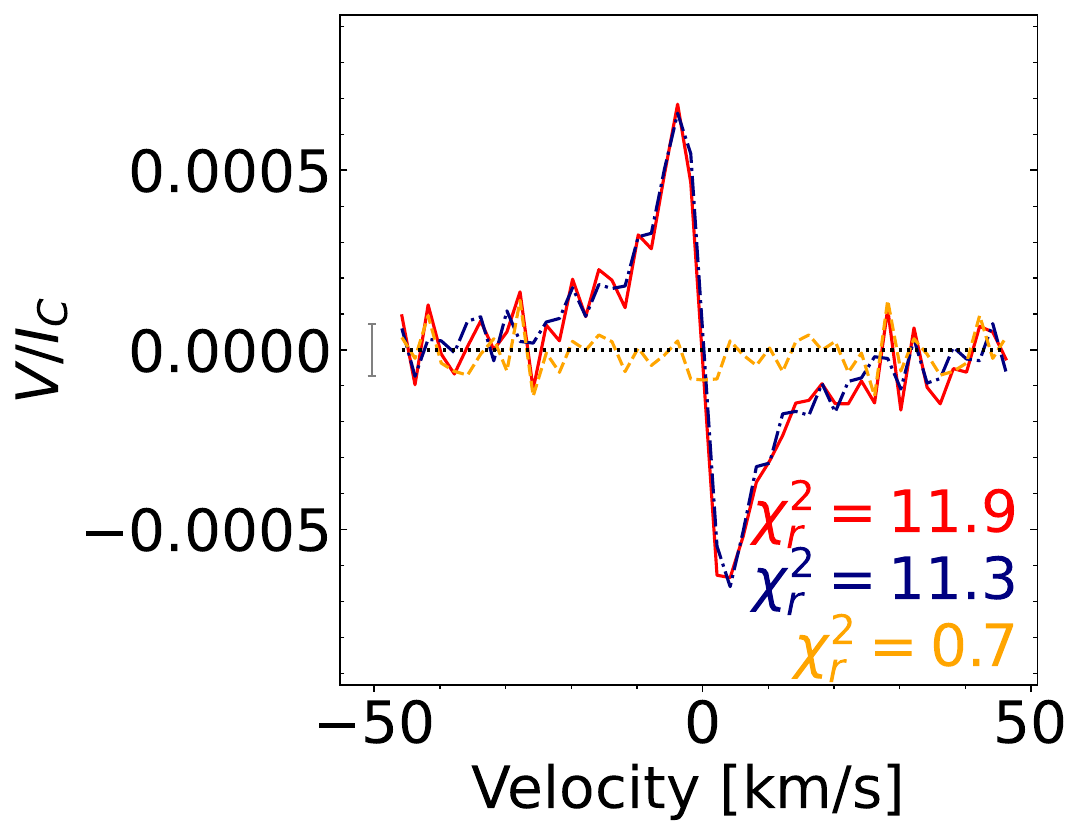}
	\includegraphics[width=0.62\columnwidth, trim={40 400 450 0}, clip]{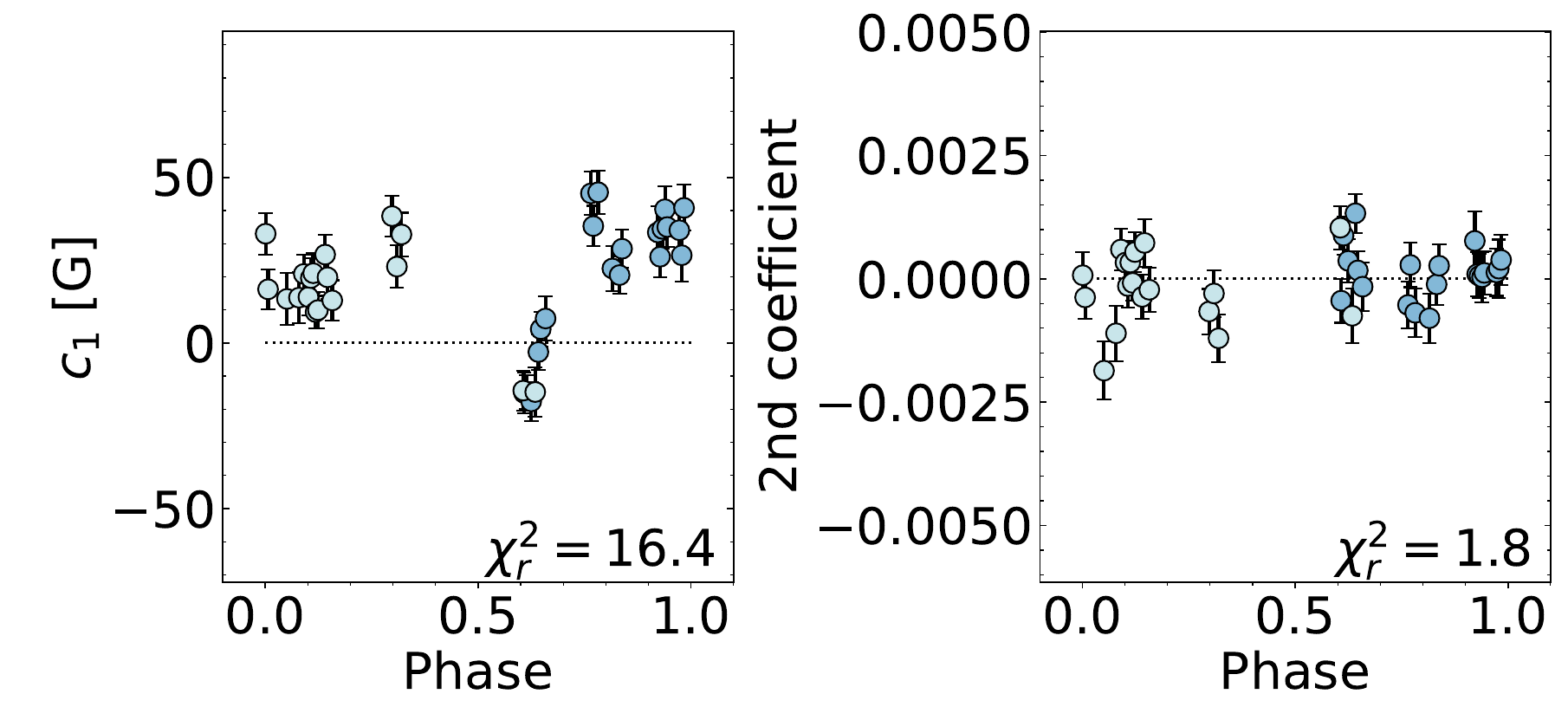}\\
		\includegraphics[width=0.35\columnwidth, trim={0 0 0 0}, clip]{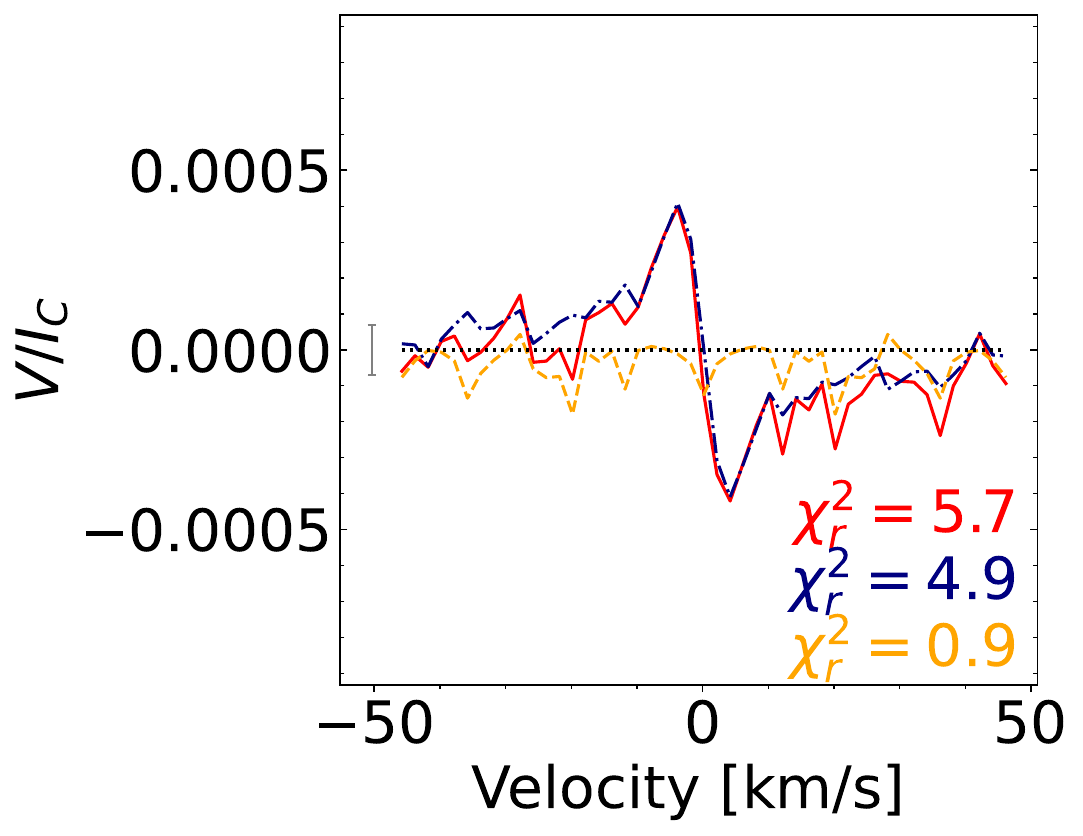}
	\includegraphics[width=0.62\columnwidth, trim={40 400 450 0}, clip]{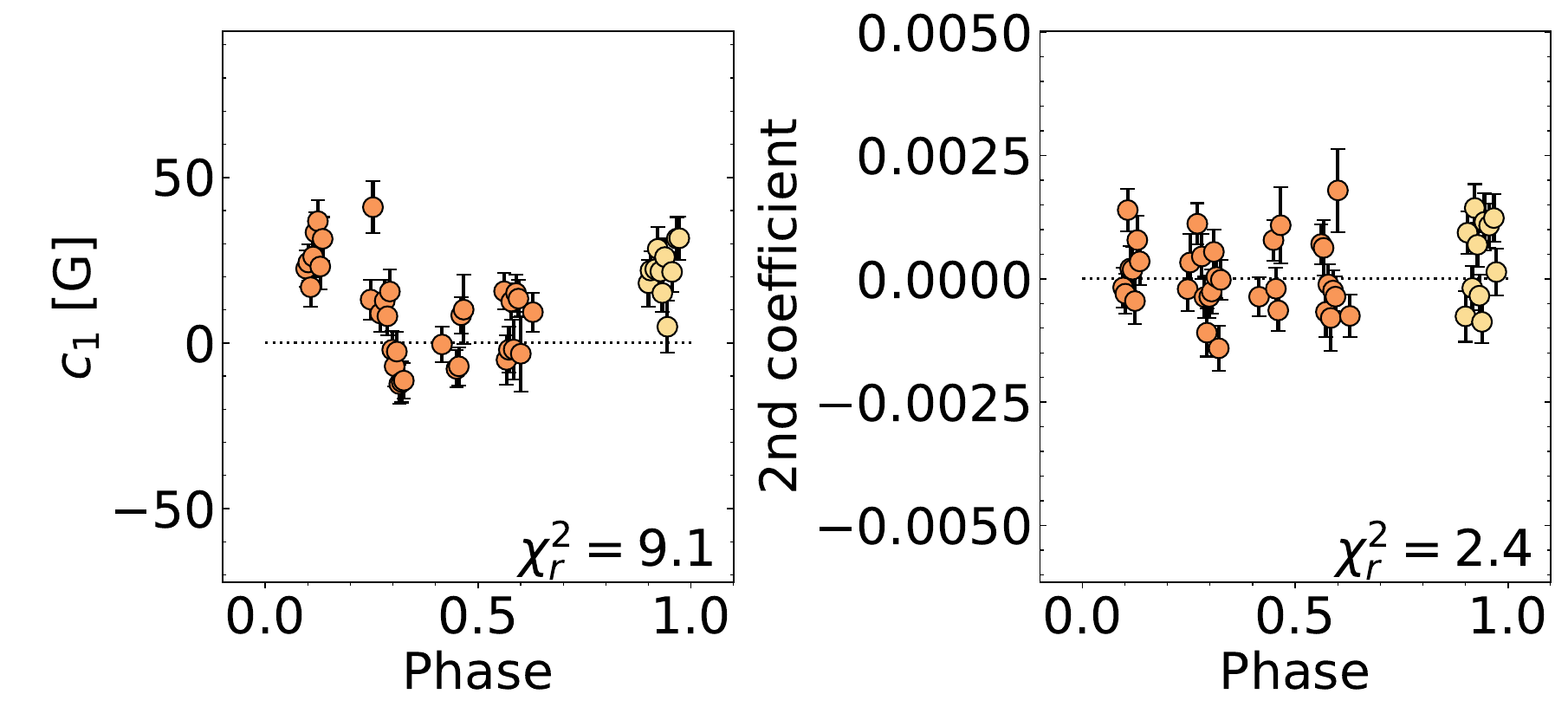}\\
    \caption{Same as Fig.~\ref{Fig:Gl905_PCA} for GJ~1286.}
    \label{Fig:GJ1286_PCA}
\end{figure}

The mean profile of GJ~1286 is antisymmetric with respect to the line centre and appears more noisy than usual but clearly indicates a purely axisymmetric poloidal field (see Fig.~\ref{Fig:GJ1286_PCA}a). The first eigenvector has an antisymmetric shape, too, and is the only one that emerges from the noise (see Fig.~\ref{Fig:GJ1286_PCA}b). 

\begin{figure}
	\centering
	\includegraphics[width=\columnwidth, trim={0 0 0 0}, clip]{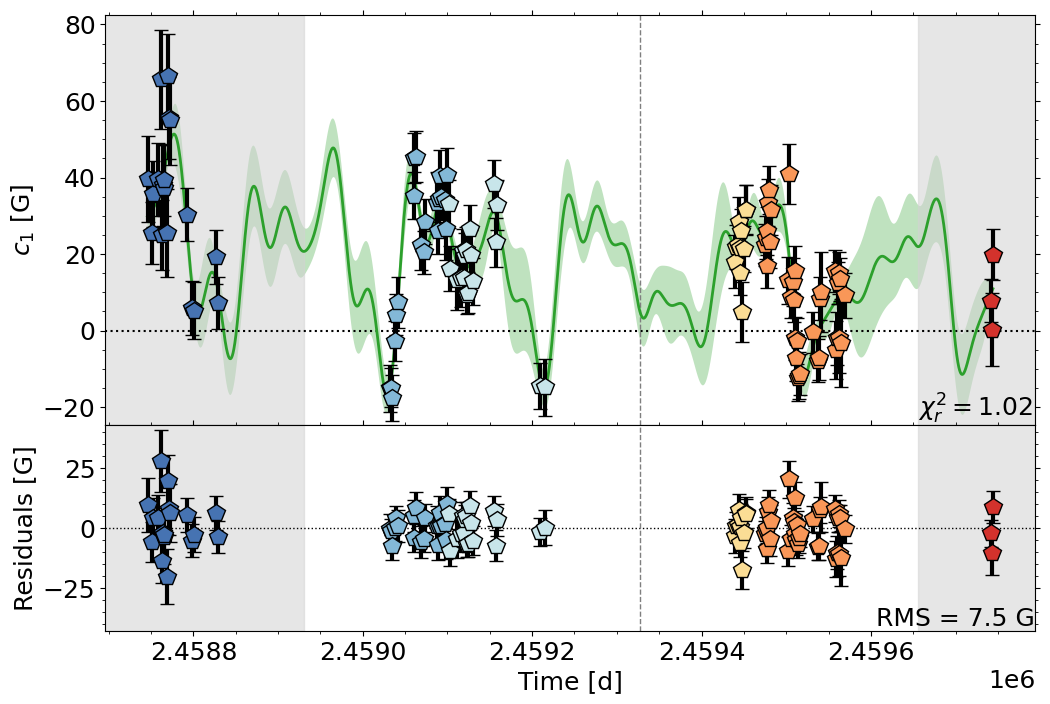}\\ 
		\includegraphics[width=\columnwidth, trim={0 0 0 0}, clip]{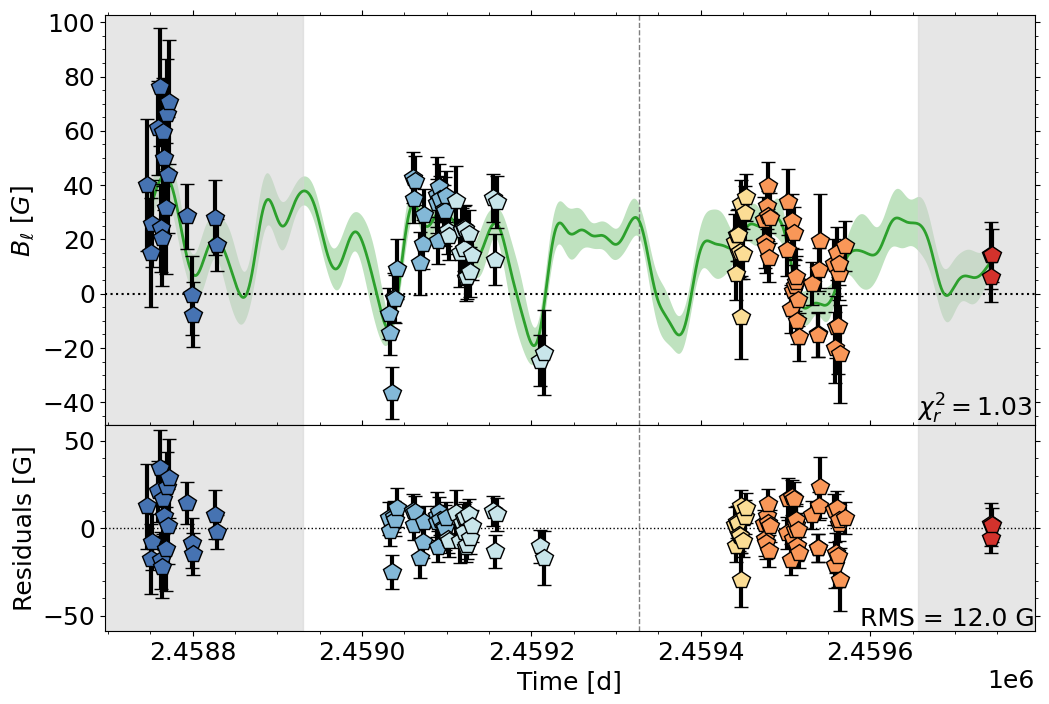}\\ 
    \caption{Same as Fig.~\ref{Fig:Gl905_CoeffvsTime} for GJ~1286.}
    \label{Fig:GJ1286_CoeffvsTime}
\end{figure}

The best-fitting model of the QP GPR for $c_1$ finds a $\Prot = 186.8^{+9.5}_{-5.8}\,\dy$ for a fixed decay time of 300\,d reaching a $\chi^2_r = 1.02$ similar to the GPR fit of $B_\ell$ (see Fig.~\ref{Fig:GJ1286_CoeffvsTime}). 
The rotation period agrees with the values found by D23 and \cite{Fouque2023} ($\Prot = 178\pm15\,\dy$ and $\Prot = 203^{+14}_{-21}\,\dy$, respectively) and our own GP result for $B_\ell$ ($\Prot = 181^{+18}_{-13}\,\dy$).

The mean profile of season 2020 is nearly twice as high as for 2021 (see Fig.~\ref{Fig:GJ1286_PCA}c). Also $c_1$ shows a lower amplitude for 2021. We can therefore conclude, that the surface averaged field decreases for 2021. The field topology becomes simpler as the $c_1$ curve gets less complex with phase for 2021.

\subsection{ZDI reconstructions of GJ~1286}

We fitted the LSD Stokes $V$ profiles of the two seasons for GJ~1286 down to $\chi^2_r \approx 1.0$, assuming $\Prot = 178\,\dy$, $f_V = 0.1, v_e \sin i = 0.03\,\kms$ and $i = 60^\circ$. 

\begin{figure}
\centering
\begin{minipage}{0.49\columnwidth}
\centering
\includegraphics[width=8cm, angle=270, trim={140 0 0 29}, clip]{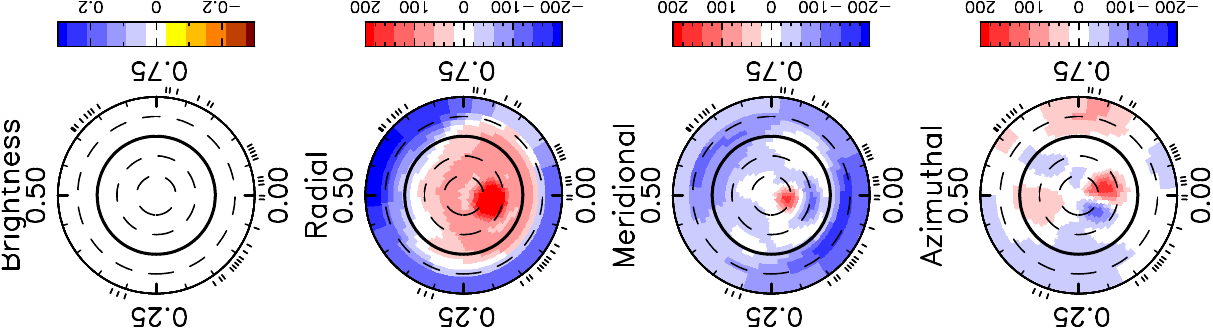} 
\end{minipage}
\begin{minipage}{0.49\columnwidth}
\centering
\includegraphics[width=8cm, angle=270, trim={140 0 0 29}, clip]{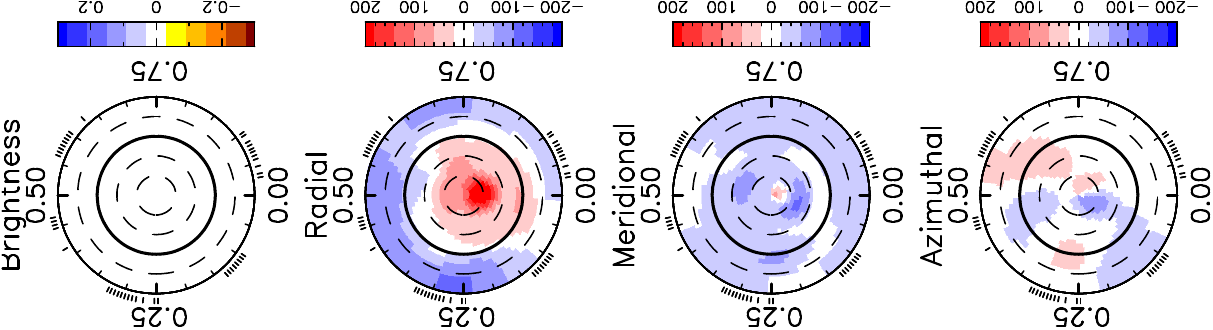} 
\end{minipage}
\includegraphics[height=0.6cm, angle=180, trim={460 130 2 0}, clip]{Figures/GJ1286_ZDIMap_JFDLSD_epo2_v11.pdf}
\vspace*{2mm}
\includegraphics[width=0.95\columnwidth, clip]{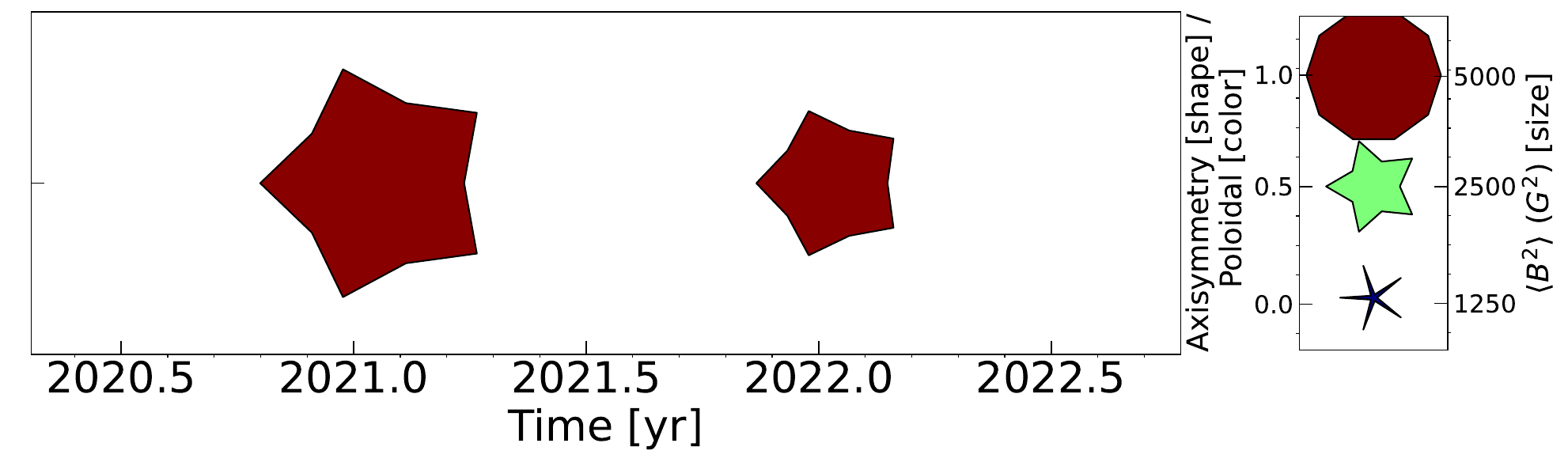}
    \caption{Same as Fig.~\ref{Fig:Gl905_ZDIMaps} for GJ~1286.}
    \label{Fig:GJ1286_ZDIMaps}
\end{figure}

As concluded from the PCA analysis, the ZDI maps confirm that the topology becomes simpler and weaker: the fractional energy of the dipole component $f_{\mrm{dip}}$ increases from 0.73 to 0.79, while \BV\ decreases by almost half from $113\,$G to $71\,$G (see Fig.~\ref{Fig:GJ1286_ZDIMaps} and Tab.~\ref{tab:MagProp_GJ1286}). The reconstructed toroidal field \Btor\ of the seasons are 9\,G and 6\,G, respectively, while the typical 1$\sigma$ error bars on the axisymmetric toroidal field are 325\,G and 300\,G, respectively.

\begin{table}
%
%
    \caption[]{Same as Table~\ref{tab:MagProp_GJ1289} for GJ~1286.}
    \label{tab:MagProp_GJ1286}
    \begin{center}
    \begin{tabular}{lcc}
        \hline
        \noalign{\smallskip}
      season & \bf{2020} & \bf{2021}  \\
      start & 2020 June & 2021 Aug \\
      end & 2020 Dec & 2021 Dec \\
        \noalign{\smallskip}
        \hline
        \noalign{\smallskip}
\BV [G] & 113 & 71 \\
$\langle B_\mrm{dip} \rangle$[G]  & 103 & 67 \\
\Btormax [G] & 325 & 300 \\
$f_{\mrm{pol}}$ & 0.99 & 0.99 \\
 $f_{\mrm{axi}}$ & 0.67 & 0.79 \\
 $f_{\mrm{dip}}$ & 0.73 & 0.79 \\
        \noalign{\smallskip}
        \hline
        \noalign{\smallskip}
dipole tilt angle &  $28^\circ$ & $25^\circ$ \\
pointing phase & 0.03 & 0.95 \\
        \noalign{\smallskip}
        \hline
        \noalign{\smallskip}
 $\chi^2_{r,V}$ & 2.21 & 1.65 \\
 $\chi^2_{r,V,\mrm{ZDI}}$ & 0.87 & 0.94 \\
 $\chi^2_{r,N}$ & 1.05 & 0.82 \\
 nb. obs & 38 & 45 \\
        \hline
    \end{tabular}
    \end{center}
\end{table}

\section{Gl~617B}

Gl~617B (EW~Dra, HIP~79762, LHS~3176) is a partly convective M~dwarf with $M = 0.45\pm0.02\,$\Msun\ \ \cc{\citep{Cristofari2022}} and was observed between 2019 Sept and 2022 June with SPIRou. Our following analysis is based on 144 LSD Stokes spectra, which we split into  three seasons (2020 Feb -- Oct, 2021 Jan -- July, 2022 Mar -- June) for the per-season analysis. As for the other stars, the first 15 spectra collected in 2019 were left out of the per-season analysis.

\subsection{PCA analysis of Gl~617B}

\begin{figure}
	\raggedright \textbf{a.} \hspace{2.7cm} \textbf{b.} \\
	\centering
	\includegraphics[width=0.35\columnwidth, trim={0 0 0 0}, clip]{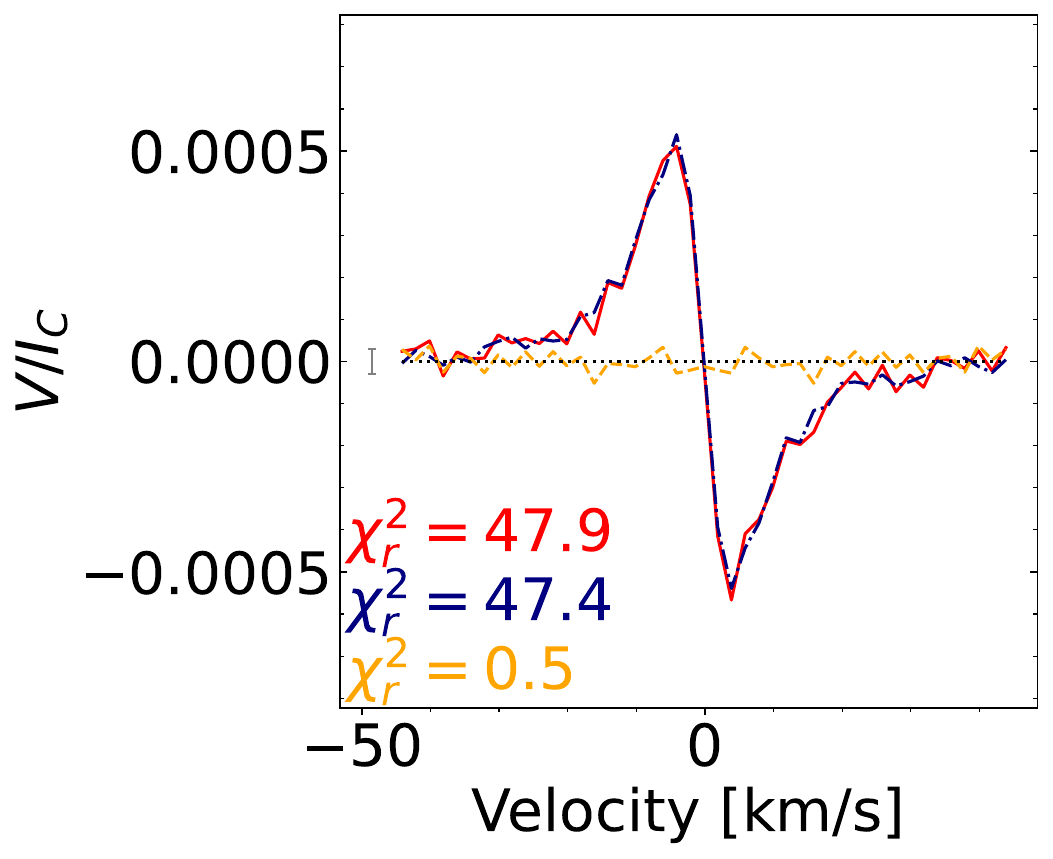}
	\includegraphics[width=0.63\columnwidth, trim={0 400 445 0}, clip]{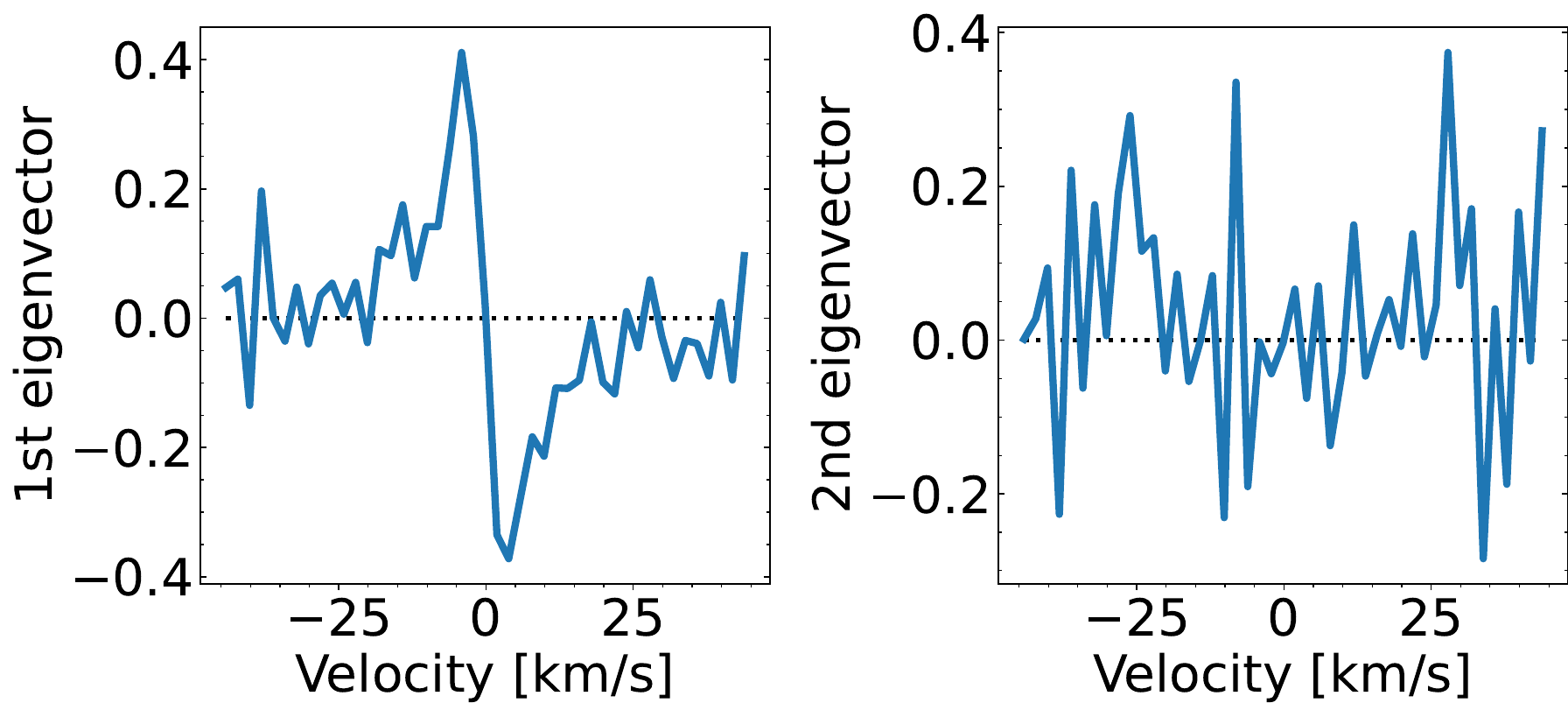}\\ 
	\rule{7cm}{0.3mm}\\
	\raggedright \textbf{c.} \\
	\centering
	\includegraphics[width=0.35\columnwidth, trim={0 0 0 0}, clip]{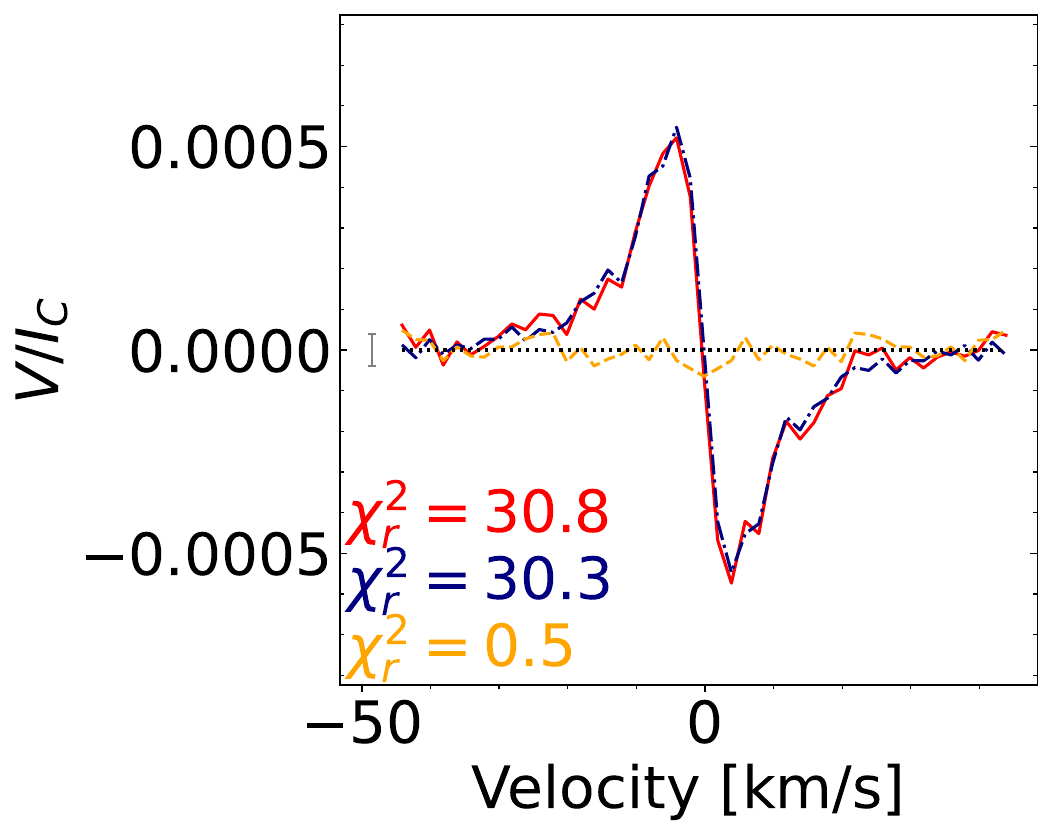}
	\includegraphics[width=0.63\columnwidth, trim={30 400 445 0}, clip]{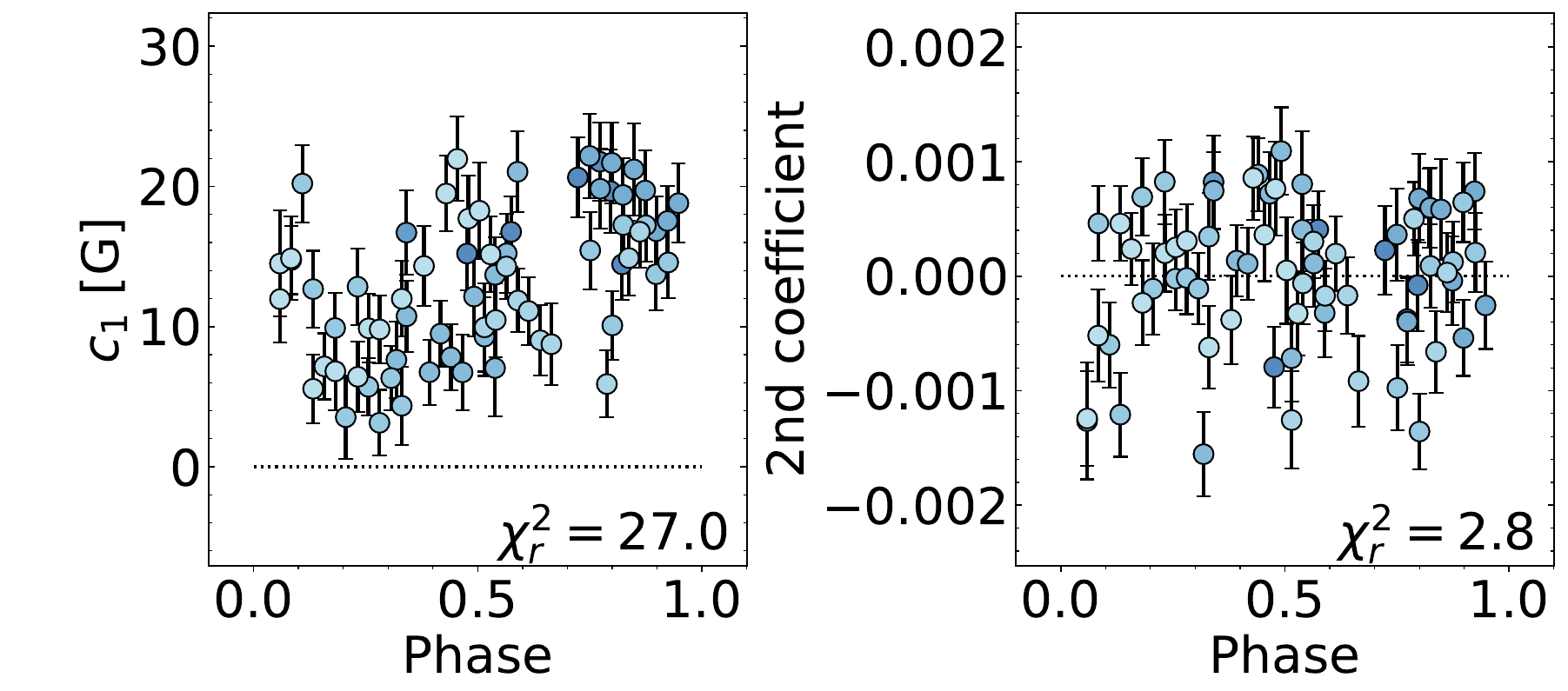}\\
		\includegraphics[width=0.35\columnwidth, trim={0 0 0 0}, clip]{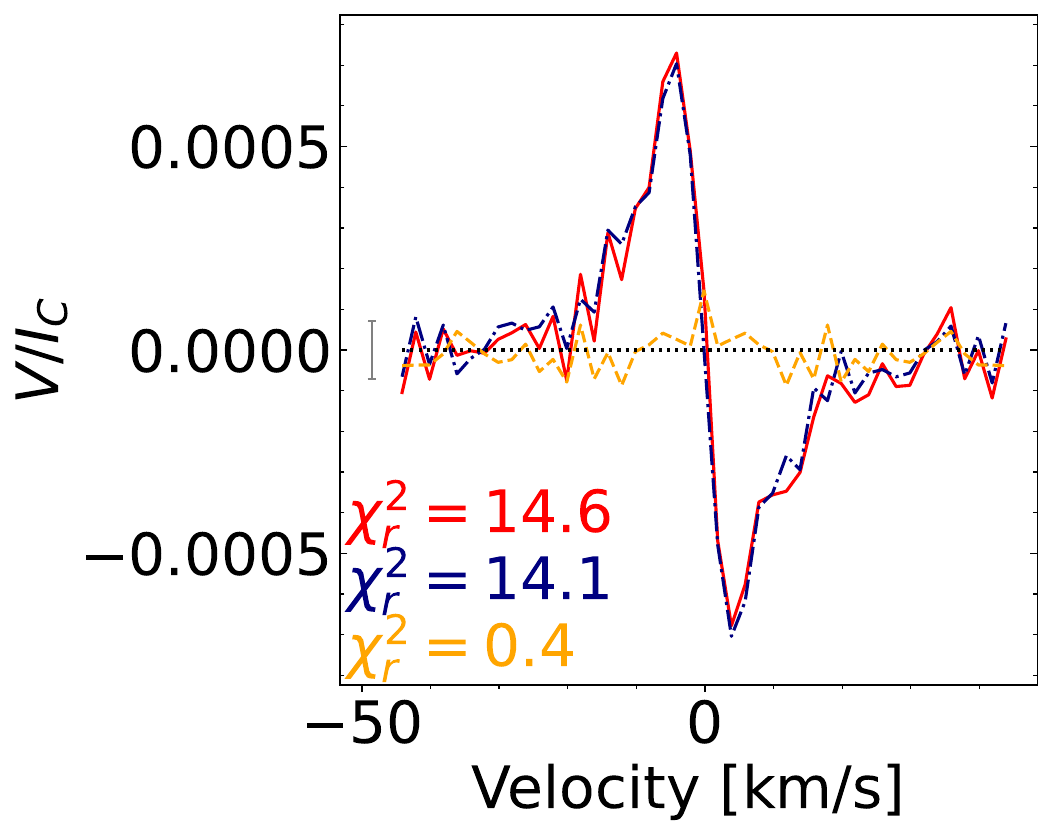}
	\includegraphics[width=0.63\columnwidth, trim={30 400 445 0}, clip]{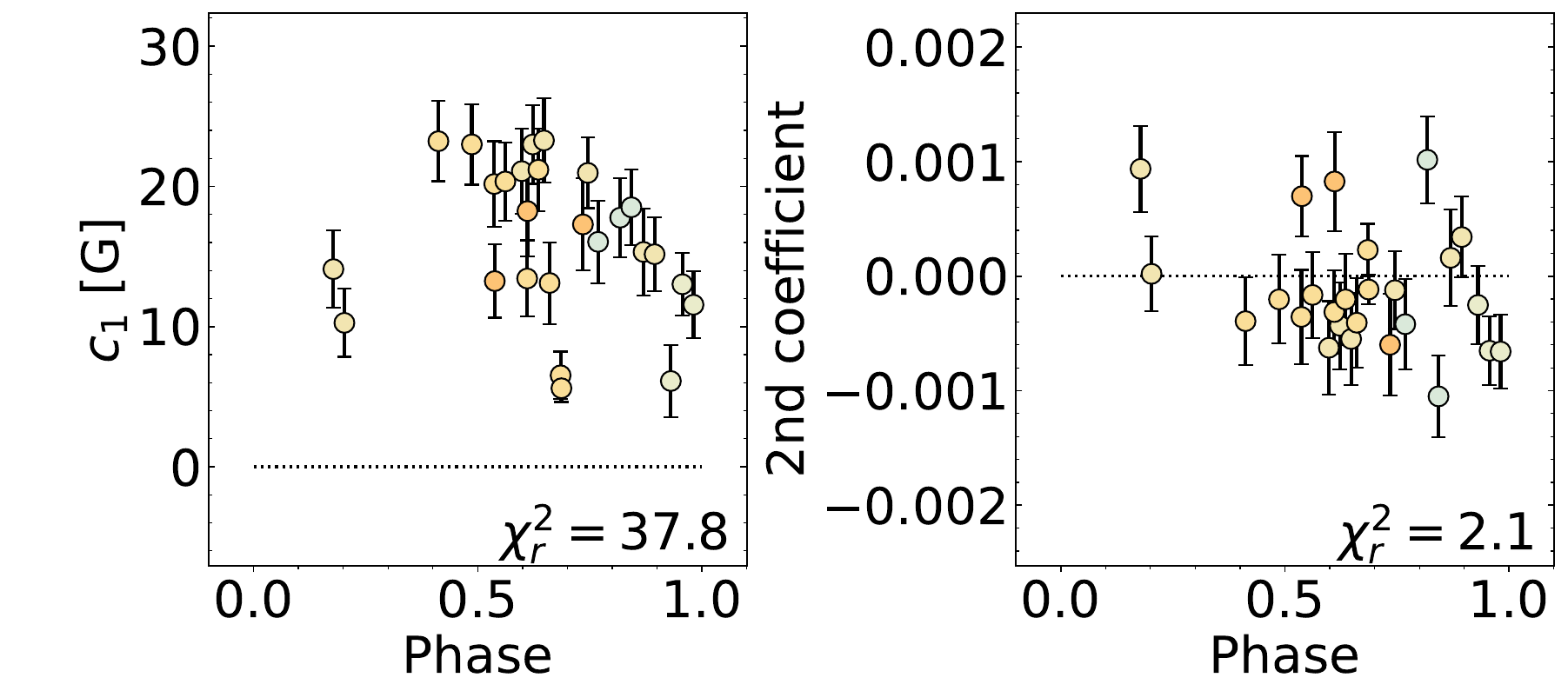}\\
		\includegraphics[width=0.35\columnwidth, trim={0 0 0 0}, clip]{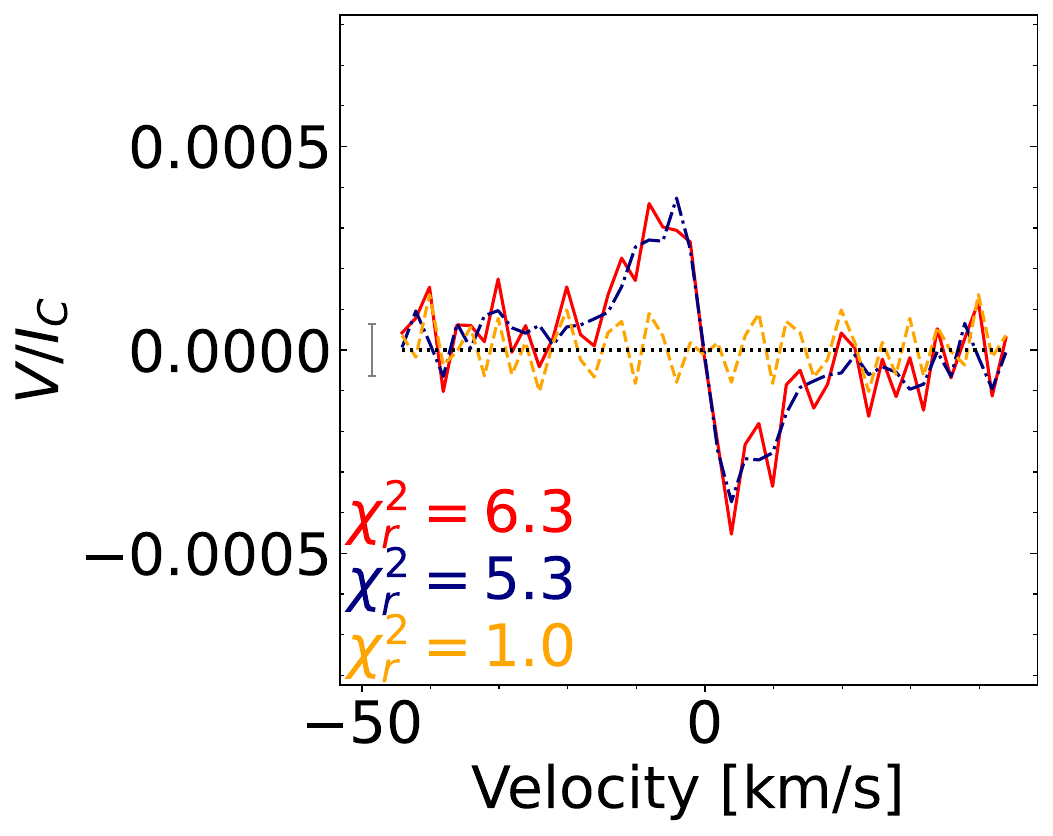}
	\includegraphics[width=0.63\columnwidth, trim={30 400 445 0}, clip]{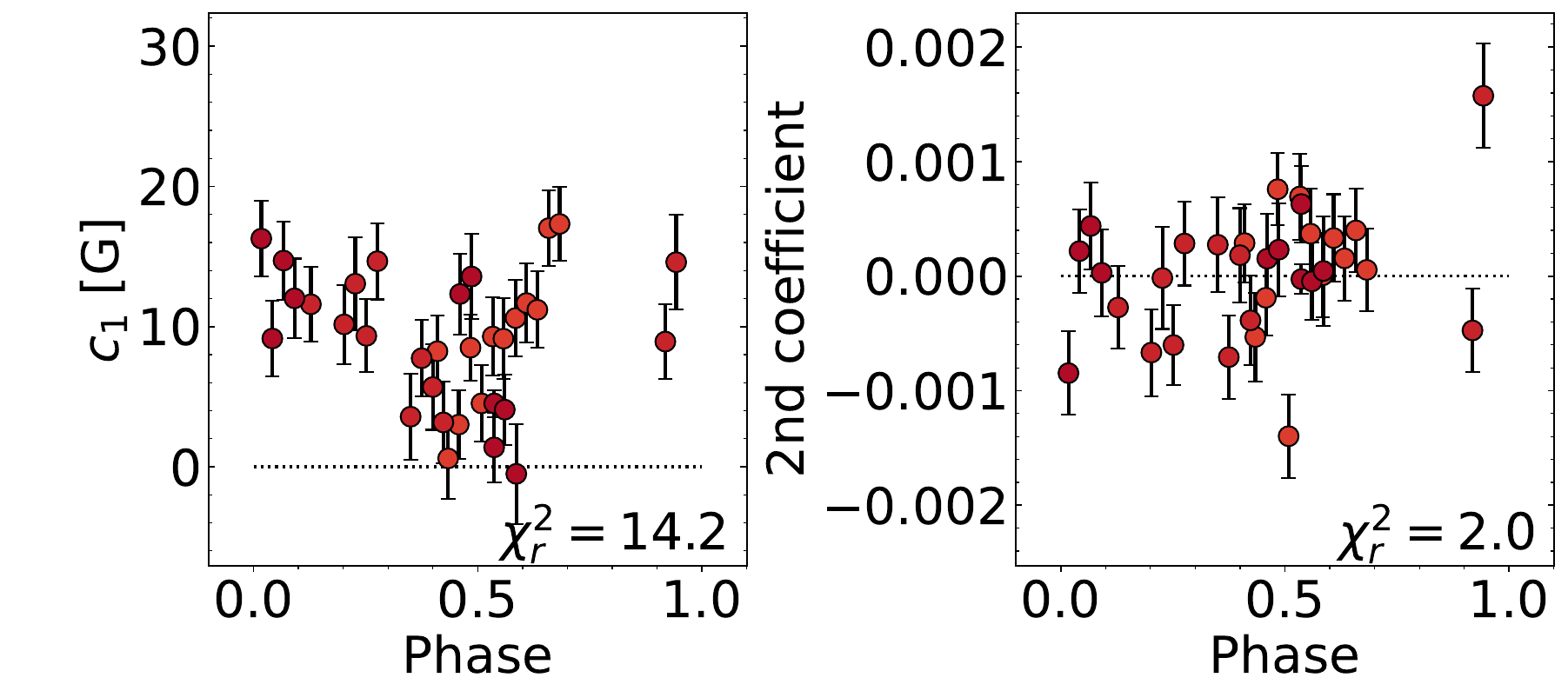}\\
    \caption{Same as Fig.~\ref{Fig:Gl905_PCA} for Gl~617B.}
    \label{Fig:Gl617B_PCA}
\end{figure}

We find, that the mean profile is much larger than the mean-subtracted Stokes~$V$ profiles indicating that the axisymmetric component of the magnetic field is dominant (see Fig.~\ref{Fig:Gl617B_PCA}a). The mean profile is again antisymmetric with respect to the line centre, indicating an axisymmetric poloidal field. The first eigenvector is already very noisy and is the only one that shows a clear signal, confirming that the field is indeed dominantly axisymmetric (see Fig.~\ref{Fig:Gl617B_PCA}b). 

\begin{figure}
	\centering
	\includegraphics[width=\columnwidth, trim={0 0 0 0}, clip]{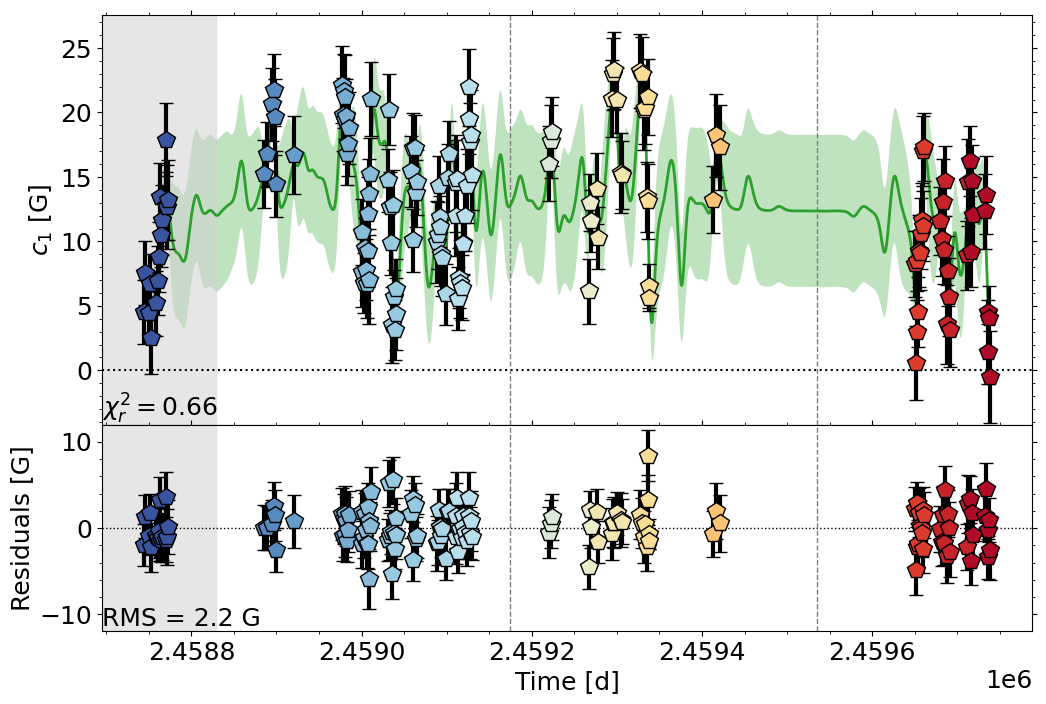}\\ 
		\includegraphics[width=\columnwidth, trim={0 0 0 0}, clip]{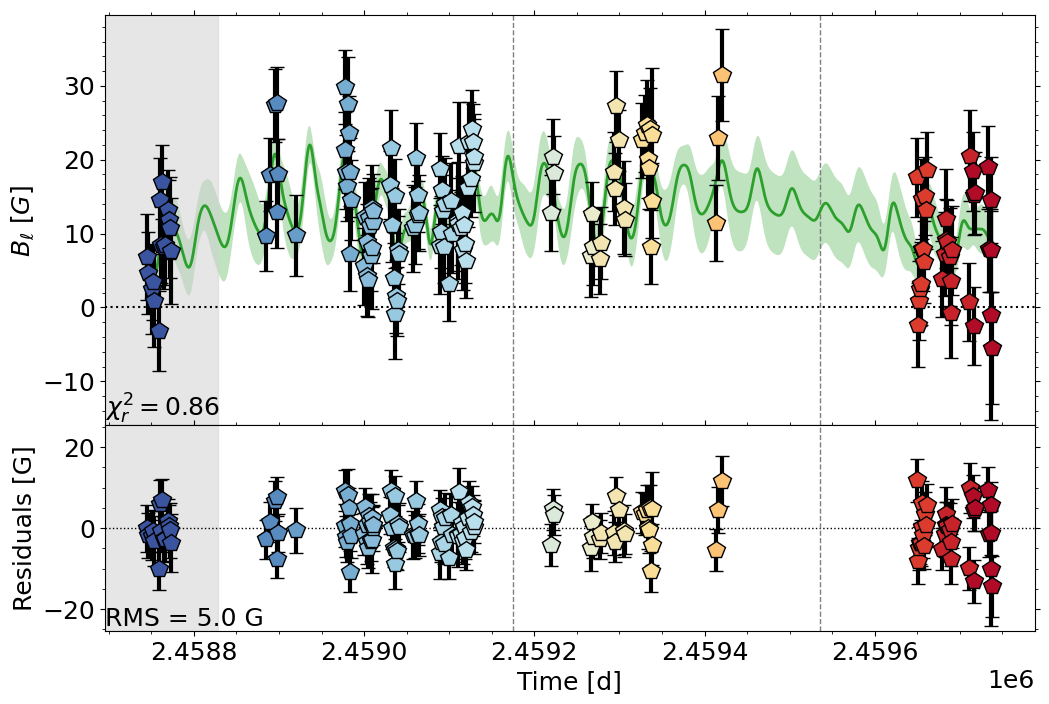}\\ 
    \caption{Same as Fig.~\ref{Fig:Gl905_CoeffvsTime} for Gl~617B.}
    \label{Fig:Gl617B_CoeffvsTime}
\end{figure}

The QP GPR model fits $c_1$ down to a $\chi^2_r = 0.66$ finding a rotation period of $37.8^{+8.5}_{-2.6}\,\dy$ in agreement with the results of D23 ($\Prot = 40.4\pm3.0\,\dy$) and our GP fit of $B_\ell$ ($\Prot = 40.6^{+2.1}_{-4.4}\,\dy$, see Fig.~\ref{Fig:Gl617B_CoeffvsTime}). However, the decay time for the GP fit of $c_1$ with $l = 35^{+8}_{-4}\,\dy$ is shorter than the results determined from the $B_\ell$ curves ($l = 69^{+35}_{-23}\,\dy$ for D23 and $l = 82^{+45}_{-30}\,\dy$ from our own fit of $B_\ell$).
\cite{Fouque2023} found no clear periodic variation using the \textsc{APERO} pipeline reduced spectra of Gl~617B.

The mean profile is antisymmetric to the line centre and therefore poloidal dominated for all three seasons, but varies in amplitude (see Fig.~\ref{Fig:Gl617B_PCA}c, left column). Nonetheless, $c_1$ traces a varying non-axisymmetric component. 
Season 2020 shows the highest range in amplitude of $c_1$, indicating the largest dipole tilt angle of all three seasons, although it will still be small ($<20^\circ$) due to the predominantly axisymmetric topology of Gl~617B.

\subsection{ZDI reconstructions of Gl~617B}

We could fit Gl~617B down to $\chi^2_r \approx 1.0$ assuming $\Prot = 40.4\,\dy, f_V = 0.1, v_e\sin i = 0.50\,\kms, i = 60^\circ$.
The ZDI maps confirm a very axisymmetric, poloidal configuration. The axisymmetry is always equal to or greater than 97\%, so variations of the non-axisymmetric field are difficult to see, but appear largest in 2021 (see Tab.~\ref{tab:MagProp_Gl617B}). The data set in season 2020 shows the largest tilt angle ($7^\circ$) as predicted by the PCA analysis. The reconstructed toroidal field \Btor\ reaches 9\,G for 2020, 6\,G for 2021 and 2\,G for 2022, while the estimated 1$\sigma$ error bar on the axisymmetric toroidal field is about 7\,G in 2020, 13\,G in 2021 and 6\,G in 2022, which is a lower uncertainty than for the other M~dwarfs thanks to the higher $v_e \sin i = 0.5\,\kms$ of Gl~617B.

We see \BV\ changing by approximately $\pm 25$\,G for the three seasons, otherwise the main properties of the maps are similar (see Tab.~\ref{tab:MagProp_Gl617B}).

For highly axisymmetric topologies, it is difficult to infer the inclination $i$. It may be that $i$ is actually lower for Gl~617B. We provide the ZDI maps for an inclination $i = 30^\circ$ and $v_e \sin i = 0.29\,\kms$, while otherwise using the same parameters (see Fig.~\ref{Fig:Gl617B_ZDIMaps_i30}). The $\chi^2_r$ values reached for the ZDI fits are slightly higher for $i=30^\circ$ than for $i = 60^\circ$.

\begin{figure}
\centering
\begin{minipage}{0.32\columnwidth}
\centering
\includegraphics[height=0.85\columnwidth, angle=270, trim={140 0 0 29}, clip]{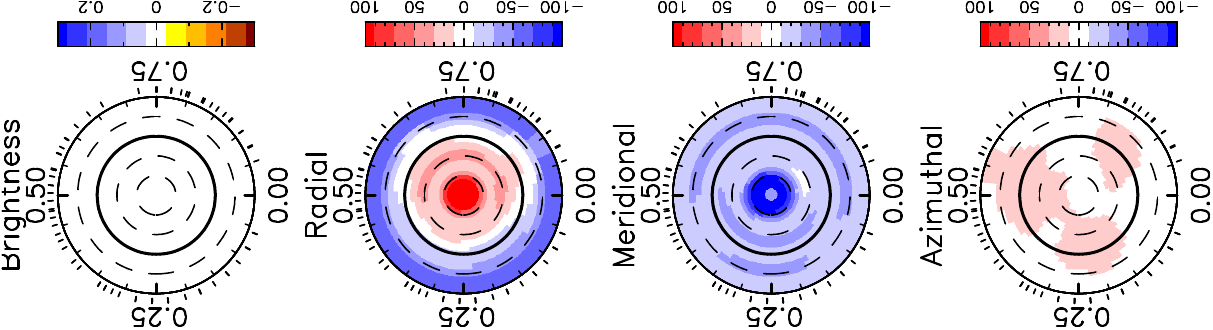} 
\end{minipage}
\begin{minipage}{0.32\columnwidth}
\centering
\includegraphics[height=0.85\columnwidth, angle=270, trim={140 0 0 29}, clip]{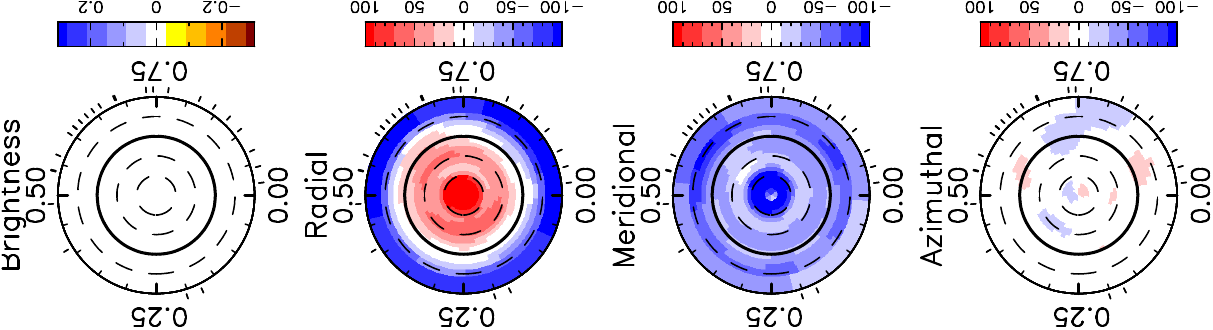} 
\end{minipage}
\begin{minipage}{0.32\columnwidth}
\centering
\includegraphics[height=0.85\columnwidth, angle=270, trim={140 0 0 29}, clip]{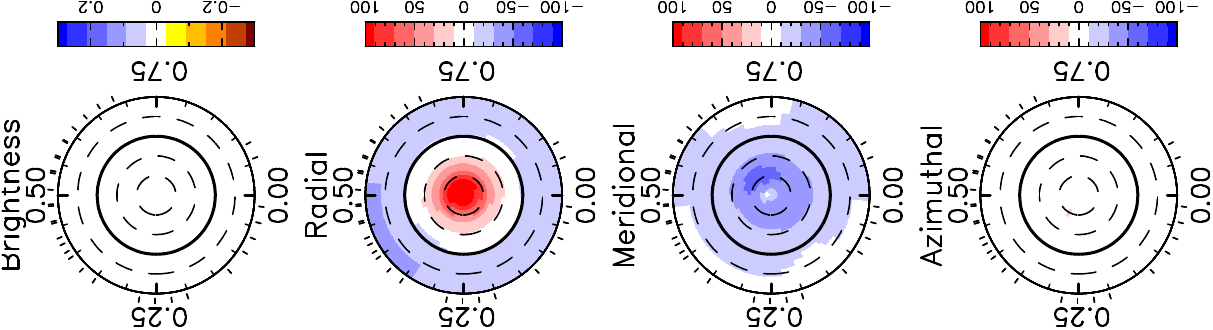} 
\end{minipage}
\includegraphics[width=0.3\columnwidth, angle=180, trim={460 130 2 0}, clip]{Figures/Gl617B_ZDIMap_JFDLSD_epo3_v10.pdf}
\vspace*{2mm}
\includegraphics[width=0.95\columnwidth, clip]{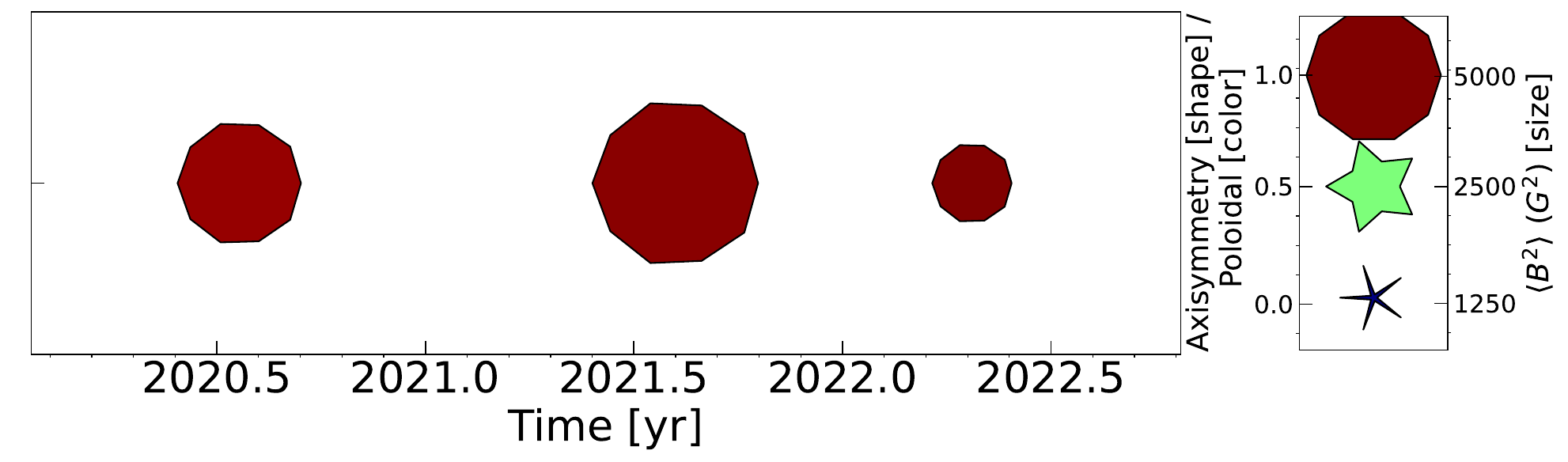}
    \caption{Same as Fig.~\ref{Fig:Gl905_ZDIMaps} for Gl~617B.}
    \label{Fig:Gl617B_ZDIMaps}
\end{figure}

\begin{table}
%
%
    \caption[]{Same as Table~\ref{tab:MagProp_GJ1289} for Gl~617B.}
    \label{tab:MagProp_Gl617B}
    \begin{center}
    \begin{tabular}{lccc}
        \hline
        \noalign{\smallskip}
    season & \bf{2020} & \bf{2021} & \bf{2022} \\
      start & 2020 Feb & 2021 Jan & 2022 Mar \\
      end & 2020 Oct & 2021 July & 2022 June \\
        \noalign{\smallskip}
        \hline
        \noalign{\smallskip}
\BV [G] & 53 & 75 & 36 \\
$\langle B_\mrm{dip} \rangle$[G]  & 52 & 73 & 35 \\
\Btormax [G] & 7 & 13 & 6 \\
$f_{\mrm{pol}}$ & 0.98 & 0.99 & 1.0 \\
 $f_{\mrm{axi}}$ & 0.98 & 0.97 & 0.98 \\
 $f_{\mrm{dip}}$ & 0.67 & 0.74 & 0.71 \\
        \noalign{\smallskip}
        \hline
        \noalign{\smallskip}
dipole tilt angle &  $7^\circ$ & $4^\circ$ & $3^\circ$ \\
pointing phase & 0.74 & 0.47 & 0.87 \\
        \noalign{\smallskip}
        \hline
        \noalign{\smallskip}
 $\chi^2_{r,V}$ & 2.37 & 3.08 & 1.59 \\
$\chi^2_{r,V,\mrm{ZDI}}$ & 0.99 & 0.95 & 1.00 \\
 $\chi^2_{r,N}$ & 1.00 & 1.06 & 0.89 \\
 nb. obs & 70 & 26 & 33 \\
        \hline
    \end{tabular}
    \end{center}
\end{table}

\section{Gl~408}
\label{Sec:Gl408}

Gl~408 (Ross~104, HIP~53767, LHS~6193) is another partly convective star, with $M = 0.38\pm0.02\,$\Msun\ , \cc{\citep{Cristofari2022}}. SPIRou observed Gl~408 between 2019 Apr and 2022 June. We use 157 Stokes~$V$ profiles for the following analysis split into three seasons (2019 Oct -- 2020 June, 2020 Oct -- 2021 July, 2021 Nov -- 2022 June) for the per-season analysis. As for the other stars, the first 17 spectra collected in early 2019 were left out of the per-season analysis.

\subsection{PCA analysis of Gl~408}

\begin{figure}
	\raggedright \textbf{a.} \hspace{2.7cm} \textbf{b.} \\
	\centering
	\includegraphics[width=0.34\columnwidth, trim={0 0 0 0}, clip]{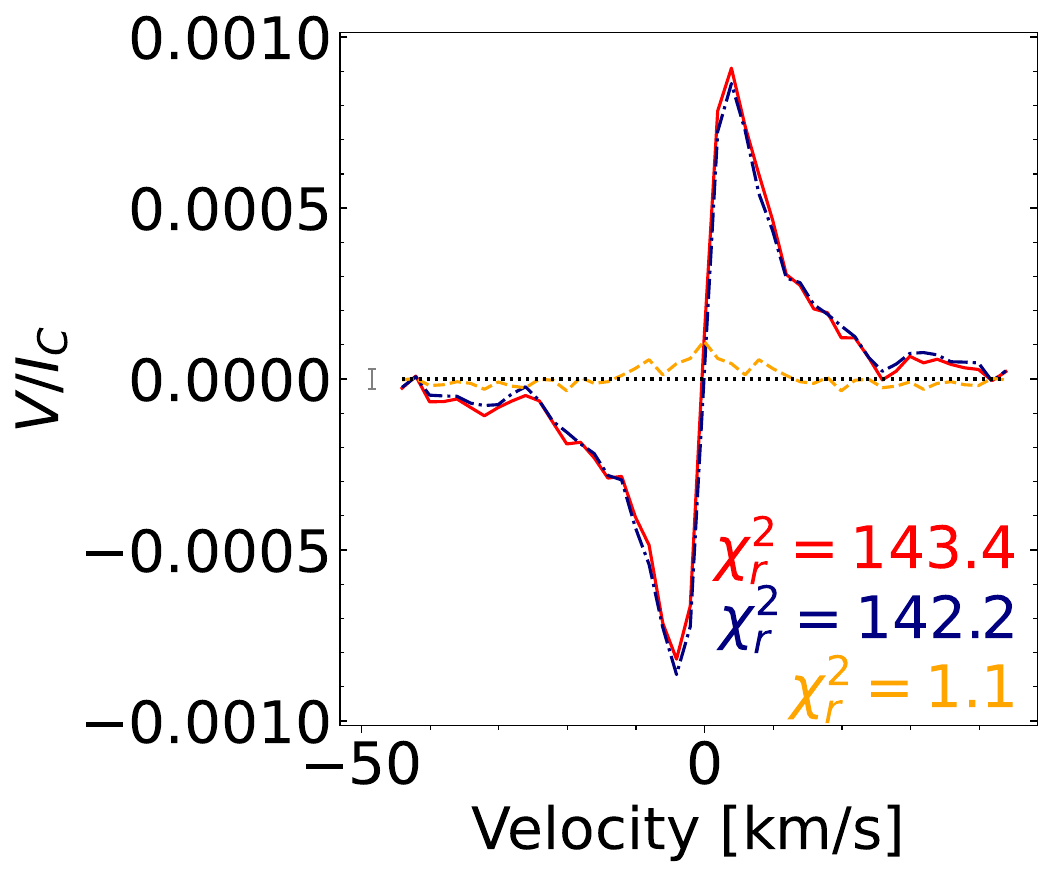}
	\includegraphics[width=0.63\columnwidth, trim={0 400 445 0}, clip]{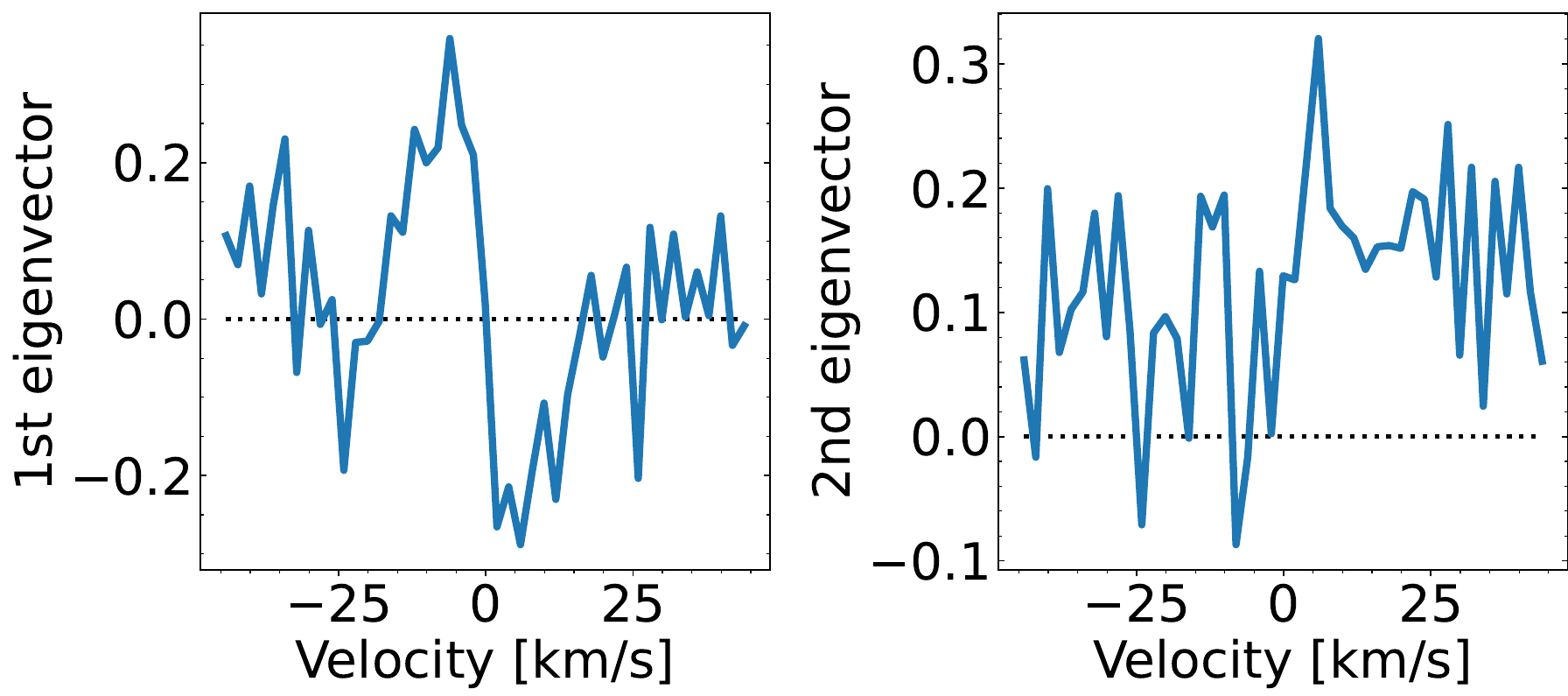}\\ 
	\rule{7cm}{0.3mm}\\
	\raggedright \textbf{c.} \\
	\centering
	\includegraphics[width=0.35\columnwidth, trim={0 0 0 0}, clip]{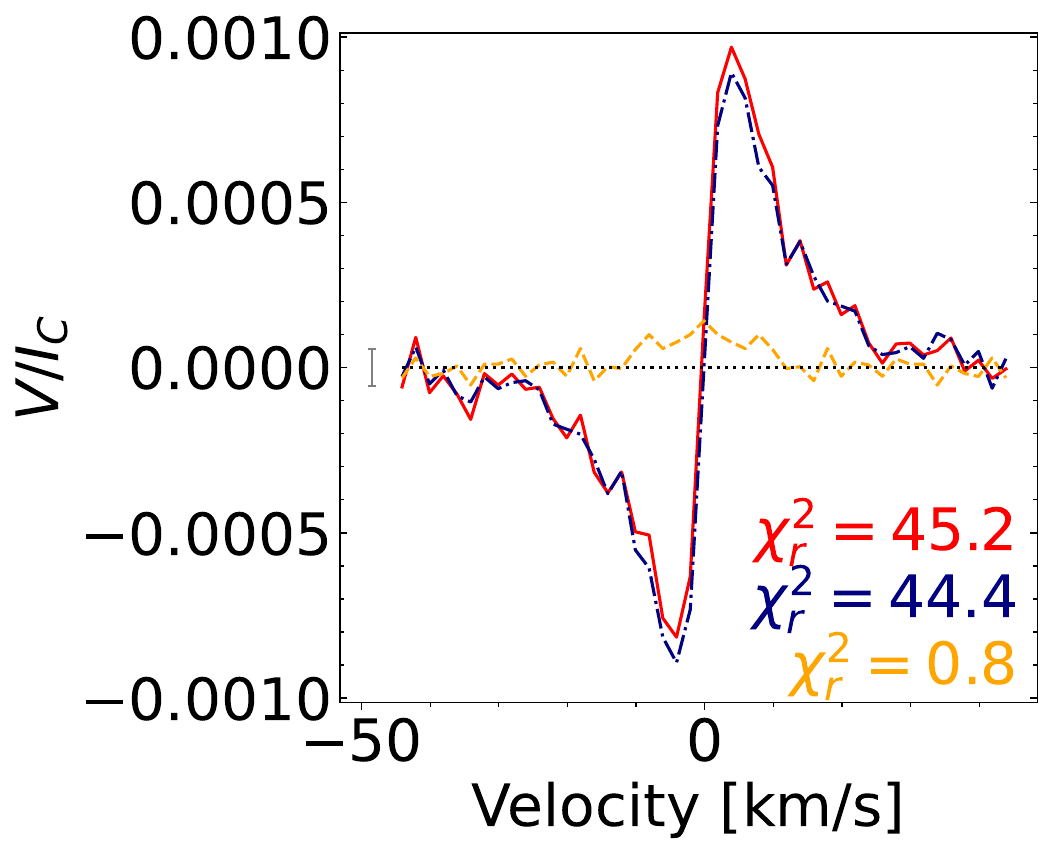}
	\includegraphics[width=0.63\columnwidth, trim={30 400 445 0}, clip]{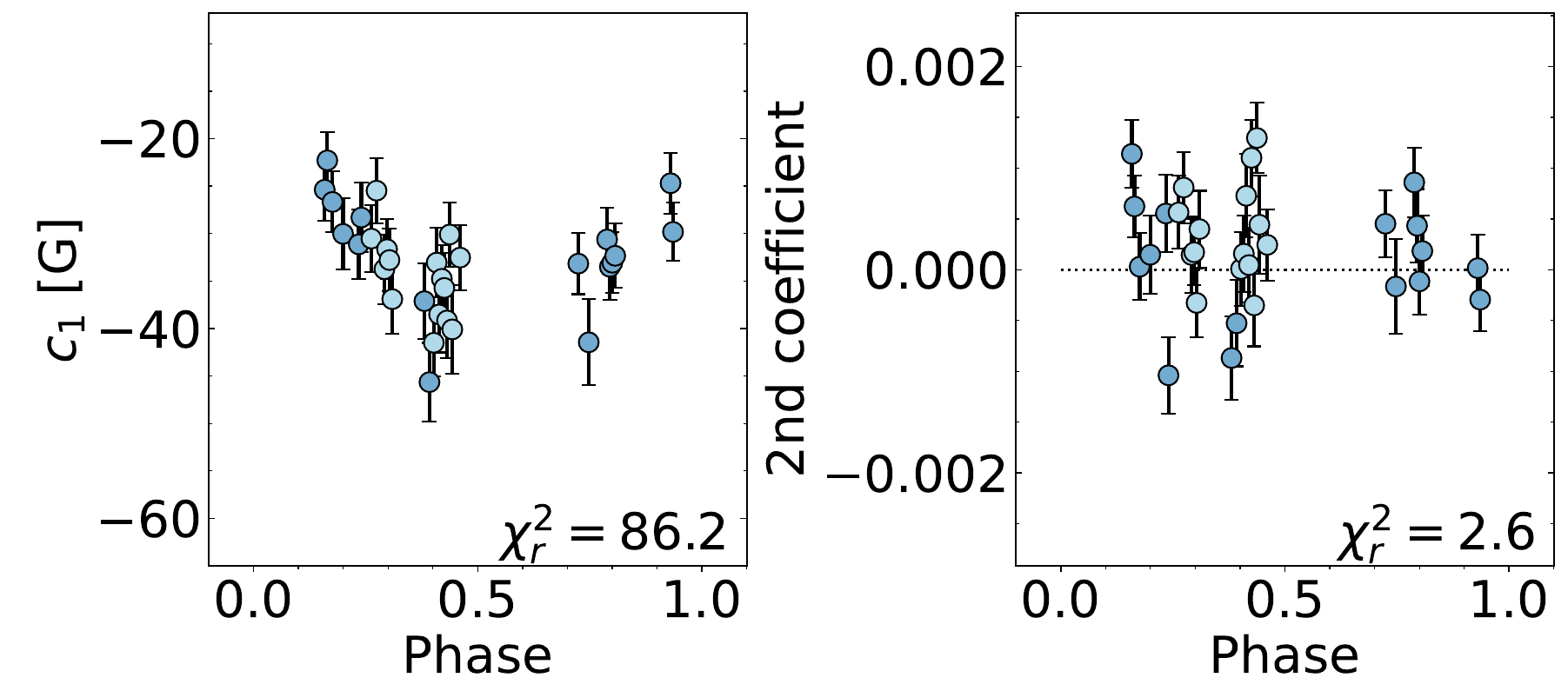}\\
		\includegraphics[width=0.35\columnwidth, trim={0 0 0 0}, clip]{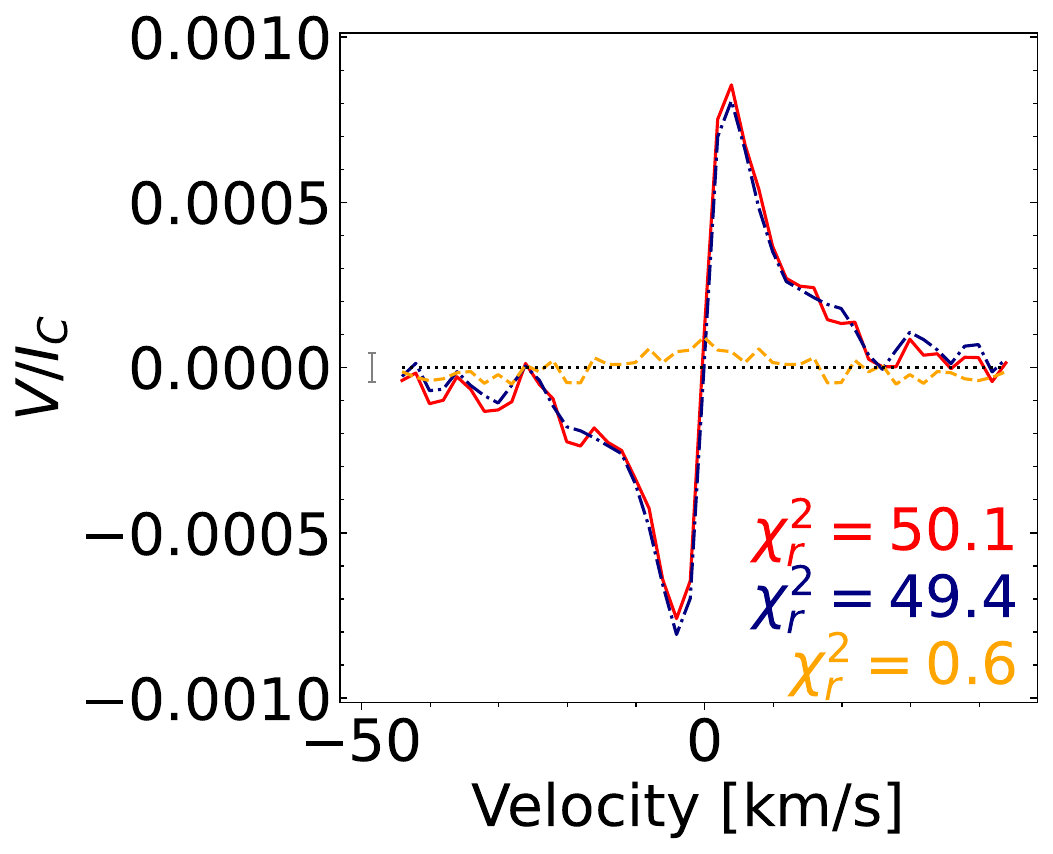}
	\includegraphics[width=0.63\columnwidth, trim={30 400 445 0}, clip]{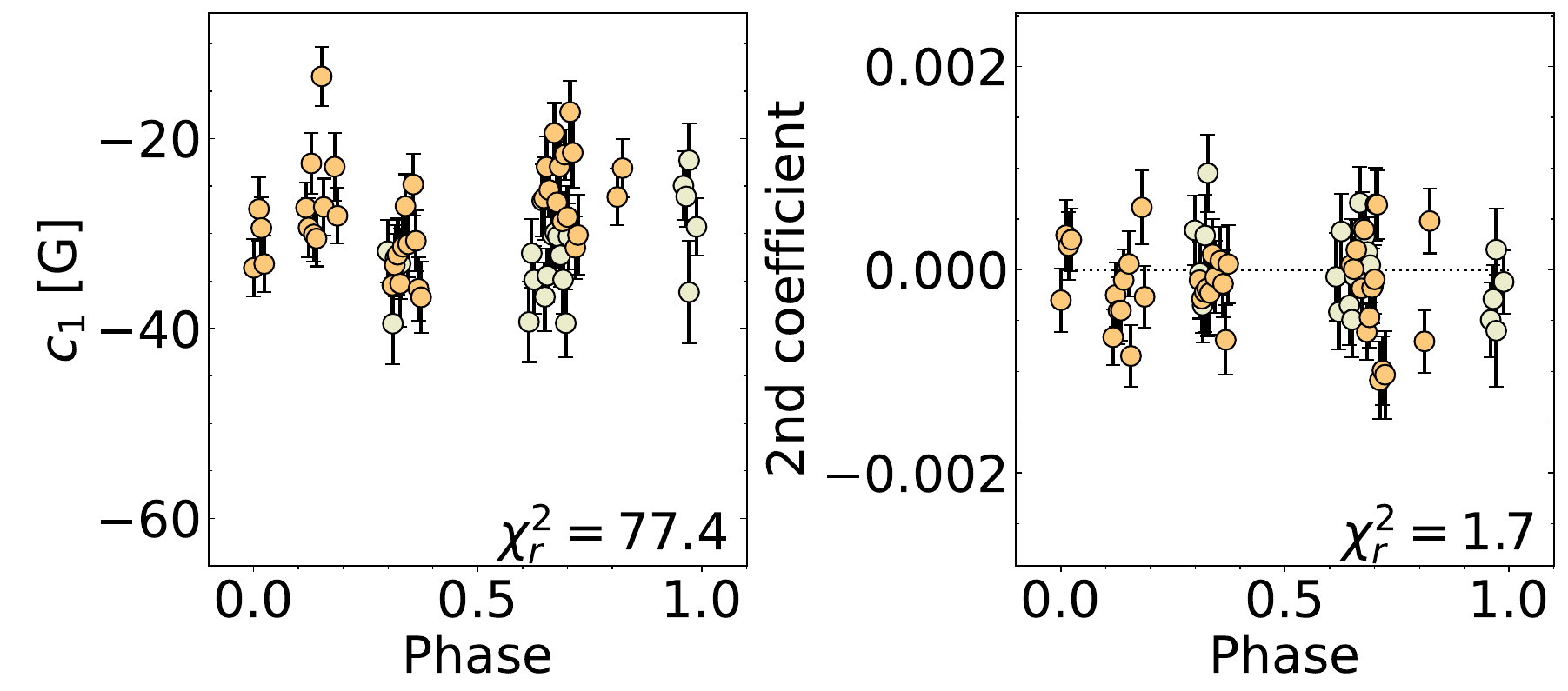}\\
		\includegraphics[width=0.35\columnwidth, trim={0 0 0 0}, clip]{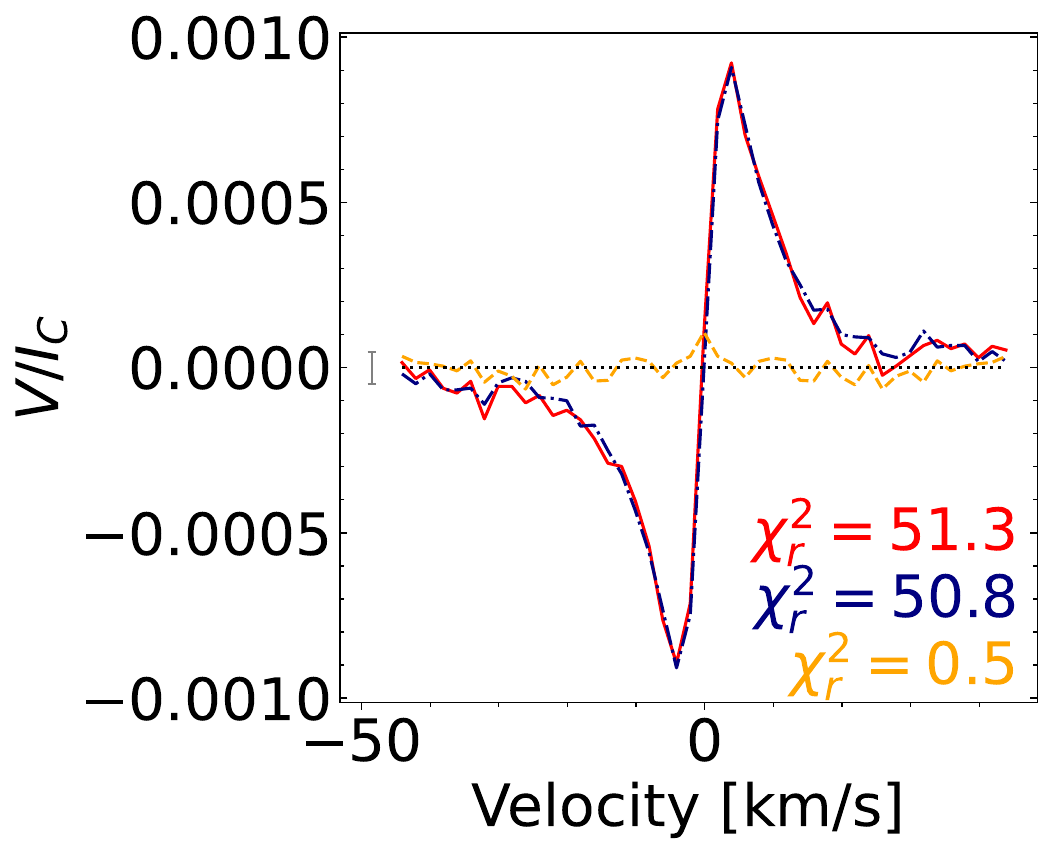}
	\includegraphics[width=0.63\columnwidth, trim={30 400 445 0}, clip]{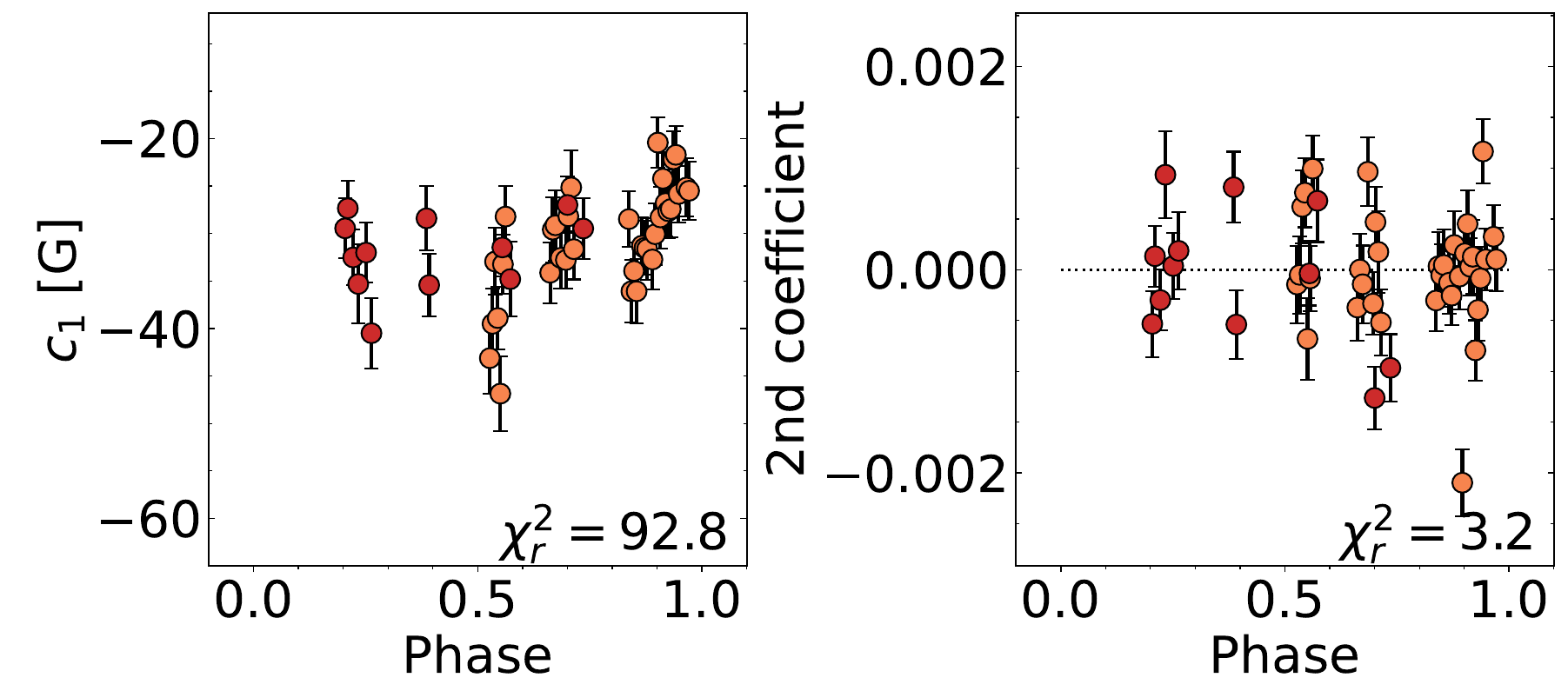}\\
    \caption{Same as Fig.~\ref{Fig:Gl905_PCA} for Gl~408.}
    \label{Fig:Gl408_PCA}
\end{figure}

Gl~408 has the strongest mean profile compared to the mean-subtracted profile in our sample (see Fig.~\ref{Fig:Gl408_PCA}a and \ref{Fig:Gl408_StVFit}). The first eigenvector of Gl~408 (the only one showing a signal) is already noisy, a strong indication of a very axisymmetric field topology (see Fig.~\ref{Fig:Gl408_PCA}b). 

\begin{figure}
	%
	%
	\centering
	\includegraphics[width=\columnwidth, trim={0 0 0 0}, clip]{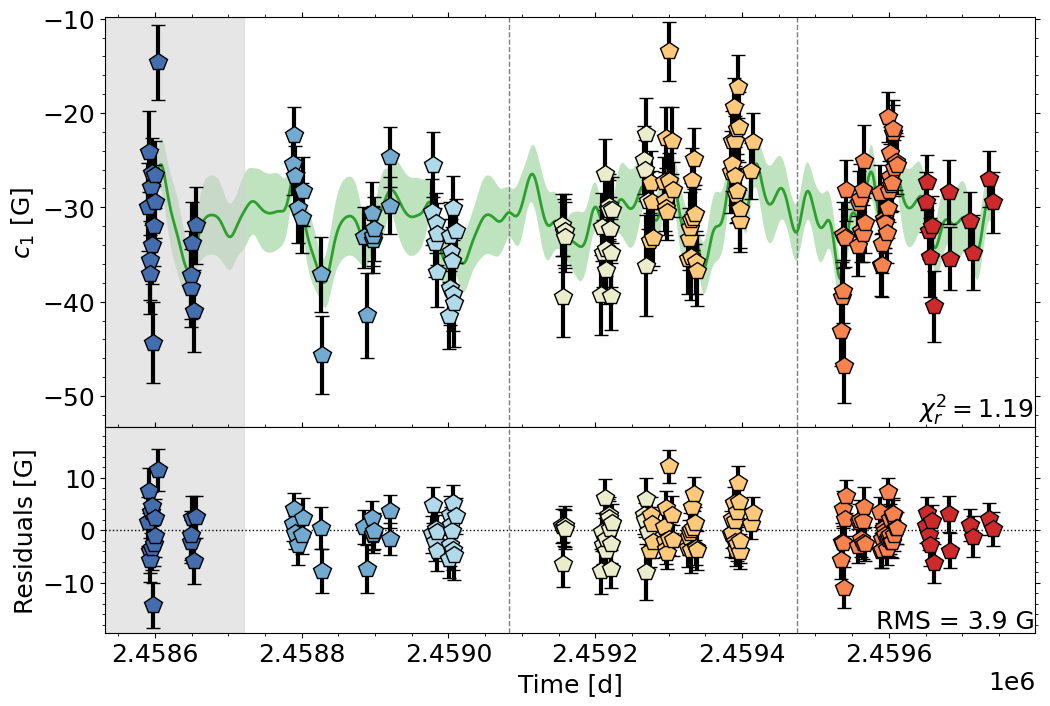}\\ 
		\includegraphics[width=\columnwidth, trim={0 0 0 0}, clip]{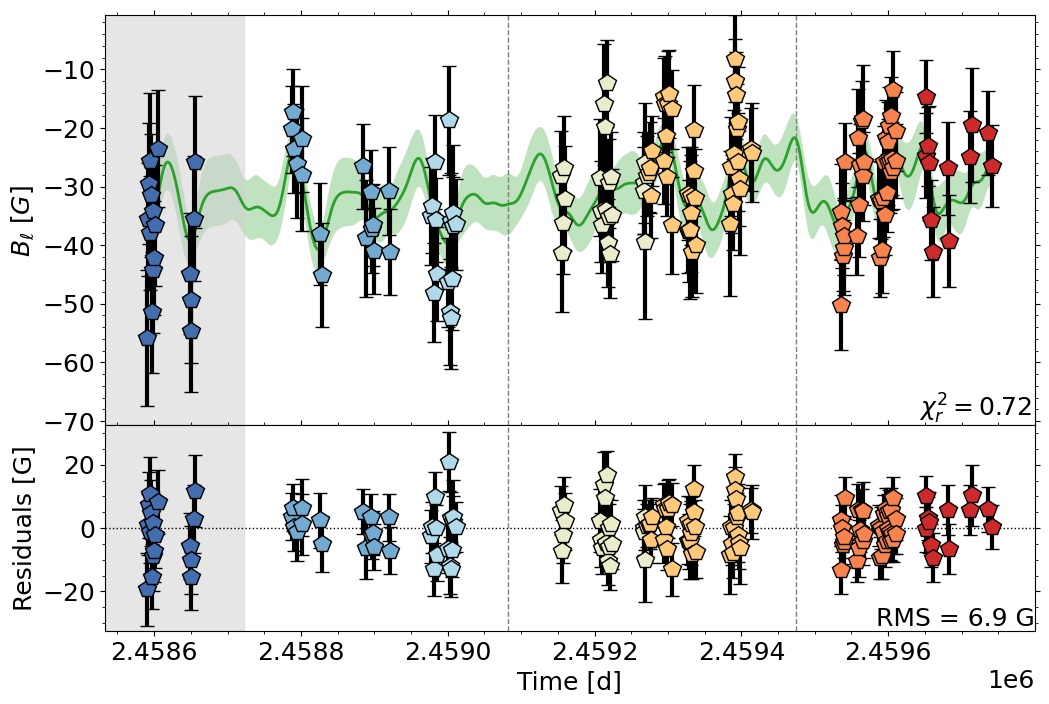}\\ 
    \caption{Same as Fig.~\ref{Fig:Gl905_CoeffvsTime} for Gl~408.}
    \label{Fig:Gl408_CoeffvsTime}
\end{figure}

Fig.~\ref{Fig:Gl408_CoeffvsTime} (top) presents the QP GPR fit of $c_1$, which gives a $\Prot = 175^{+12}_{-14}\,\dy$ similar to D23 determining $\Prot = 171.0\pm8.4\,\dy$.  Fitting $B_\ell$ with our GP routines, we derive a $\Prot = 170.7^{+7.1}_{-9.8}\,\dy$ (see Fig.~\ref{Fig:Gl408_CoeffvsTime} bottom). The decay time was fixed at 200\,d for both variables following D23. However, we find a decay time of $\approx 200\pm70\,\dy$ but higher $\chi^2$ for GPR fits without fixing the decay time. The \textsc{APERO} reduced spectra of Gl~408 did not allow \cite{Fouque2023} to determine a rotation period.

In Fig.~\ref{Fig:Gl408_PCA}c, we see that $c_1$ is mostly flat for all three seasons, again indicating a highly axisymmetric topology. 
All mean profiles are antisymmetric with respect to the line centre and show an axisymmetric poloidal large-scale field.

\subsection{ZDI reconstructions of Gl~408}

\begin{figure}
\centering
\begin{minipage}{0.32\columnwidth}
\centering
\includegraphics[height=0.85\columnwidth, angle=270, trim={140 0 0 29}, clip]{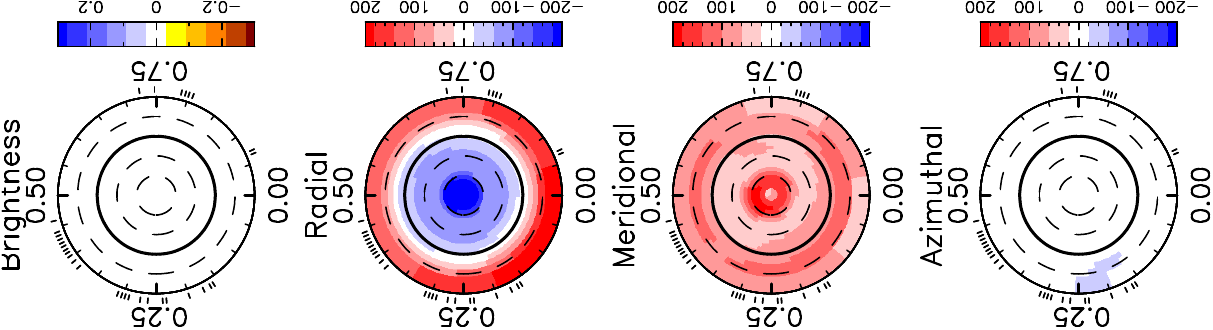} 
\end{minipage}
\begin{minipage}{0.32\columnwidth}
\centering
\includegraphics[height=0.85\columnwidth, angle=270, trim={140 0 0 29}, clip]{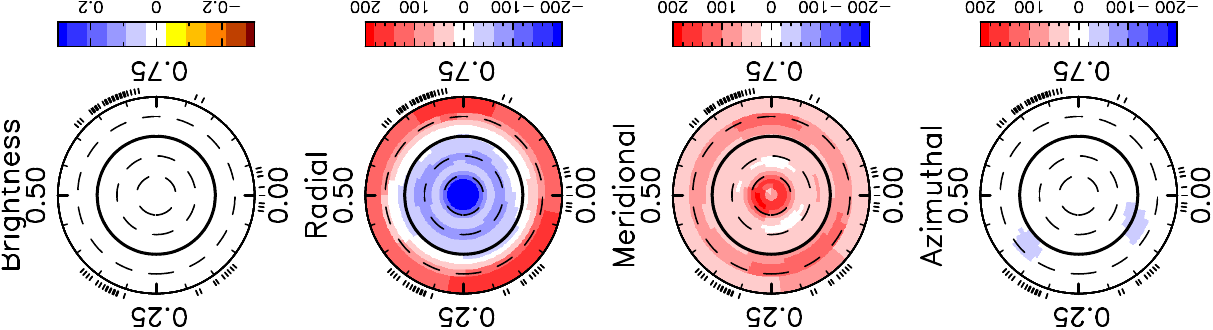} 
\end{minipage}
\begin{minipage}{0.32\columnwidth}
\centering
\includegraphics[height=0.85\columnwidth, angle=270, trim={140 0 0 29}, clip]{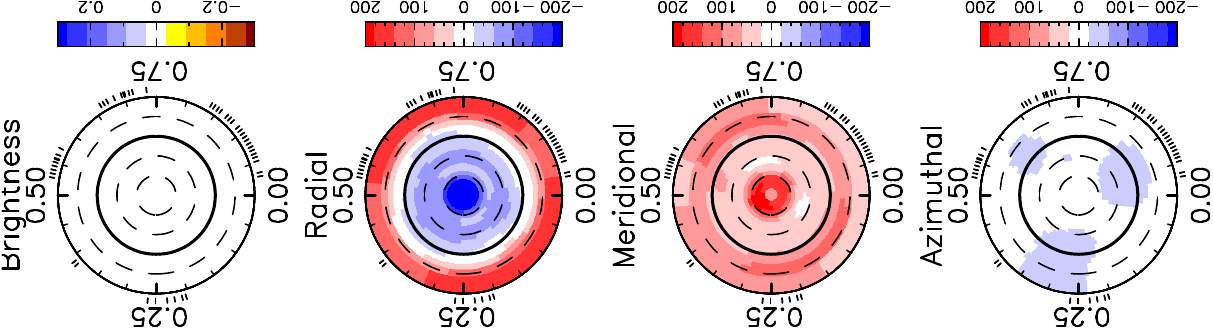} 
\end{minipage}
\includegraphics[width=0.3\columnwidth, angle=180, trim={460 130 2 0}, clip]{Figures/Gl408_ZDIMap_JFDLSD_epo3_v10.pdf}
\vspace*{2mm}
\includegraphics[width=0.95\columnwidth, clip]{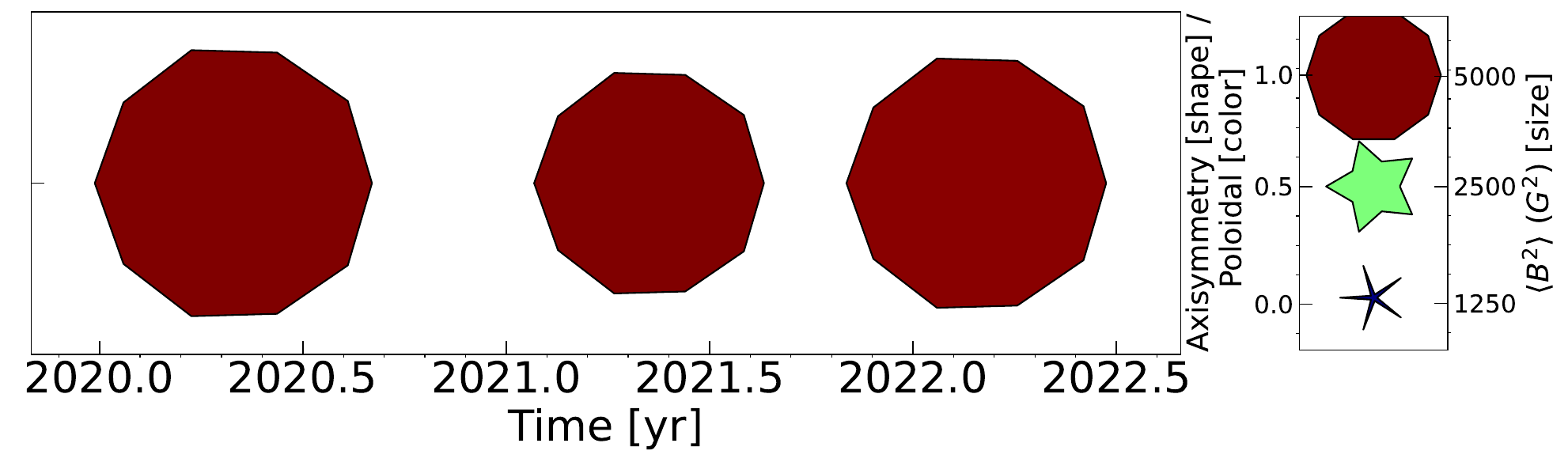}
    \caption{Same as Fig.~\ref{Fig:Gl905_ZDIMaps} for Gl~408.}
    \label{Fig:Gl408_ZDIMaps}
\end{figure}

All three seasons could be fitted down to $\chi^2 \approx 1.0$ assuming $\Prot = 171.0\,\dy, f_V = 0.1, v_e \sin i = 0.10\,\kms, i = 60^\circ$. The topology changes little over the three seasons and is characterised by a strong, axisymmetric, poloidal dipole of negative polarity (see Fig.~\ref{Fig:Gl408_ZDIMaps}). It is the most stable topology in our sample and only \BV\ varies marginally between $106-130$\,G (see Tab.~\ref{tab:MagProp_Gl408}). We find a 1$\sigma$ error bar on the axisymmetric toroidal field of $55-81$\,G for Gl~408 whereas the reconstructed \Btor\ ranges between $4-13\,$G.

Similar to Gl~617B, we also determine the ZDI maps for an inclination of $i = 30^\circ$ and $v_e \sin i = 0.06\,\kms$ (see Fig.~\ref{Fig:Gl408_ZDIMaps_i30}). The $\chi^2_r$ values of the ZDI fits are again slightly higher for the lower inclination $i=30^\circ$ than for $i = 60^\circ$.

\begin{table}
%
%
    \caption[]{Same as Table~\ref{tab:MagProp_Gl905} for Gl~408.}
    \label{tab:MagProp_Gl408}
    \begin{center}
    \begin{tabular}{lccc}
        \hline
        \noalign{\smallskip}
      season & \bf{2019/20} & \bf{2020/21} & \bf{2021/22} \\
      start & 2019 Oct & 2020 Oct & 2021 Nov \\
      end & 2020 June & 2021 July & 2022 June \\
        \noalign{\smallskip}
        \hline
        \noalign{\smallskip}
\BV & 130 & 106 & 120 \\
$\langle B_\mrm{dip} \rangle$[G]  & 129 & 104 & 117 \\
\Btormax [G] & 81 & 55 & 57 \\
$f_{\mrm{pol}}$ & 1.0 & 1.0 & 0.99 \\
 $f_{\mrm{axi}}$ & 0.98 & 0.98 & 0.98 \\
 $f_{\mrm{dip}}$ & 0.82 & 0.77 & 0.78 \\
        \noalign{\smallskip}
        \hline
        \noalign{\smallskip}
dipole tilt angle &  $6^\circ$ & $4^\circ$ & $4^\circ$ \\
pointing phase & 0.53 & 0.44 & 0.47 \\
        \noalign{\smallskip}
        \hline
        \noalign{\smallskip}
 $\chi^2_{r,V}$ & 5.63 & 4.13 & 5.66 \\
 $\chi^2_{r,V,\mrm{ZDI}}$ & 1.01 & 1.02 & 1.01 \\
 $\chi^2_{r,N}$ & 0.89 & 0.99 & 1.02 \\
 nb. obs & 31 & 62 & 47 \\
        \hline
    \end{tabular}
    \end{center}
\end{table}

\section{Summary, discussion \& conclusions}
\label{Sec:Conclusions}

In this paper, we study the large-scale magnetic field of six slowly rotating mid to late M~dwarfs observed with SPIRou at the CFHT as part of the SLS from 2019 to 2022. The 3.5-yr time series, including $\approx100-200$ polarimetric spectra for each of our six M~dwarfs, allowed us to confirm their rotation periods and to investigate their magnetic field topology using both our PCA analysis and ZDI.

We use the reduced observations from D23 but different analysis tools to redetermine the rotation period. Our estimate of the rotation periods using $c_1$, i.e., the scaled and translated first coefficient of the PCA analysis (see Sec.~\ref{SubSec:GP}), agrees with the results of D23 and \cite{Fouque2023}. 
We confirm that both Gl~617B and Gl~408, for which \cite{Fouque2023} did not recover a rotation period, host very axisymmetric topologies with $f_{\mrm{axi}} \geq 0.97$ between 2019 and 2022. The higher the axisymmetry of the large-scale field, the smaller are the variations of $B_\ell$ or $c_1$ with time, and the harder it is to determine a rotation period. For the highly axisymmetric topologies, we find that the $\chi^2_r$ of the GPR fits increases, reflecting that in such cases, the $B_\ell$ curves are more sensitive to intrinsic variability and less to rotational modulation, and thereby reducing the ability at measuring rotation periods (see e.g.\ Fig.~\ref{Fig:Corner_Gl617B} or \ref{Fig:Corner_Gl408} and Tab.~\ref{tab:GPFitParams}). 

Using the PCA analysis, we derive information about axisymmetry and complexity directly from the LSD Stokes~$V$ time series, which are in agreement with the results obtained from the ZDI maps for all six M~dwarfs, while PCA does not rely on any assumptions about stellar parameters such as $v_e \sin i$, inclinations, etc. 

We find evidence for a polarity reversal of the large-scale field (via sign changes of $B_\ell$, $c_1$ or in the mean profiles) taking place on GJ~1151 and possibly also on Gl~905, for which the axisymmetric component collapsed during the last season (to be confirmed with new, ongoing, observations). For most stars, PCA traces the time-evolving field topologies using only the first eigenvector. For GJ~1289 we even detect two evolving field components directly from the Stokes~$V$ time series. This highlights that we are able to reliably detect topological complexity in the magnetic fields of  slowly rotating M~dwarfs directly from the observed LSD Stokes~$V$ profiles. 
The lower the $v_e \sin i$, the higher the 1$\sigma$ error bar on the toroidal field. The typical 1$\sigma$ error bars on the toroidal field ranges from 6 to $450$\,G depending on SNR and $v_e \sin i$. 
 
We determined the ZDI maps for each season of our targets, obtaining a total of 17 vector magnetic field maps. 
The ZDI maps of GJ~1151 and Gl~905 confirm the polarity switches that were diagnosed with PCA, and further show that GJ~1151 may have been in a magnetically quiescent state until it became more magnetic in 2022, switching polarity at the same time. 

\begin{figure}
\centering
\includegraphics[width=\columnwidth, clip]{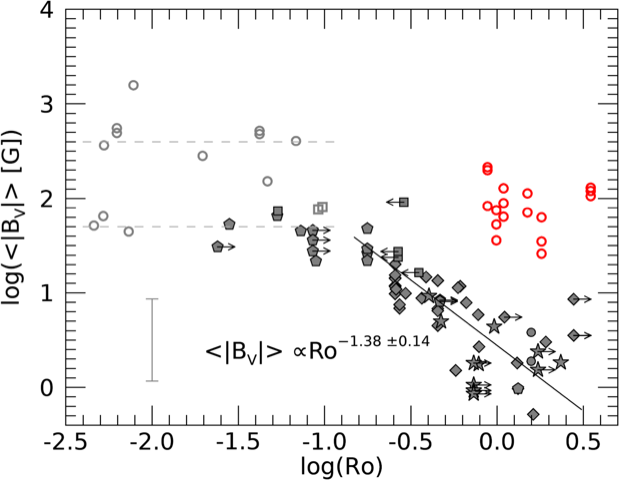}
\caption{The averaged unsigned magnetic field strength $\langle B_V \rangle$ versus Rossby number $Ro$ as shown by \citet{Vidotto2014} (in grey scales) including our sample of slowly rotating M~dwarfs (red circles). Note that for this figure only, we determined $Ro$ using \citet{Wright2011} for consistency with \citet{Vidotto2014}. For further details and the coloured version of the original symbols and annotations see Fig.~4a of \citet{Vidotto2014}.}
    \label{Fig:AlinesFig}
\end{figure}

The slowly rotating M~dwarfs \cc{of our sample} show large-scale field strengths in the range \BV\ $\approx 20- 200\,\mrm{G}$. They show similar \BV\ to faster rotating M~dwarfs in the saturated regime. We add our sample to Fig.~4a of \cite{Vidotto2014}, which originally shows \BV\ versus Rossby number $Ro$ for 73 stars, including stars in the mass range $\sim 0.1-2$\,\Msun\ \ (see Fig.~\ref{Fig:AlinesFig}). The grey open circles depict the \cc{mid- and late-type rapidly} rotating M~dwarfs of \cite{Vidotto2014}, while the red circles show our slow rotating M~dwarfs. The solar-like G--K~dwarfs (grey diamonds and pentagons) follow a decreasing trend with increasing $Ro$, while our M~dwarfs show stronger \BV\ values than expected given their $Ro$. This is in agreement with the results of \cite{Medina2022}, showing that M~dwarfs can remain extremely active (flaring) even when their rotation period increases beyond 100\,days.
Besides, it implies a harsher interplanetary environment for potential close-in planets (e.g., \citealt{Kavanagh2021}).

\begin{figure*}
\centering
\includegraphics[width=\textwidth, clip]{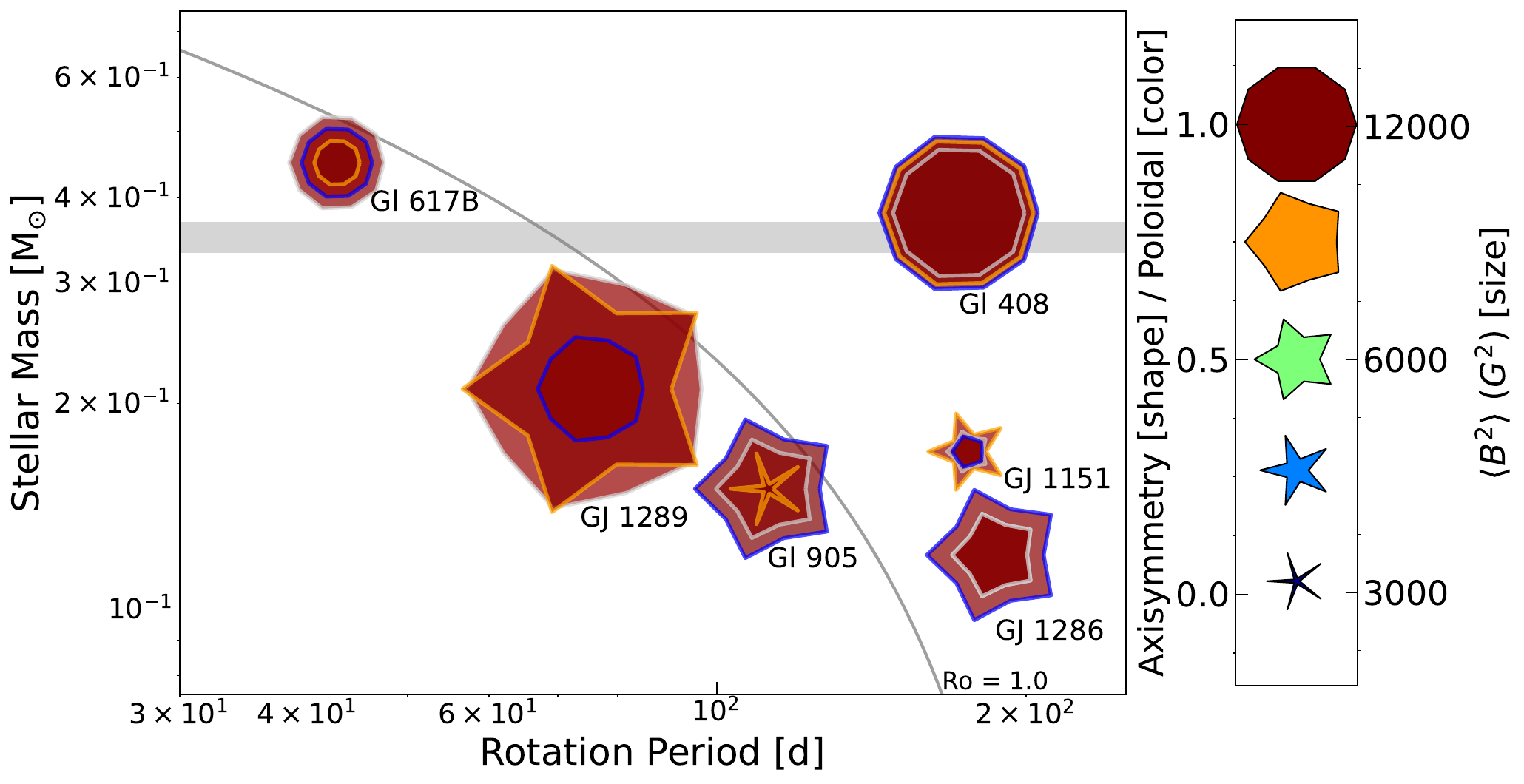}
\caption{A summary of the magnetic properties of our M~dwarf sample as a function of stellar mass and rotation period. The symbol size indicates the magnetic energy, the colour the poloidal fractional energy $f_{\mrm{pol}}$ and the shape the axisymmetric fractional energy $f_{\mrm{axi}}$ (see the legend on the right). For each star, we display two to three symbols representing the individual seasons (blue, grey, orange border for the first, second and third season) of each M~dwarf. The thin grey line indicates $Ro=1$ determined using the empirical relation of \citet{Wright2018}. The thick grey line marks the fully convection limit at $M \approx 0.35\,$\Msun\ .}
    \label{Fig:Confusogram_Mdwarfs_Epoch}
\end{figure*}

\cc{We stress that our paper focused on the most magnetic M~dwarfs of the SLS sample (e.g., D23), whereas the other (less magnetic) stars of this sample will presumably be more in line with (and fill the gap between) the high-$Ro$ stars of the \cite{Vidotto2014} sample. This will be the subject of forthcoming studies.
Beside, our results suggests that the large-scale fields of the very slowly rotating M~dwarfs of our sample are likely generated through dynamo processes} operating in a different regime than those of the faster rotators that have been magnetically characterized so far.
Fig.~\ref{Fig:Confusogram_Mdwarfs_Epoch} summarises the properties of the large-scale magnetic field topology for our six M~dwarfs displaying all seasons on top of each other. It can be seen that the two partly convective M~dwarfs (Gl~617B and Gl~408) show a smaller range of variations compared to the fully convective stars. The fully convective M~dwarfs host large-scale fields that evolve on timescales comparable to their rotation periods. Our small sample suggests that fully-convective, slowly-rotating M~dwarfs tend to have large-scale fields that are less axisymmetric than their more massive counterparts. 

In conclusion, we have analysed six slowly rotating M~dwarfs observed by the SLS over 3.5 years. We find, that the large-scale magnetic field of these M~dwarfs is unusually strong despite their slow rotation (40--190\,d) and suggest that the efficiency of the dynamo for mid and late M~dwarfs depends on $Ro$ in a different way than that reported in the literature for faster rotators. Furthermore, we find that the large-scale magnetic field topology of the fully convective M~dwarfs exhibit a larger range of variations than those of the two partly convective targets of our sample. Given this, it may be useful in the future to apply the time-dependent ZDI \citep{Finociety2022}, which has only been tested for faster rotating stars up to now. We detected a polarity reversal on one (GJ~1151) and possibly two (Gl~905) of the 4 fully-convective stars of our sample, suggesting that magnetic cycles may indeed be occurring in such stars, as initially suggested by \cite{Route2016} from radio observations. Further long-term observations of the same type are needed to document in a more systematic fashion the long-term evolution of the large-scale magnetic fields of M~dwarfs, and whether these field topologies are varying cyclically like for the Sun or in a more random fashion.

\section*{Acknowledgements}

We acknowledge funding from the European Research Council (ERC) under the H2020 research \& innovation programme (grant agreement \#740651 NewWorlds). AC acknowledge funding from the French ANR under contract number ANR\-18\-CE31\-0019 (SPlaSH). This work is supported by the French National Research Agency in the framework of the Investissements d'Avenir program (ANR-15-IDEX-02), through the funding of the ``Origin of Life" project of the Grenoble-Alpes University. AAV acknowledges funding from the European Research Council (ERC) under the European Union's Horizon 2020 research and innovation programme (grant agreement No 817540, ASTROFLOW).
Our study is based on data obtained at the CFHT, which is operated by the CNRC (Canada), INSU/CNRS (France) and the University of Hawaii. The authors wish to recognise and acknowledge the very significant cultural role and reverence that the summit of Maunakea has always had within the indigenous Hawaiian community. We are very fortunate to have the opportunity to conduct observations from this mountain and gratefully acknowledge the CFHT QSO observers. This work has used Astropy \citep{astropy:2013, astropy:2018, astropy:2022}, NumPy \citep{Numpy}, Matplotlib \citep{Matplotlib}, SciPy \citep{Scipy} and benefited from the \href{http://simbad.u-strasbg.fr/simbad}{SIMBAD CDS database} and the \href{https://ui.adsabs.harvard.edu}{ADS system}.

\section*{Data Availability}

All data underlying this paper are part of the SLS, and will be publicly available from the Canadian Astronomy Data Center by February 2024.



\bibliographystyle{mnras}
\bibliography{Lehmann2023_Slow_Rot_Mdwarfs} 




\appendix

\section{Additional GPR fits information and figures}
\label{Sec:AddGPFigures}

Table~\ref{tab:Ephemeris} presents the ephemeris used for our M~dwarf sample.

\begin{table}
    \caption[]{The ephemeris data used to determine the phases ($\mathrm{phase} = \frac{T_0 -T_{\mrm{obs}}}{\Prot}$) for the six M~dwarfs of our sample. For $T_0$ we used the barycentric Julian date of the first SPIRou observation of the target (second column). The last column indicate the rotation period $\Prot$ used for the phase determination. }
    \label{tab:Ephemeris} 
    \begin{center}
    \begin{tabular}{lcc}
        \hline
        \noalign{\smallskip}
star & $T_0$ [JD] & $\Prot$ [d] \\
        \noalign{\smallskip}
        \hline
        \noalign{\smallskip}
       	Gl~905 & 2458600.1348284 & 114.3\\
       	GJ~1289 & 2458648.9872293 & 73.66\\
       	GJ~1151 & 2458828.1234878  & 175.6\\
        GJ~1286 & 2458745.8452793 & 178.0 \\
       	Gl~617B & 2458744.743105 & 40.4\\
       	Gl~408 & 2458590.0269338 &171.0 \\
		\hline
    \end{tabular}
    \end{center}
   \end{table}

For the GPR fits we used uniform $\mathcal{U}$, normal $\mathcal{N}$ and Jeffreys $\mathcal{J}$ prior distributions. Tab.~\ref{tab:Proirs} indicate the prior used for $\Prot$ and the decay time $l$. 
For the smoothing factor, we use in general $\mathcal{U}(0.1,1.0)$ beside for the GPR fits of GJ~1151 ($\mathcal{U}(0.2,0.6)$) and Gl~617B ($\mathcal{U}(0.4,0.8)$). For the amplitude $\alpha$, we applied $\mathcal{U}(0,\infty)$ with an exception for the GPR fits of GJ~1289 ($\mathcal{N}(25,10)$) and GJ~1286 ($\mathcal{N}(16.4,5)$). For the white noise, we used in general $\mathcal{U}(0,\infty)$ beside for Gl~905, where we applied a Jeffreys priori distribution $\mathcal{J}(1.5,\infty)$ for the GP fits.

\begin{table}
    \caption[]{The prior distribution used for the QP GPR fits for the rotation period $\Prot$ and decay time $l$. }
    \label{tab:Proirs} 
    \begin{center}
    \begin{tabular}{lcc}
        \hline
        \noalign{\smallskip}
star & prior $\Prot$ [d] & prior $l$ [d] \\
        \noalign{\smallskip}
        \hline
        \noalign{\smallskip}
       	Gl~905 & $\mathcal{N}(115,20)$ & $\mathcal{N}(130,25)$\\
       	GJ~1289 & $\mathcal{N}(75,5)$ & $\mathcal{U}(50,1000)$\\
       	GJ~1151 & $\mathcal{N}(175,20)$ & \textit{300} \\
       	GJ~1286 & $\mathcal{N}(170,20)$ & \textit{300} \\
       	Gl~617B & $\mathcal{N}(40.4,5)$ & $\mathcal{U}(30,1000)$\\
        Gl~408 & $\mathcal{N}(170,15)$ & \textit{200} \\
        \noalign{\smallskip}
		\hline
    \end{tabular}
    \end{center}
   \end{table}

In following, we present the corner plots of all QP GPR fits applied in this paper.

\begin{figure}
\centering
\includegraphics[width=\columnwidth, trim={0 0 0 0}, clip]{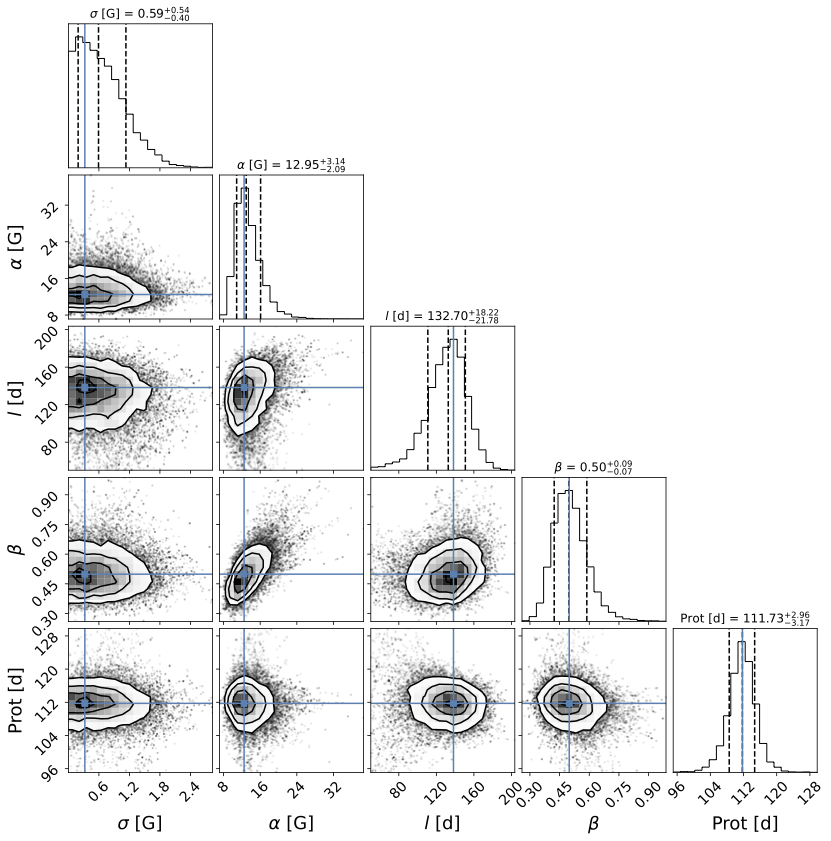}
\includegraphics[width=\columnwidth, trim={0 0 0 0}, clip]{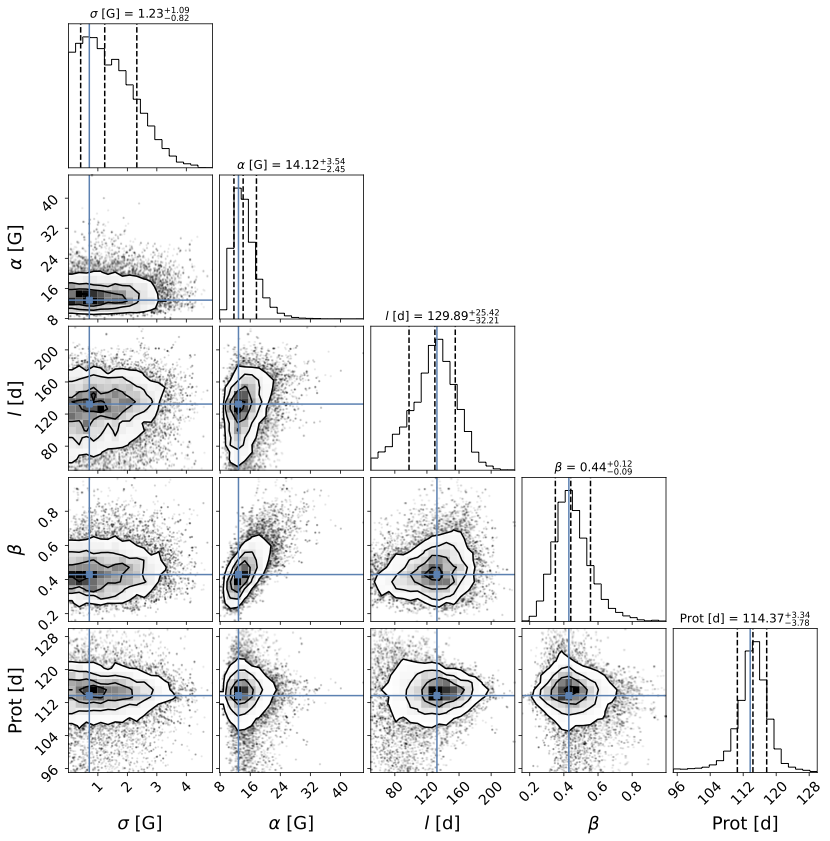}
\caption{The posterior density resulting from the MCMC sampling of the QP best-fitting GPR model for $c_1$ (top) and $B_\ell$ (bottom) for Gl~905. The concentric circle within each panel indicate the 1, 2 and 3$\sigma$ contours of the distribution. The blue lines mark the mode (maximum) of the posterior distribution and the black dashed lines the median and the 16\% and 84\% percentils of the posterior probability density function (PDF).}
\label{Fig:Corner_Gl905}
\end{figure}

\begin{figure}
%
%
\centering
\includegraphics[width=\columnwidth, trim={0 0 0 0}, clip]{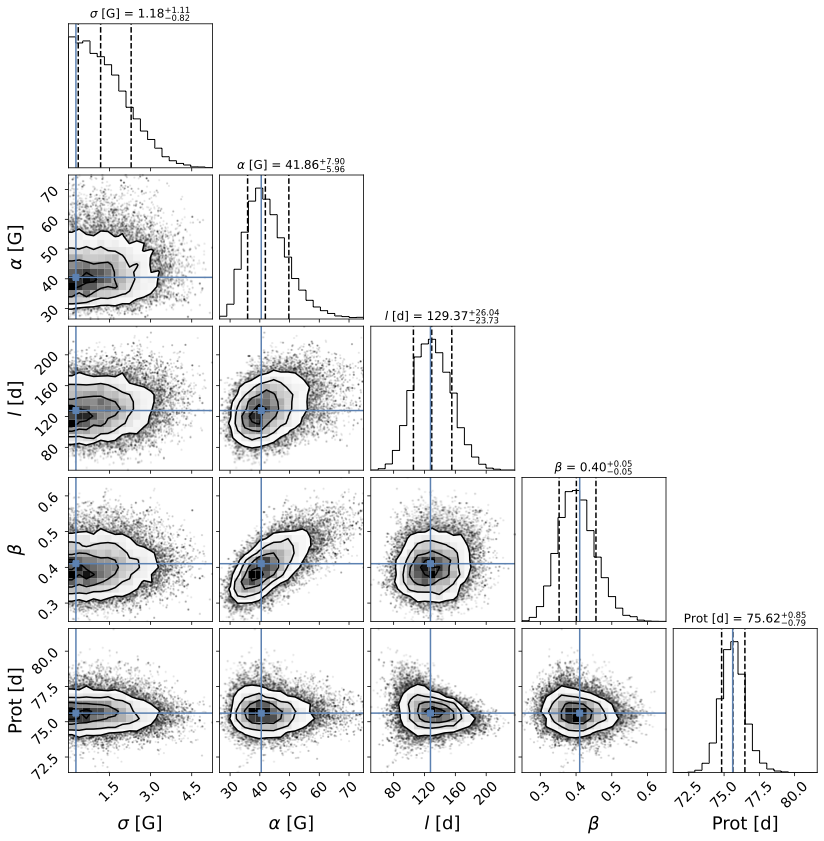}
\includegraphics[width=\columnwidth, trim={0 0 0 0}, clip]{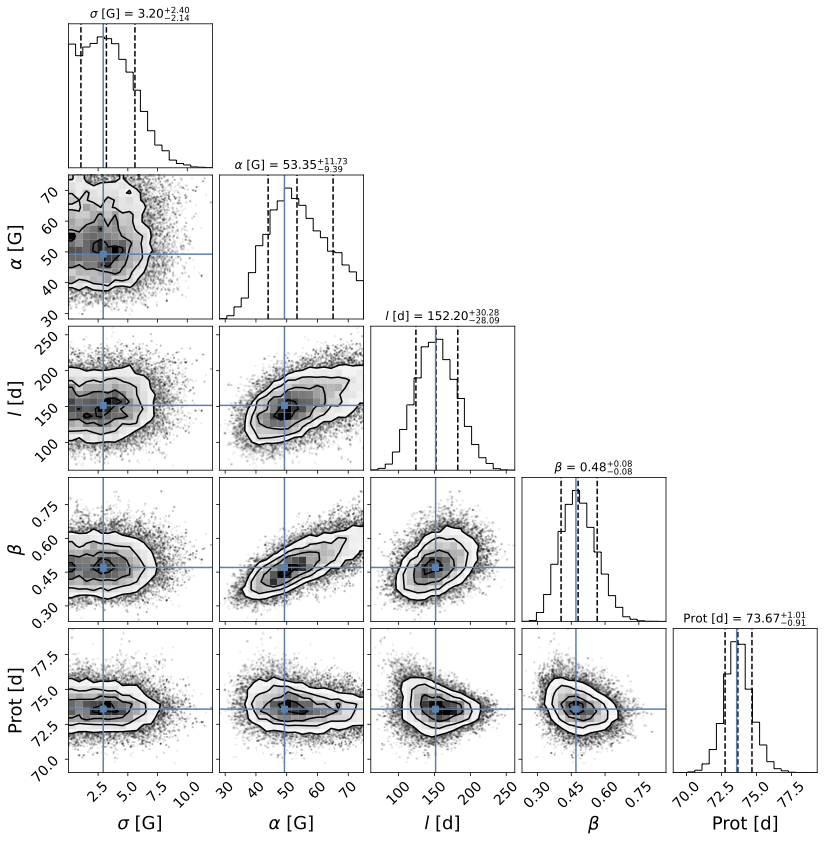}
\caption{Same as Fig.~\ref{Fig:Corner_Gl905} for GJ~1289.}
\label{Fig:Corner_GJ1289}
\end{figure}

\begin{figure}
%
%
\centering
\includegraphics[width=\columnwidth, trim={0 0 0 0}, clip]{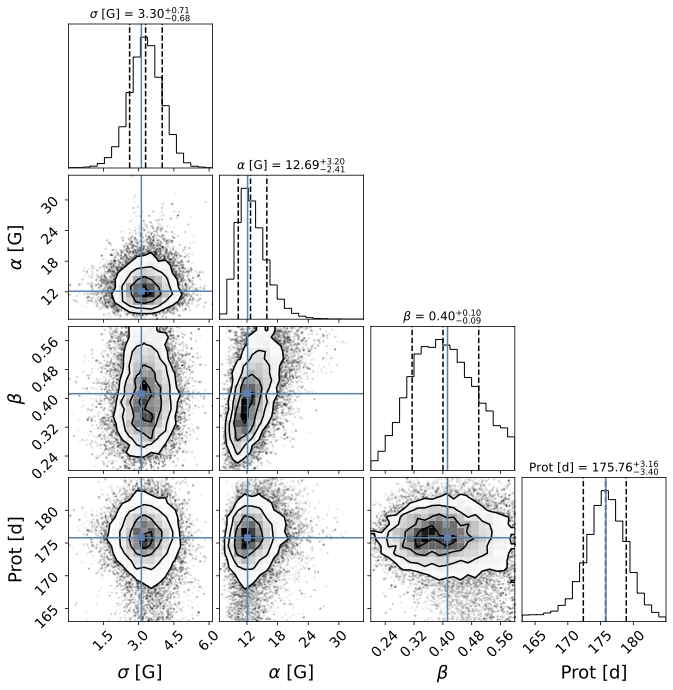}
\includegraphics[width=\columnwidth, trim={0 0 0 0}, clip]{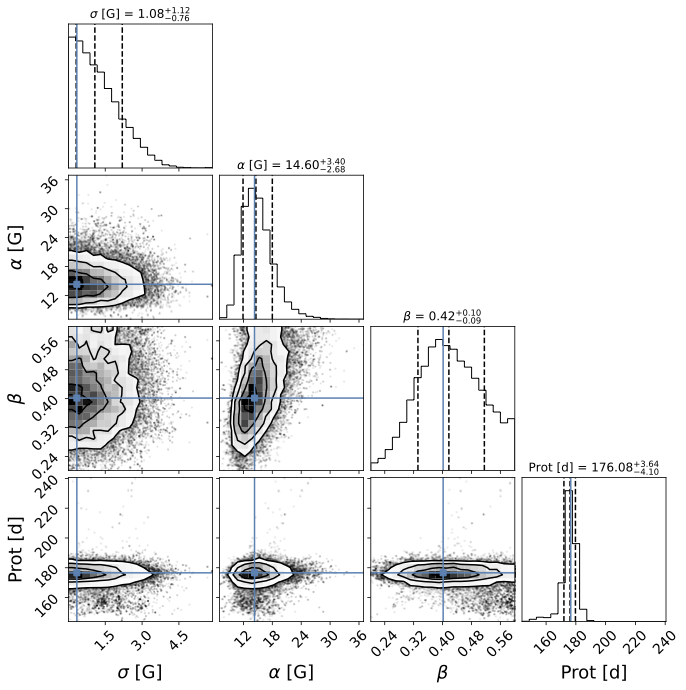}
\caption{Same as Fig.~\ref{Fig:Corner_Gl905} for GJ~1151.}
\label{Fig:Corner_GJ1151}
\end{figure}

\begin{figure}
%
%
\centering
\includegraphics[width=\columnwidth, trim={0 0 0 0}, clip]{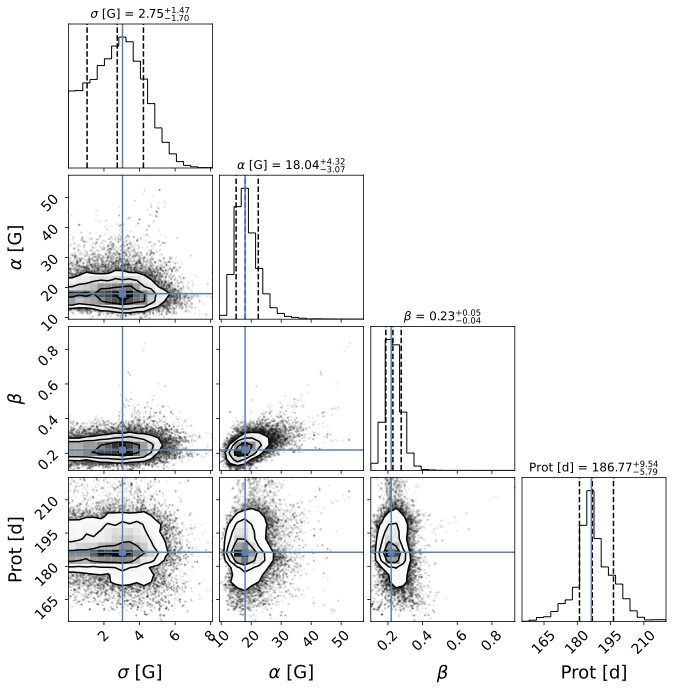}
\includegraphics[width=\columnwidth, trim={0 0 0 0}, clip]{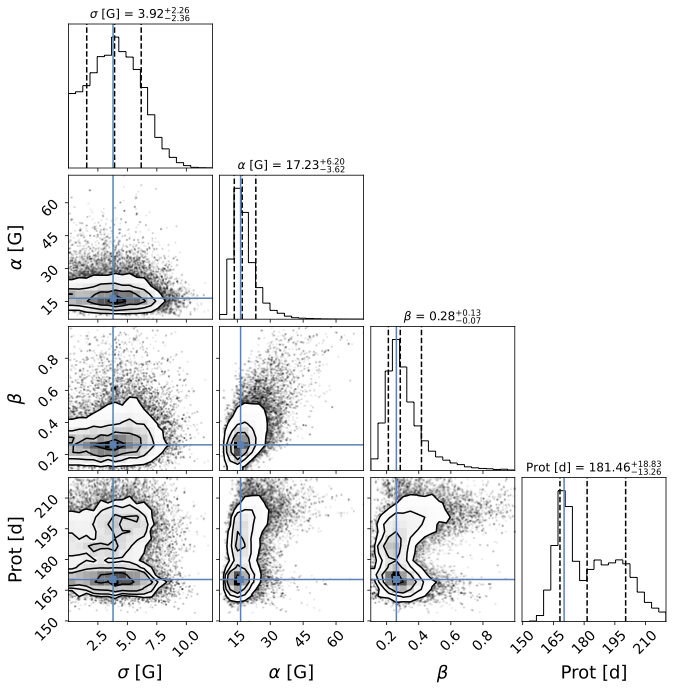}
\caption{Same as Fig.~\ref{Fig:Corner_Gl905} for GJ~1286.}
\label{Fig:Corner_GJ1286}
\end{figure}

\begin{figure}
%
%
\centering
\includegraphics[width=\columnwidth, trim={0 0 0 0}, clip]{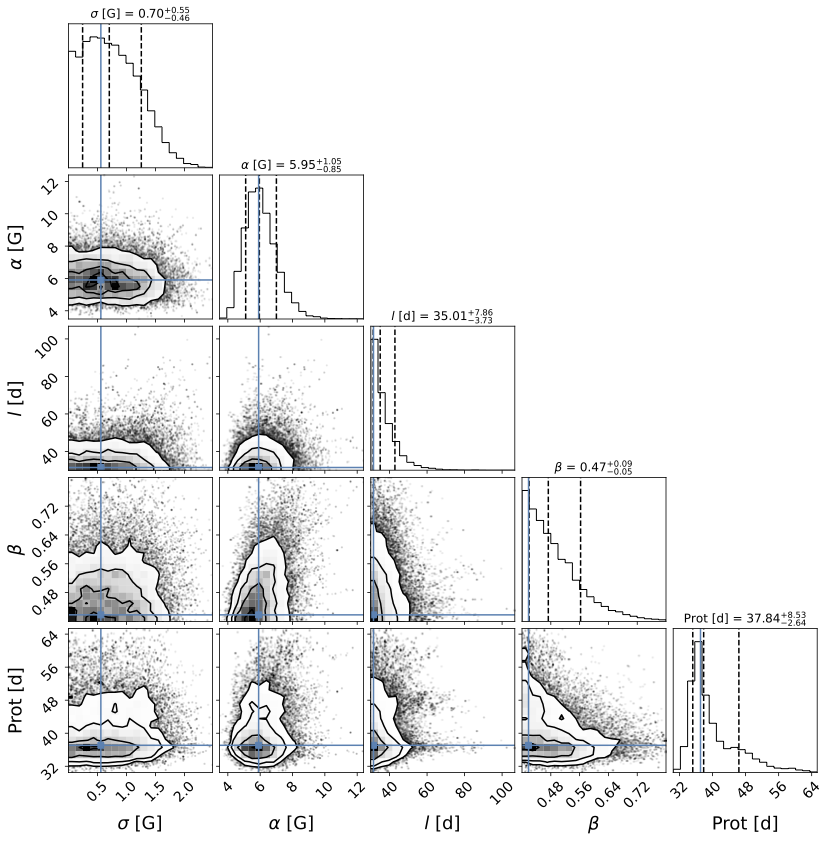}
\includegraphics[width=\columnwidth, trim={0 0 0 0}, clip]{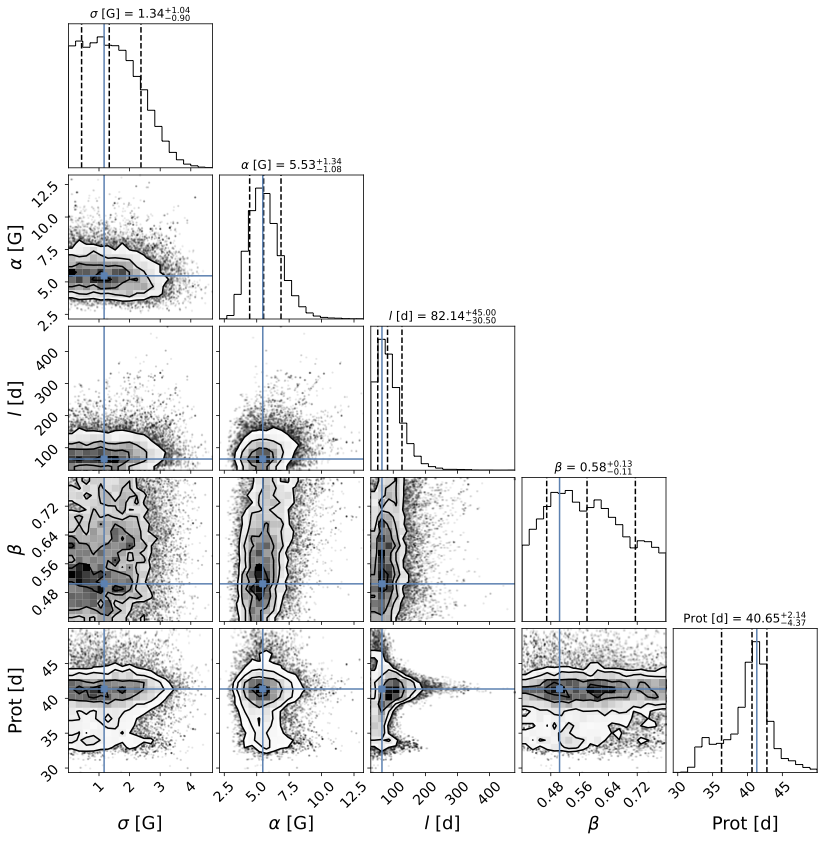}
\caption{Same as Fig.~\ref{Fig:Corner_Gl905} for Gl~617B.}
\label{Fig:Corner_Gl617B}
\end{figure}

\begin{figure}
%
%
\centering
\includegraphics[width=\columnwidth, trim={0 0 0 0}, clip]{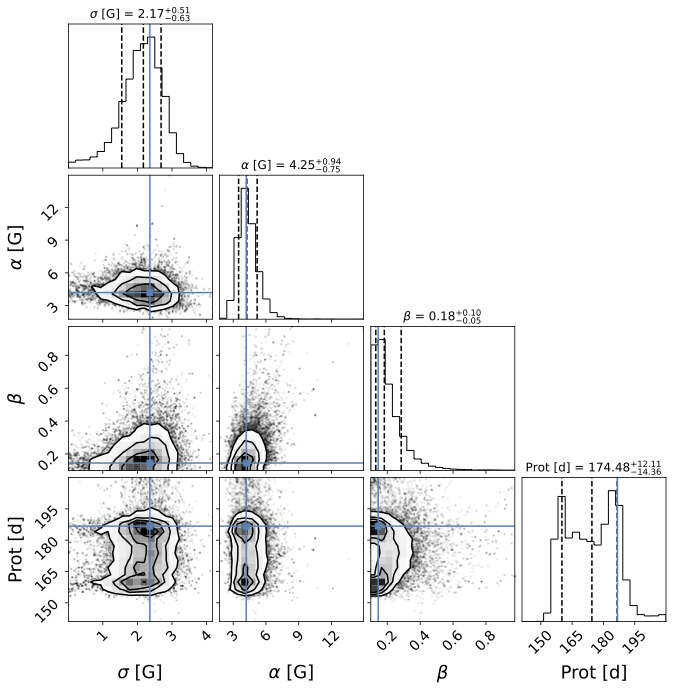}
\includegraphics[width=\columnwidth, trim={0 0 0 0}, clip]{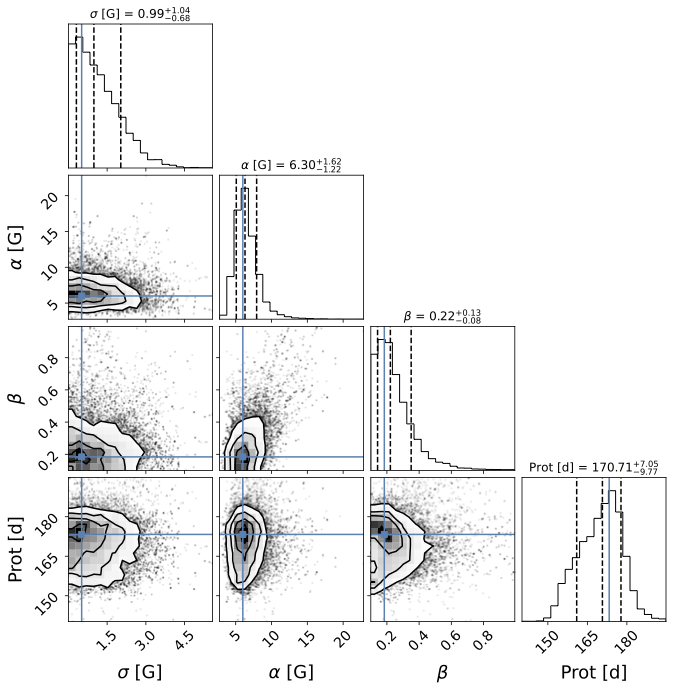}
\caption{Same as Fig.~\ref{Fig:Corner_Gl905} for Gl~408.}
\label{Fig:Corner_Gl408}
\end{figure}

\section{The GPR fits of the longitudinal field values}
\label{App:GPComp} 

\begin{table*}
    \caption[]{Summary of the best-fitting parameters of the QP GPR fits applied to $c_1$ and $B_\ell$ for the six M~dwarfs in our sample and the comparison with the corresponding GPR results for $B_\ell$ by \citet{Donati2023} marked as D23 in the second column.}
    \label{tab:GPFitParams}
    \setlength{\extrarowheight}{.3em}
    \begin{tabular}{lcccccccc}
        \hline
        \hline
        \noalign{\smallskip}
 star & data & rotation period & decay time & smoothing factor & amplitude & white noise & rms & $\chi^2_{\rm r}$  \\
 & & $P_{\rm rot}$ [d] & $l$ [d] & $\beta$ & $\alpha$ [G] & $\sigma$ [G] & [G] & \\
        \noalign{\smallskip}
        \hline
        \hline
        \noalign{\smallskip}
Gl~905 & $c_1$ & $111.7^{+3.0}_{-3.2}$ & $133^{+18}_{-22}$ & $0.50^{+0.09}_{-0.07}$ & $12.9^{+3.1}_{-2.1}$ & $0.6^{+0.5}_{-0.4}$ & 3.8 & 0.79 \\
Gl~905 & $B_\ell$ & $114.4^{+3.5}_{-2.4}$ & $130^{+25}_{-32}$ & $0.44^{+0.12}_{-0.09}$ & $14.1^{+3.5}_{-2.4}$ & $1.2^{+1.1}_{-0.8}$ & 6.9 & 0.87 \\
Gl~905 & D23 & $114.3\pm2.8$ & $129^{+25}_{-21}$ & $0.43\pm0.09$ & $13.3^{+2.5}_{-2.1}$ & $1.7^{+1.2}_{-0.7}$ & 6.2& 0.84 \\
        \noalign{\smallskip}
        \hline
        \noalign{\smallskip}
GJ~1289 & $c_1$ & $75.62^{+0.85}_{-0.79}$ & $129^{+26}_{-24}$ & $0.40\pm 0.05$ & $41.9^{+7.9}_{-6.0}$ &  $1.2^{+1.1}_{-0.8}$ & 6.9 & 0.72 \\
GJ~1289 & $B_\ell$ & $73.67^{+1.01}_{-0.91}$ & $152^{+30}_{-28}$ & $0.48\pm0.08$ & $53.3^{+11.7}_{-9.4}$ &  $3.2^{+2.4}_{-2.1}$ & 14.3 & 0.84 \\
GJ~1289 & D23 & $73.66\pm0.92$ & $152^{+32}_{-27}$ & $0.48\pm0.09$ & $53.2^{+12.4}_{-10.1}$ & $4.2^{+2.7}_{-1.6}$ & 13.9 & 0.82 \\
        \noalign{\smallskip}
        \hline
        \noalign{\smallskip}
GJ~1151 & $c_1$ & $175.8^{+3.2}_{-3.4}$ & \textit{300} & $0.40^{+0.10}_{-0.09}$ & $12.7^{+3.2}_{-2.4}$ &  $3.3\pm0.7$ & 5.8 & 0.99 \\
GJ~1151 & $B_\ell$ & $176.1^{+3.6}_{-4.1}$ & \textit{300}& $0.42^{+0.10}_{-0.09}$ & $14.6^{+3.4}_{-2.7}$ &  $1.1^{+1.1}_{-0.8}$ & 6.7 & 0.73 \\
GJ~1151 & D23 & $175.6\pm4.9$ & \textit{300} & $0.43\pm0.11$ & $14.9^{+4.2}_{-3.3}$ &  $1.6^{+1.3}_{-0.7}$ & 6.7 & 0.72 \\
        \noalign{\smallskip}
        \hline
        \noalign{\smallskip}
GJ~1286 & $c_1$ & $186.8^{+9.5}_{-5.8}$ & \textit{300} & $0.23^{+0.05}_{-0.04}$ & $18.0^{+4.3}_{-3.1}$  & $2.7^{+1.5}_{-1.7}$ & 7.5 & 1.02 \\
GJ~1286 & $B_\ell$ & $181^{+18}_{-13}$ & \textit{300} & $0.28^{+0.13}_{-0.07}$ & $17.2^{+6.2}_{-3.6}$ & $3.9^{+2.3}_{-2.4}$ & 12.0 & 1.03 \\
GJ~1286 & D23 & $178\pm15$ & \textit{300} & $0.29\pm0.09$ & $16.7^{+4.6}_{-3.6}$ &  $4.6^{+2.6}_{-1.6}$ & 10.1 & 1.02 \\
        \noalign{\smallskip}
        \hline
        \noalign{\smallskip}
Gl~617B & $c_1$ & $37.8^{+8.5}_{-2.6}$ & $35^{+8}_{-4}$ & $0.47^{+0.09}_{-0.05}$ & $5.9^{+1.2}_{-0.8}$ & $0.7^{+0.6}_{-0.5}$ & 2.2 & 0.66 \\
Gl~617B & $B_\ell$ & $40.6^{+2.1}_{-4.4}$ & $82^{+45}_{-30}$ & $0.58^{+0.13}_{-0.11}$  & $5.5^{+1.3}_{-1.1}$ & $1.3^{+1.0}_{-0.9}$ & 5.0 & 0.86 \\
Gl~617B & D23 & $40.4\pm3.0$ & $69^{+35}_{-23}$  & $0.60\pm0.22$ & $5.4^{+1.3}_{-1.0}$ &  $1.7^{+1.1}_{-0.7}$ & 4.9 & 0.86 \\
        \noalign{\smallskip}
        \hline
        \noalign{\smallskip}
Gl~408 & $c_1$ & $175^{+12}_{-14}$ & \textit{200} & $0.18^{+0.10}_{-0.05}$ & $4.2^{+0.9}_{-0.8}$ & $2.2^{+0.5}_{-0.6}$ & 3.9 & 1.19 \\
Gl~408 & $B_\ell$ & $170.7^{+7.1}_{-9.8}$ & \textit{200} & $0.22^{+0.13}_{-0.08}$  & $6.3^{+1.6}_{-1.2}$ & $1.0^{+1.0}_{-0.7}$ & 6.9 & 0.72 \\
Gl~408 & D23 & $171.0\pm8.4$ & \textit{200} & $0.21\pm0.10$ & $6.3^{+1.5}_{-1.2}$ &  $1.5^{+1.2}_{-0.7}$ & 6.3 & 0.66 \\
          \noalign{\smallskip}
        \hline
    \end{tabular}
  \end{table*}
 
Table~\ref{tab:GPFitParams} provides a comparison of the GP results for $c_1$ and $B_\ell$ obtained with the GPR framework presented in Sec.~\ref{SubSec:GP} and the GP results of D23 for the $B_\ell$ values (marked by D23 in the second column). For the GP fits the mean value is fixed to the $B_\ell$ mean values determined by D23 for both $c_1$ and $B_\ell$. Within the error, all three GP results agree for $\Prot$, $l$, $\beta$ and $\alpha$ for each M~dwarf, confirming $\Prot$, $l$, $\beta$ and $\alpha$ using two different variables and calculation routines with one exception for Gl~617B, where the decay time determined from $c_1$ differs from the GP results of $B_\ell$. The additional white noise $\sigma$ is also consistent for both $B_\ell$, while $c_1$ often shows lower $\sigma$ values. Our GP routines fit the $B_\ell$ data with a slightly higher or equal $\chi^2_r$ compared to the results of D23, while obtaining lower $\sigma$ values. Comparing the results for $c_1$ and $B_\ell$, we find that $c_1$ has a lower RMS. The RMS of $c_1$ and $B_\ell$ is for all M~dwarfs lower than the corresponding averaged error of $c_1$ and $B_\ell$. For M~dwarfs with significant non-axisymmetric field (e.g.\ Gl~905, GJ~1289, GJ~1286), $c_1$ is fitted with lower or equal $\chi^2_r$ and can provide smaller errors for $\Prot$ (e.g.\ GJ~1289, GJ~1286). For M~dwarfs with highly axisymmetric fields (e.g.\ Gl~617B, Gl~408), $B_\ell$ gives smaller errors on $\Prot$.

\section{Additional PCA and ZDI Figures}

\begin{figure}
\centering
	\includegraphics[width=0.5\columnwidth, trim={0 0 0 0}, clip]{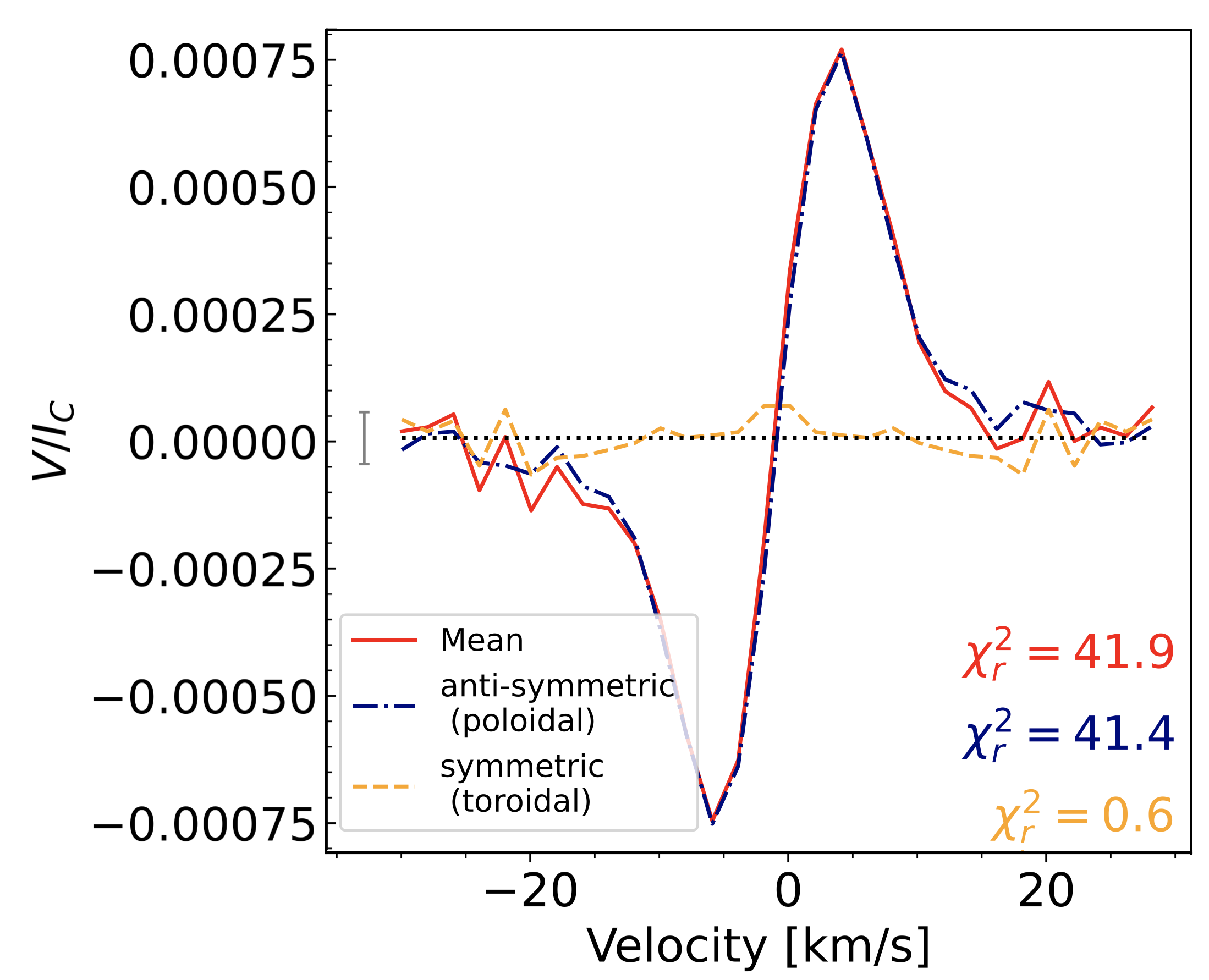}
    \caption{The mean profile and its decompositions obtained from 24 uniformly phased synthetic Stokes~$V$ LSD profiles of the 2020/21 ZDI map of Gl~905. The same format as in Fig.~\ref{Fig:Gl905_PCA}a is used.}
    \label{Fig:Gl905_EvenPhase}
\end{figure}

We can confirm, that the symmetric component (with respect to the line centre) seen in the mean profile of the 2020/21 season for Gl~905 is due to the irregular phase coverage of the observations \cc{and does not reflect an axisymmetric toroidal field}. We simulated 24 equally spaced Stokes~$V$ LSD~profiles from the 2020/21 ZDI map (Fig.~\ref{Fig:Gl905_ZDIMaps} middle column). The symmetric component of the resulting mean profile disappears for the uniform phase coverage (see Fig.~\ref{Fig:Gl905_EvenPhase}). \cc{Nonetheless, we might still miss toroidal field due to the low $v_e \sin i$, see the \Btormax\ estimation in Section~\ref{Subsec:Gl905ZDI}.}

\begin{figure}
\centering
\begin{minipage}{0.32\columnwidth}
\centering
\includegraphics[height=0.85\columnwidth, angle=270, trim={140 0 0 29}, clip]{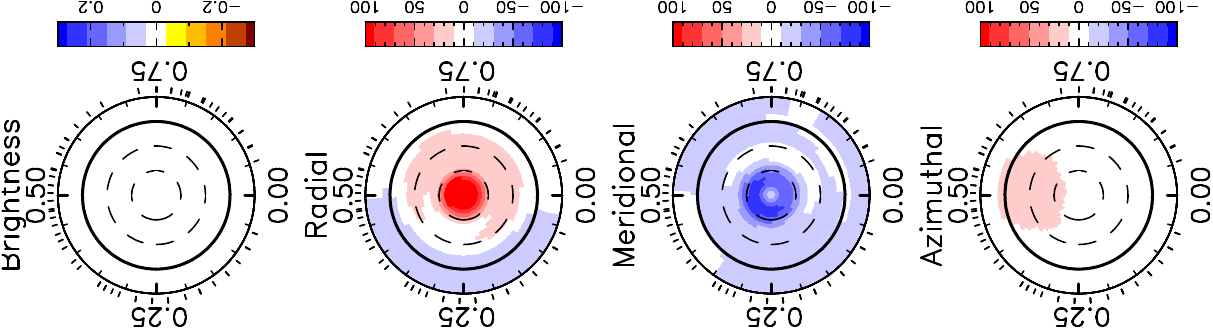} 
\end{minipage}
\begin{minipage}{0.32\columnwidth}
\centering
\includegraphics[height=0.85\columnwidth, angle=270, trim={140 0 0 29}, clip]{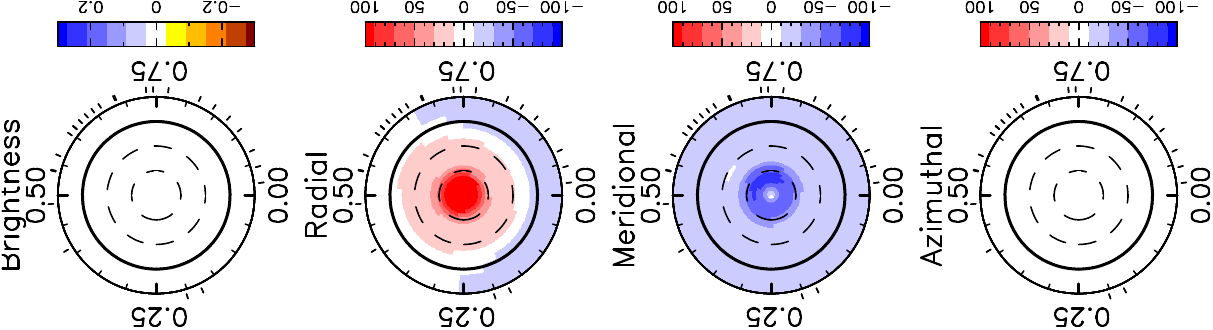} 
\end{minipage}
\begin{minipage}{0.32\columnwidth}
\centering
\includegraphics[height=0.85\columnwidth, angle=270, trim={140 0 0 29}, clip]{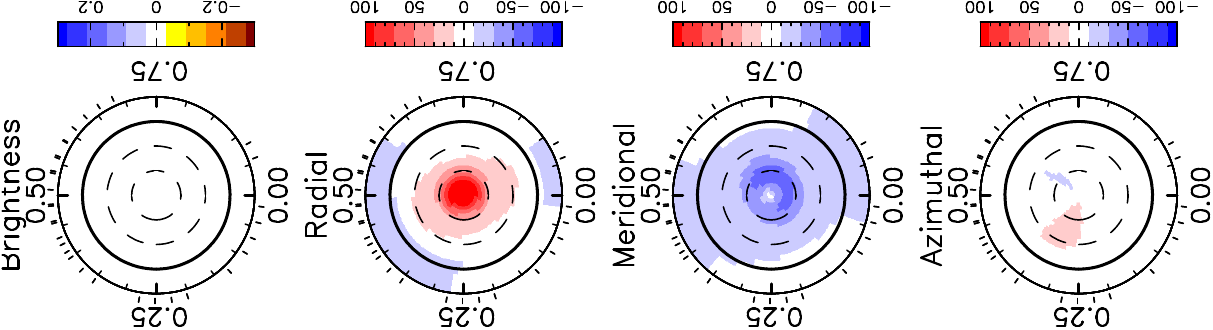} 
\end{minipage}
\includegraphics[width=0.3\columnwidth, angle=180, trim={460 130 2 0}, clip]{Figures/Gl617B_ZDIMap_JFDLSD_epo3_v11.pdf}
    \caption{The magnetic fields maps of Gl~617B using an inclination $i=30^\circ$ presented in the same format as in Fig.~\ref{Fig:Gl905_ZDIMaps}.}
    \label{Fig:Gl617B_ZDIMaps_i30}
\end{figure}

\begin{figure}
\centering
\begin{minipage}{0.32\columnwidth}
\centering
\includegraphics[height=0.85\columnwidth, angle=270, trim={140 0 0 29}, clip]{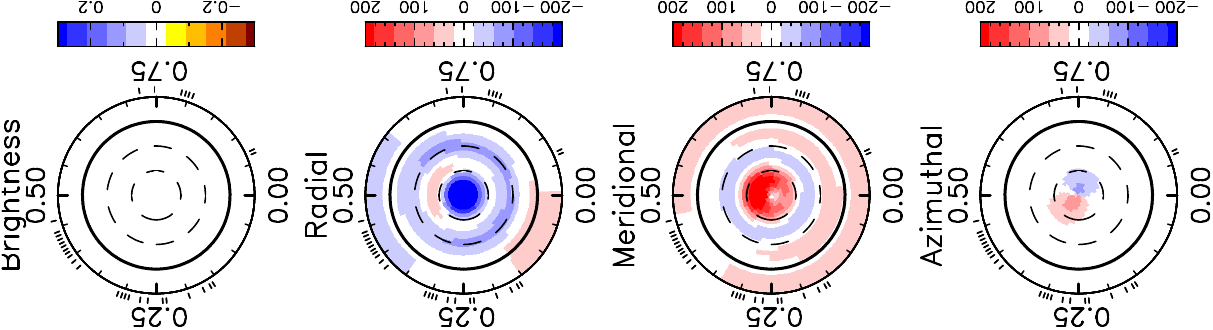} 
\end{minipage}
\begin{minipage}{0.32\columnwidth}
\centering
\includegraphics[height=0.85\columnwidth, angle=270, trim={140 0 0 29}, clip]{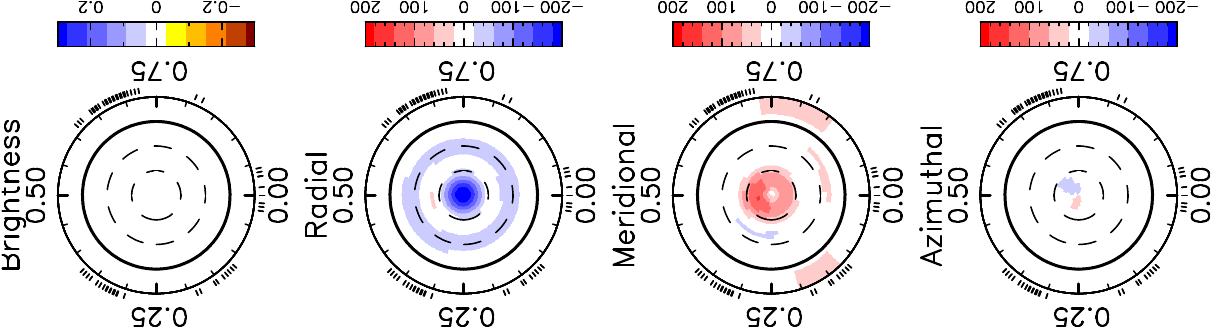} 
\end{minipage}
\begin{minipage}{0.32\columnwidth}
\centering
\includegraphics[height=0.85\columnwidth, angle=270, trim={140 0 0 29}, clip]{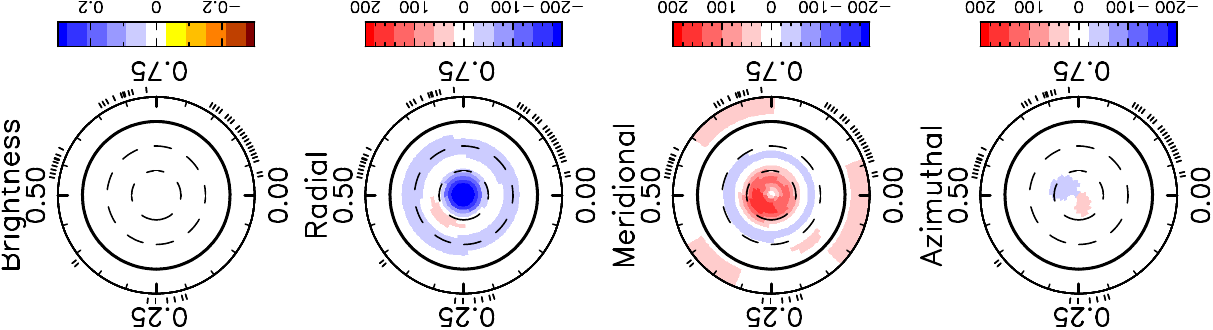} 
\end{minipage}
\includegraphics[width=0.3\columnwidth, angle=180, trim={460 130 2 0}, clip]{Figures/Gl408_ZDIMap_JFDLSD_epo3_v11.pdf}
    \caption{The magnetic fields maps of Gl~408 using an inclination $i=30^\circ$ presented in the same format as in Fig.~\ref{Fig:Gl905_ZDIMaps}.}
    \label{Fig:Gl408_ZDIMaps_i30}
\end{figure}

Fig.~\ref{Fig:Gl617B_ZDIMaps_i30} and \ref{Fig:Gl408_ZDIMaps_i30} show the ZDI maps of Gl~617B and Gl~408 for an alternative inclination $i=30^\circ$.

The following figures display the observed Stokes~$V$ LSD profiles (black) and their ZDI fits (red) split by season for the six M~dwarfs.

\begin{figure}
\centering
	\includegraphics[height=\columnwidth, trim={0 0 0 0}, clip, angle = 270]{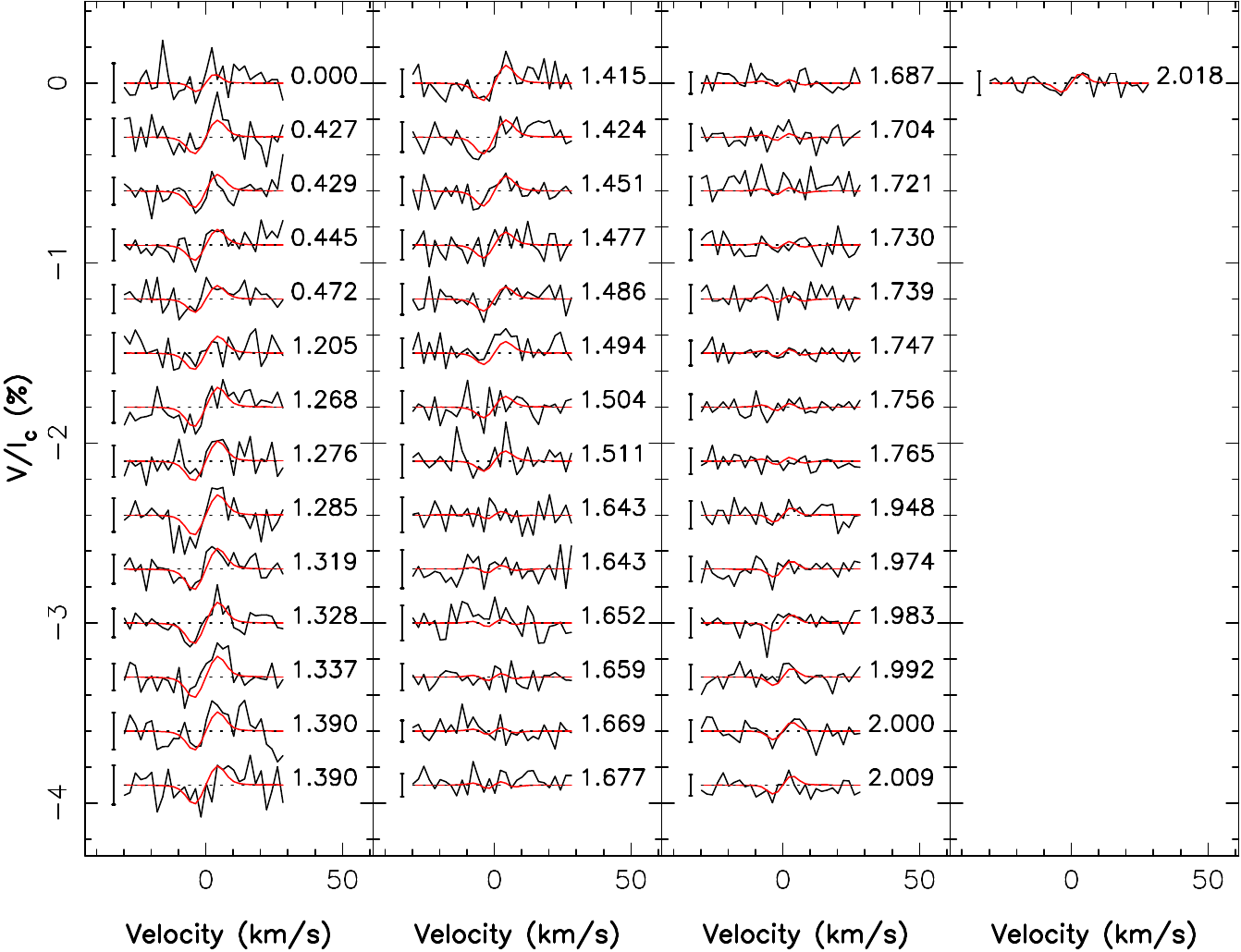}
\includegraphics[height=\columnwidth, trim={0 0 0 0}, clip, angle = 270]{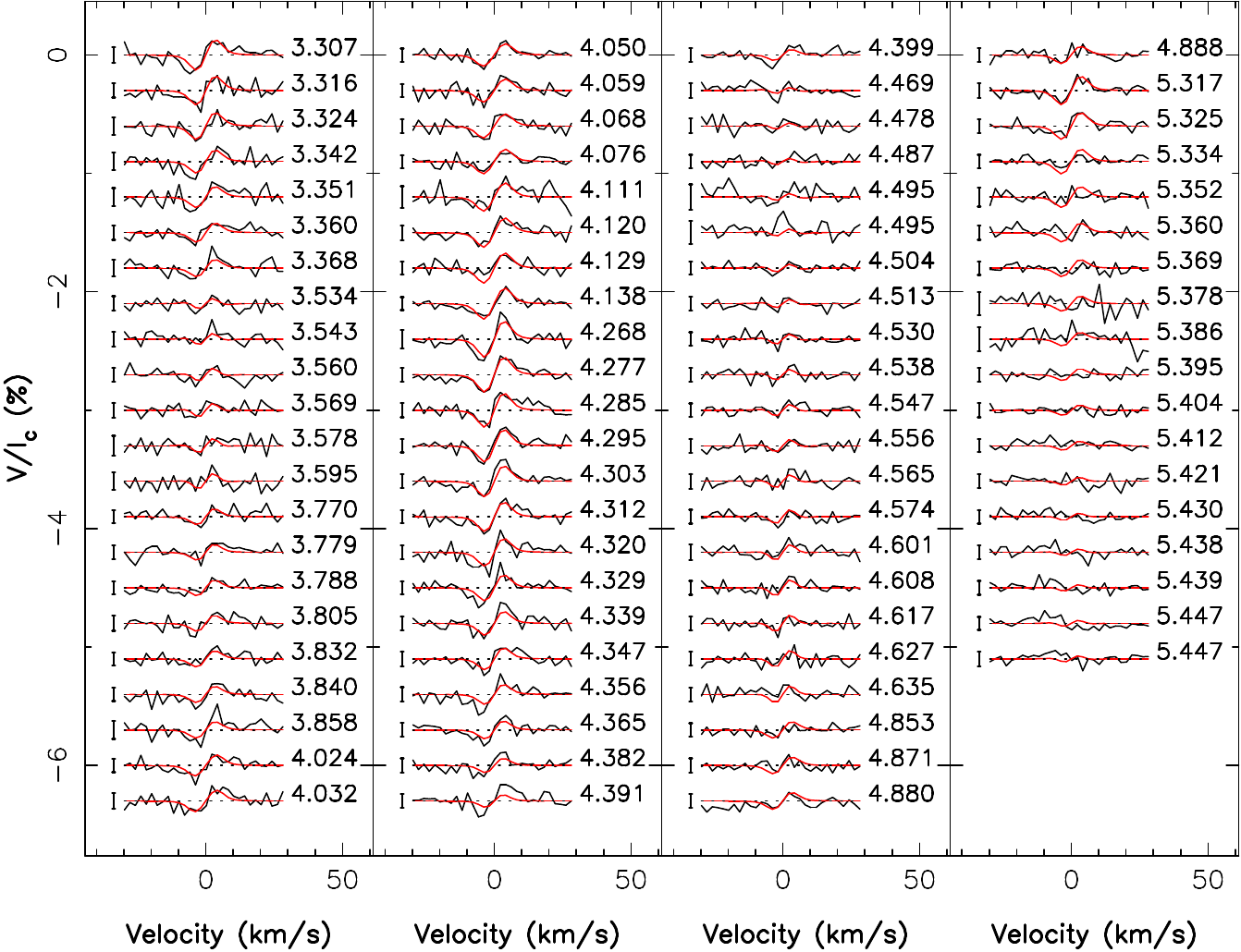}
\includegraphics[height=\columnwidth, trim={0 0 0 0}, clip, angle = 270]{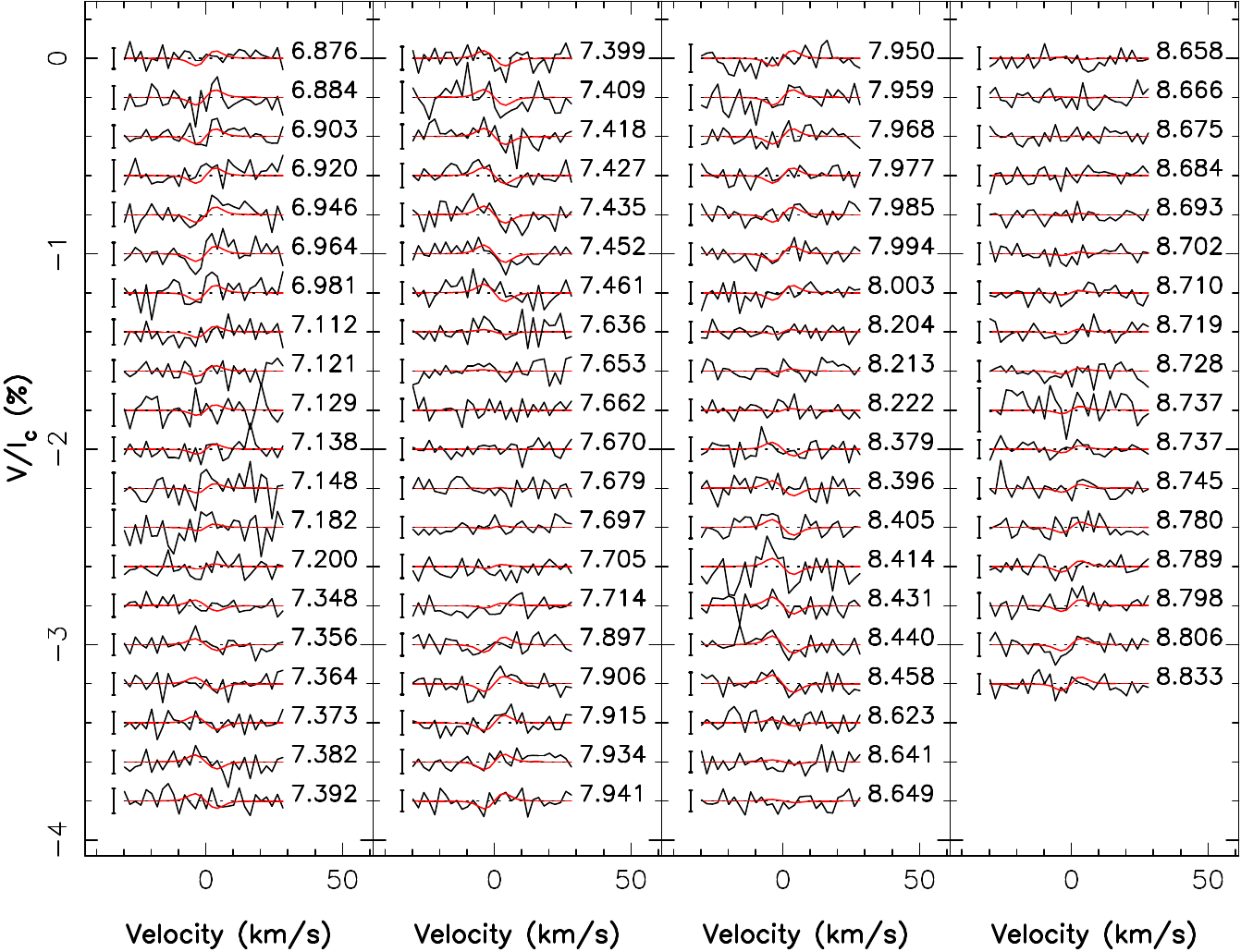}
    \caption{The SPIRou observed LSD Stokes~$V$ profiles (black) and the ZDI fit (red) for Gl~905. The rotation phase is indicated to the right of the profile and the error to the left. Each panel corresponds to one season.}
    \label{Fig:Gl905_StVFit}
\end{figure}

\begin{figure}
\centering
	\includegraphics[height=\columnwidth, trim={0 0 0 0}, clip, angle = 270]{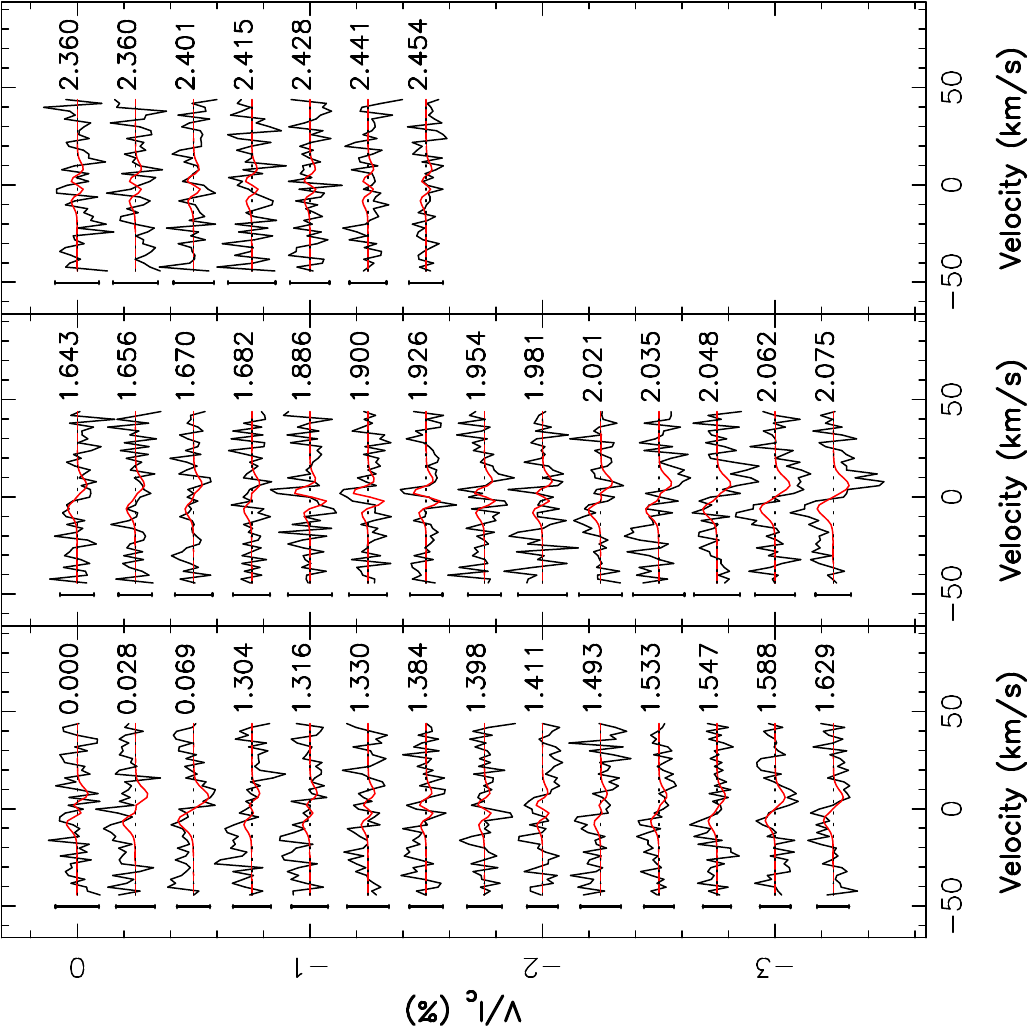}
\includegraphics[height=\columnwidth, trim={0 0 0 0}, clip, angle = 270]{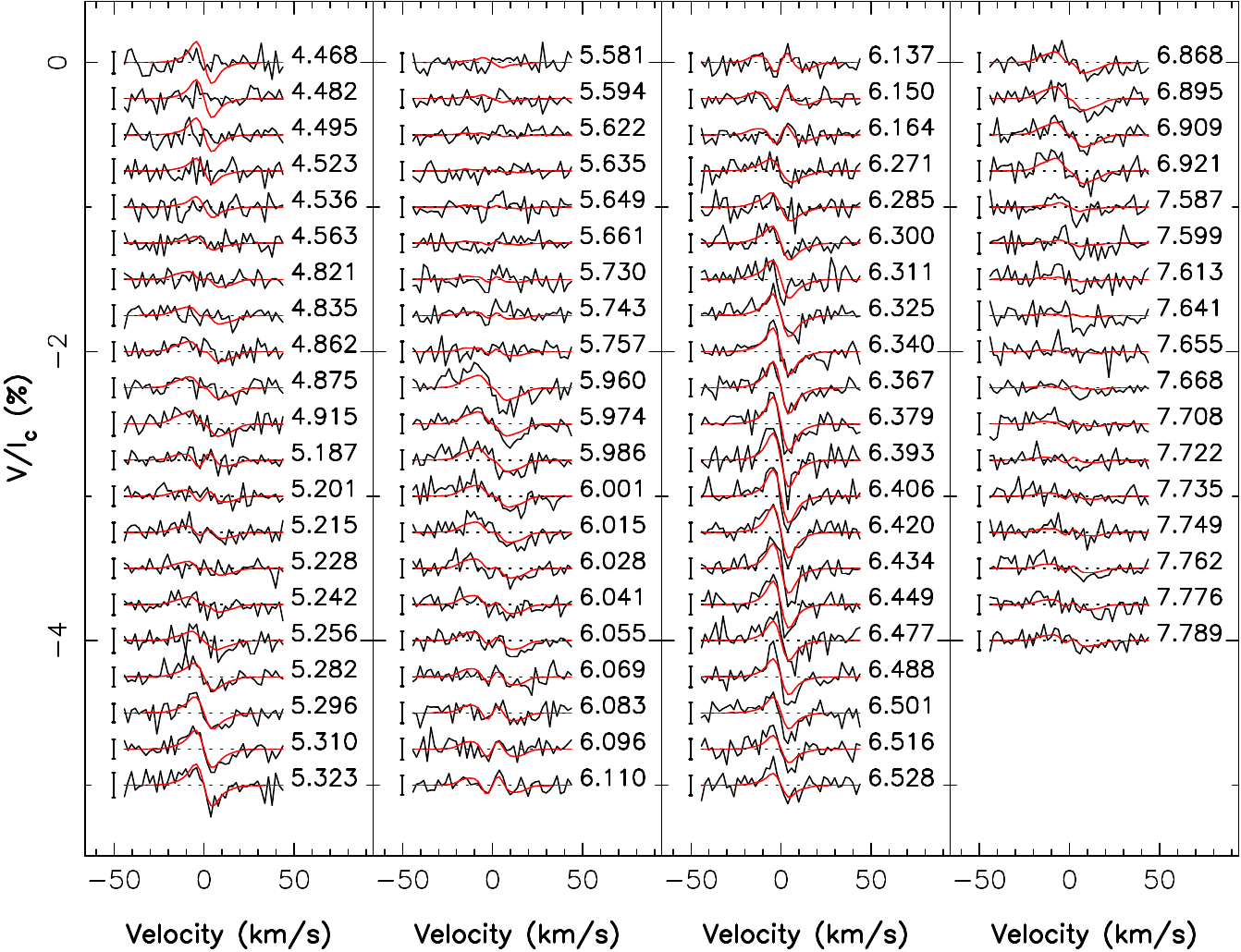}
\includegraphics[height=\columnwidth, trim={0 0 0 0}, clip, angle = 270]{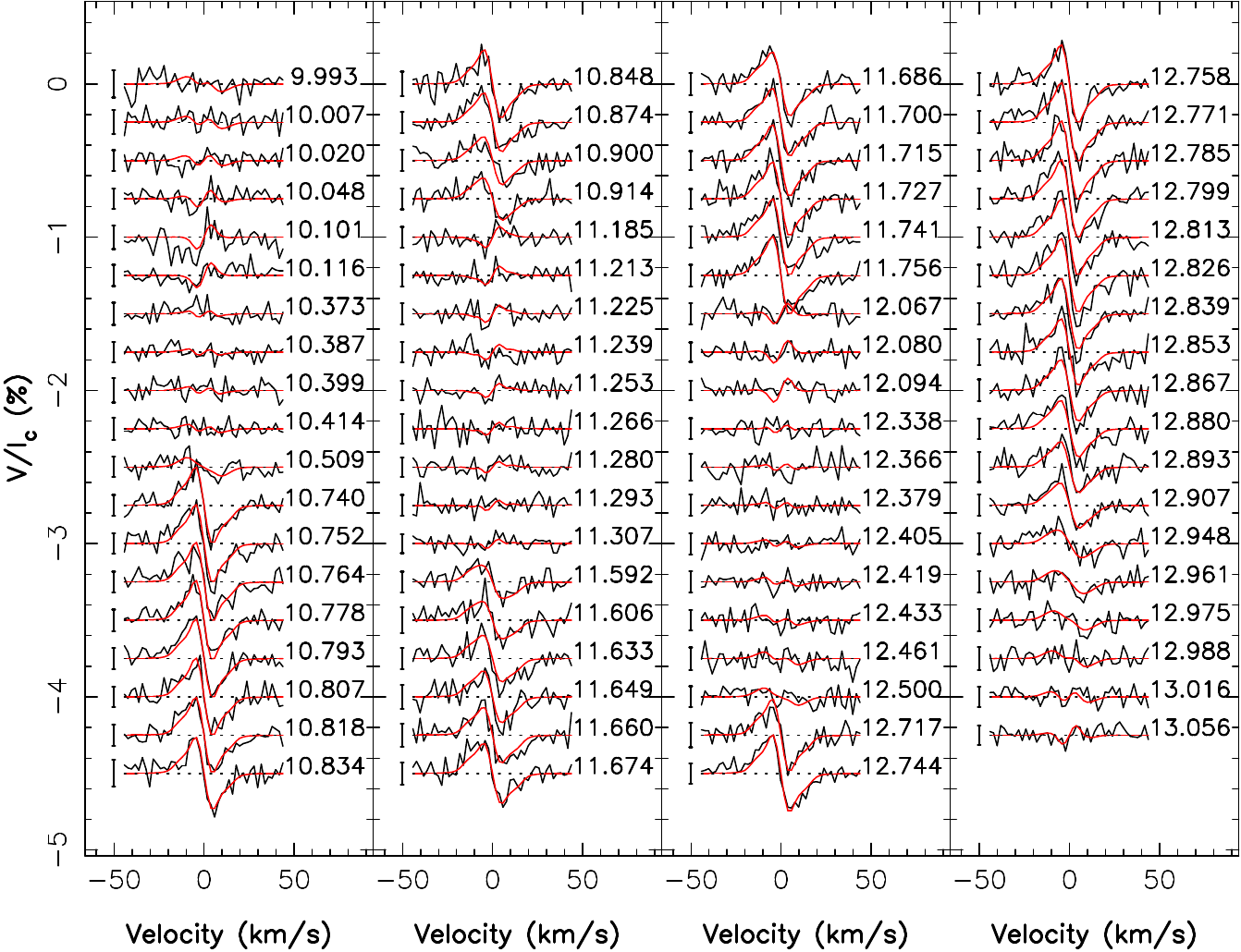}
    \caption{Same as Fig.~\ref{Fig:Gl905_StVFit} for GJ~1289.}
    \label{Fig:GJ1289_StVFit}
\end{figure}

\begin{figure}
\centering
	\includegraphics[height=\columnwidth, trim={0 0 0 0}, clip, angle = 270]{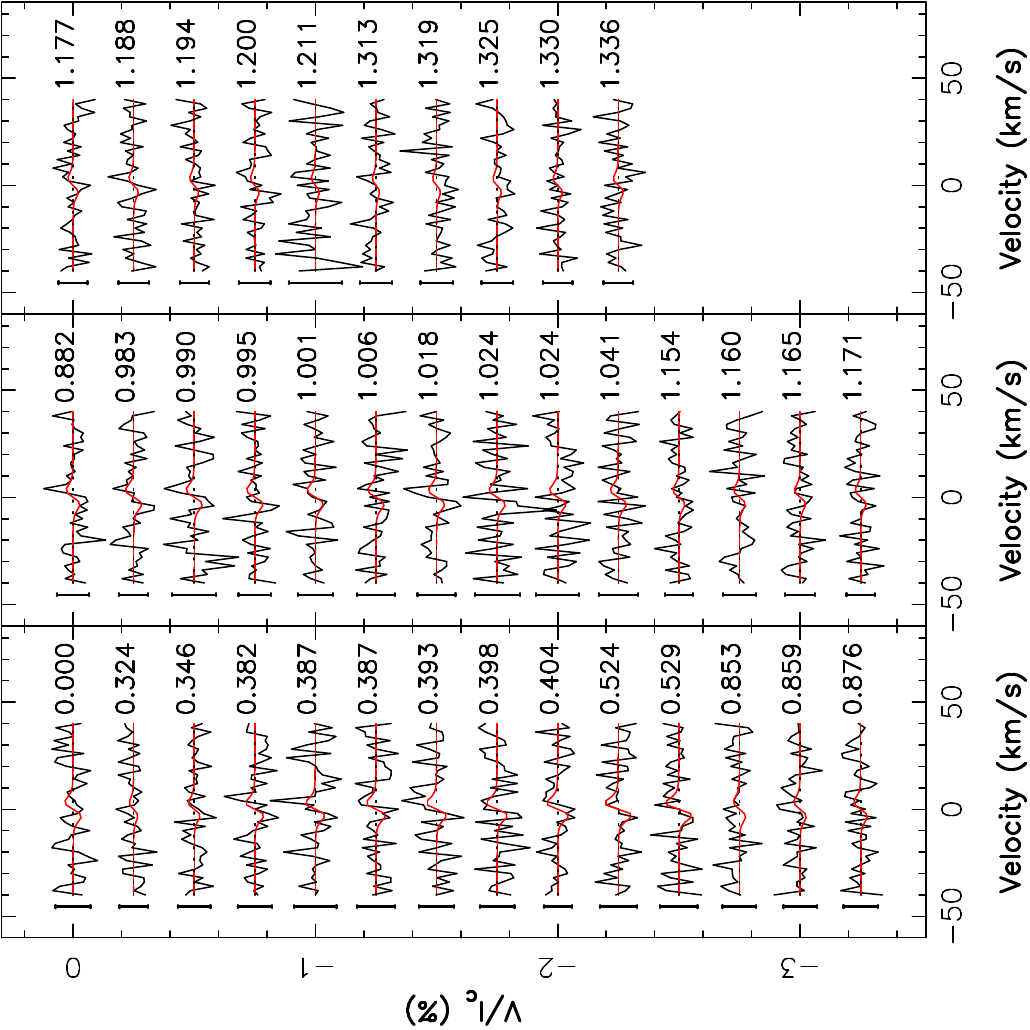}
\includegraphics[height=\columnwidth, trim={0 0 0 0}, clip, angle = 270]{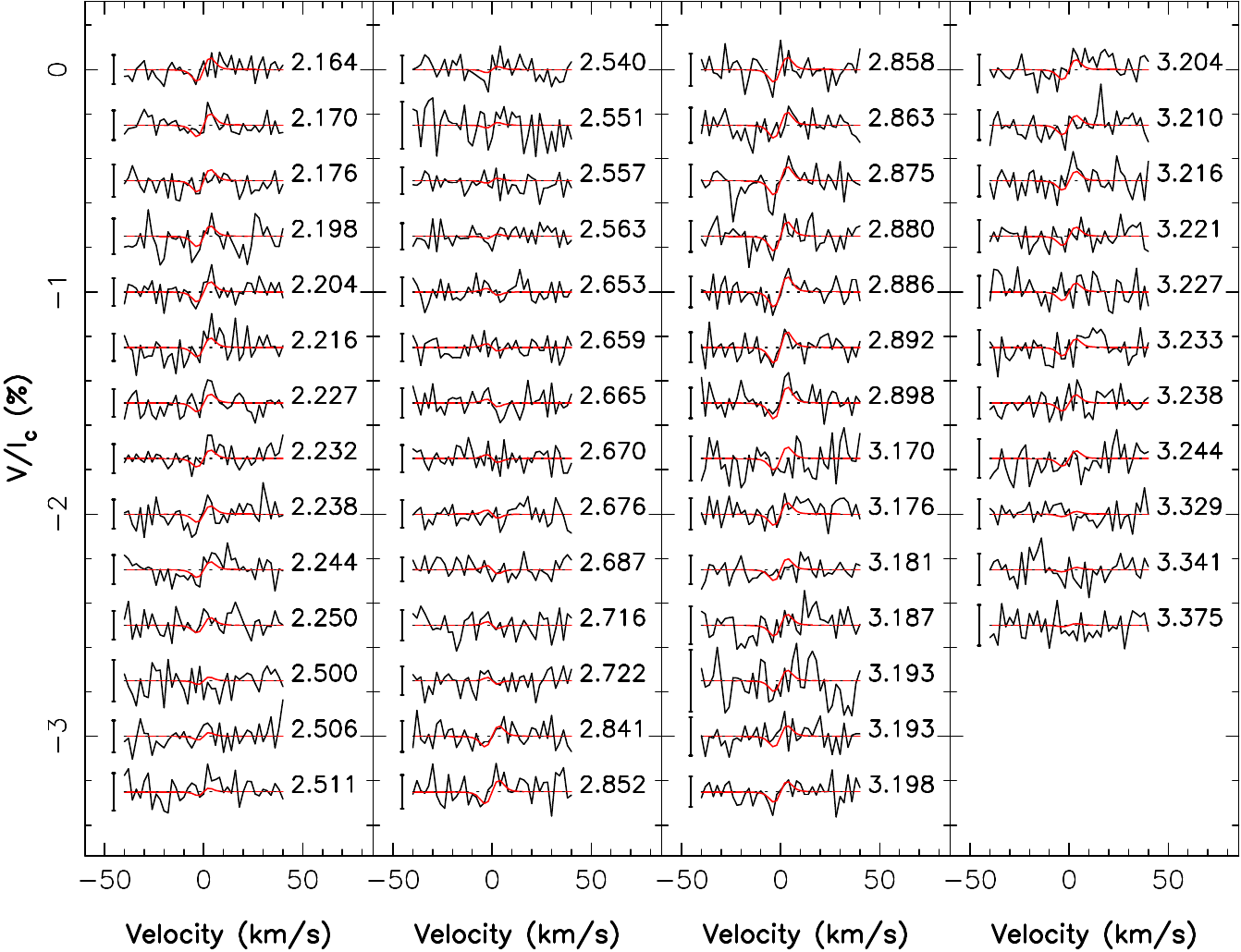}
\includegraphics[height=\columnwidth, trim={0 0 0 0}, clip, angle = 270]{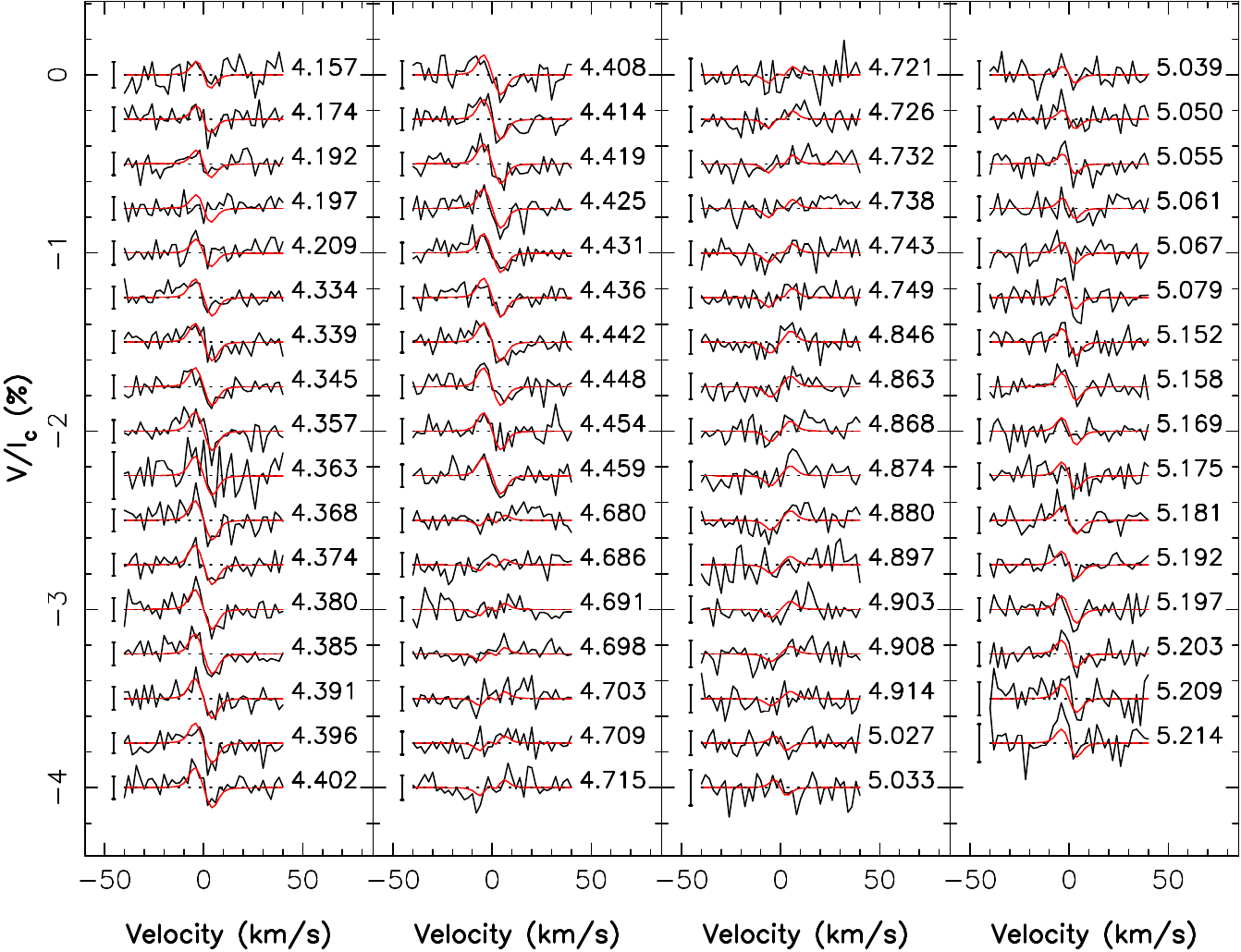}
    \caption{Same as Fig.~\ref{Fig:Gl905_StVFit} for GJ~1151.}
    \label{Fig:GJ1151_StVFit}
\end{figure}

\begin{figure}
\centering
	\includegraphics[height=\columnwidth, trim={0 0 0 0}, clip, angle = 270]{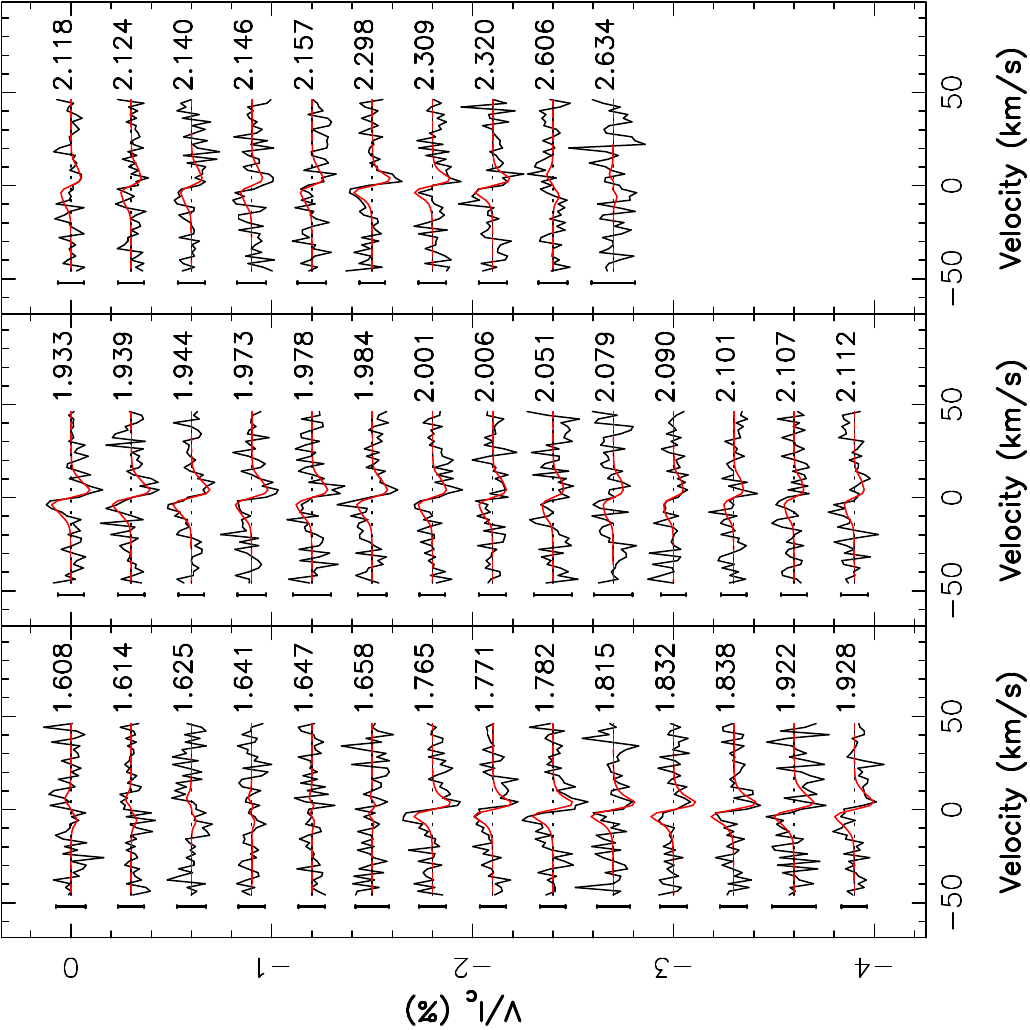}
\includegraphics[height=\columnwidth, trim={0 0 0 0}, clip, angle = 270]{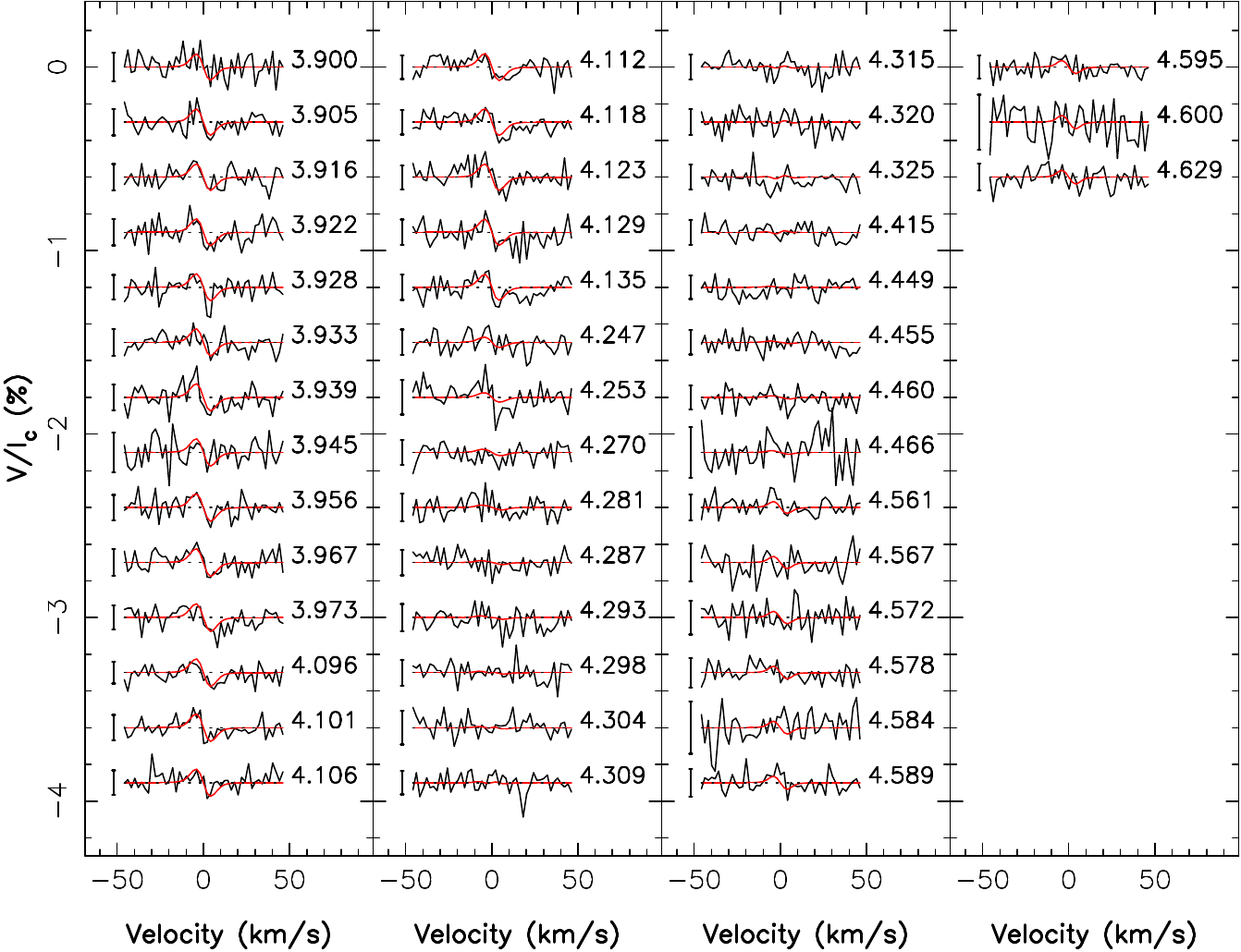}
    \caption{Same as Fig.~\ref{Fig:Gl905_StVFit} for GJ~1286.}
    \label{Fig:GJ1286_StVFit}
\end{figure}

\begin{figure}
\centering
	\includegraphics[height=\columnwidth, trim={0 0 0 0}, clip, angle = 270]{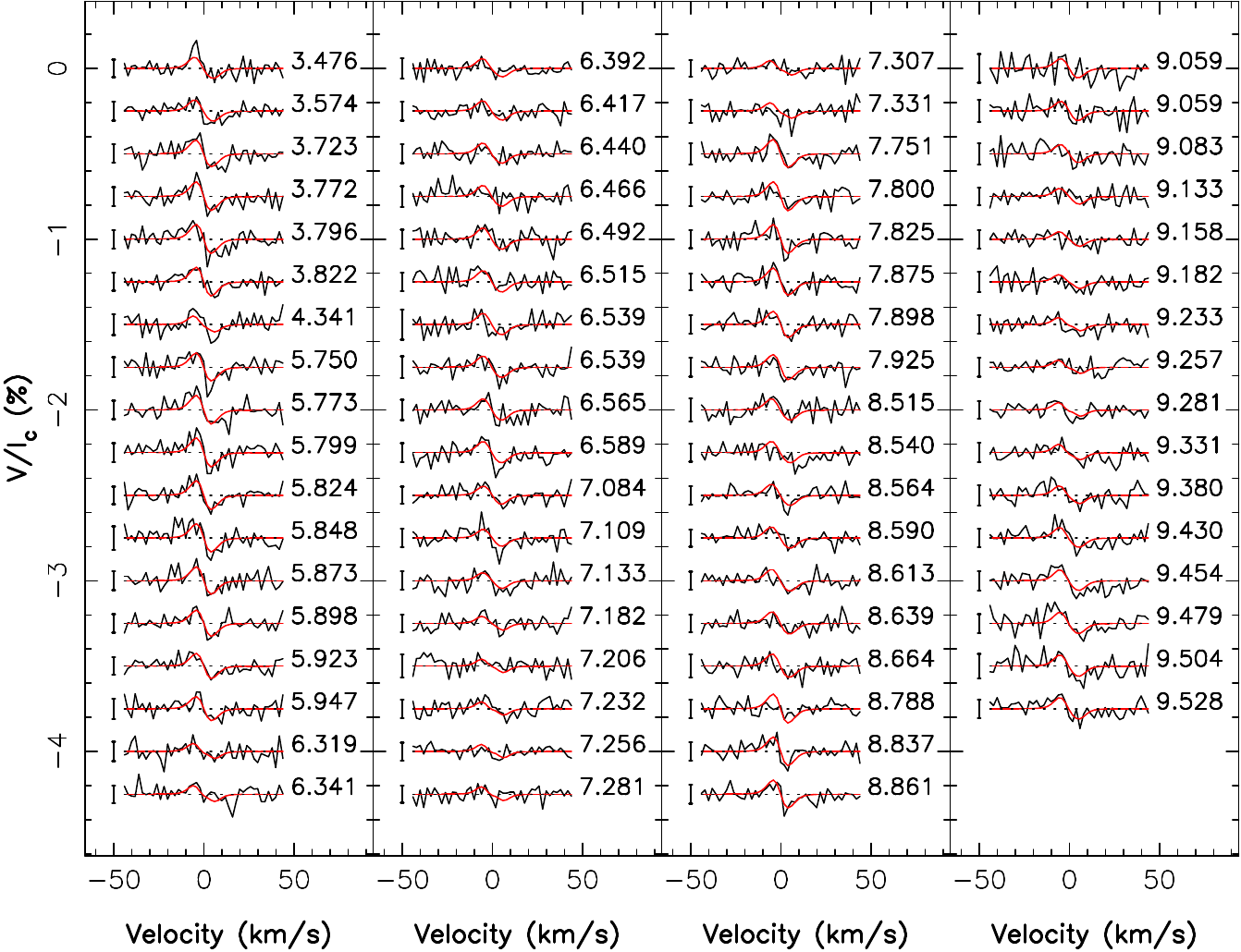}
\includegraphics[height=0.6\columnwidth, trim={0 0 0 0}, clip, angle = 270]{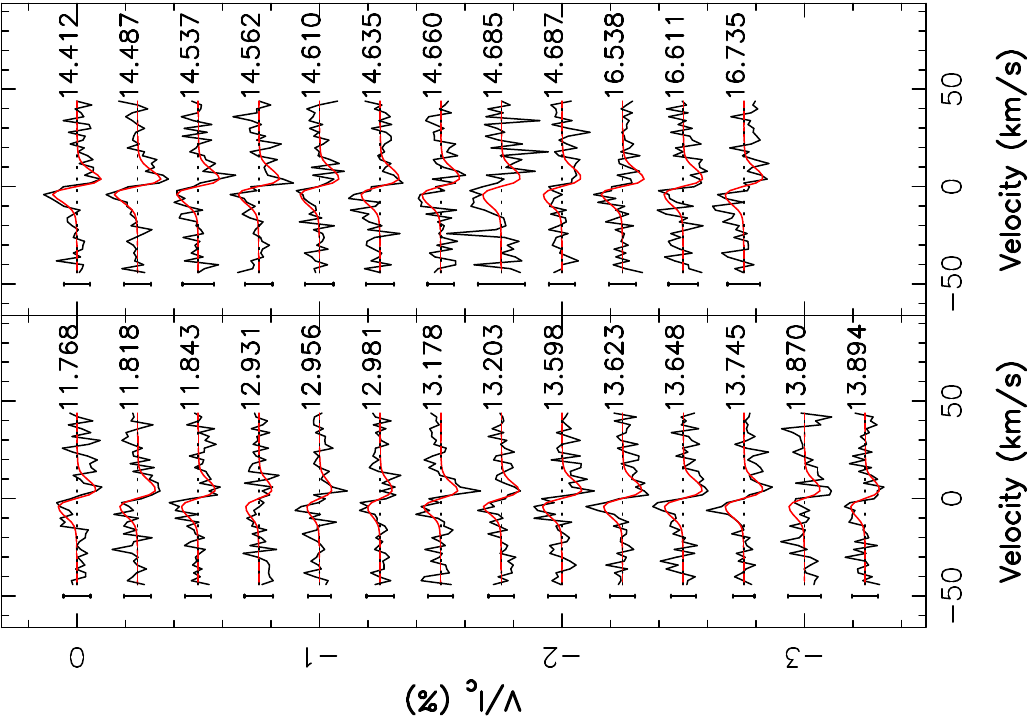}
\includegraphics[height=\columnwidth, trim={0 0 0 0}, clip, angle = 270]{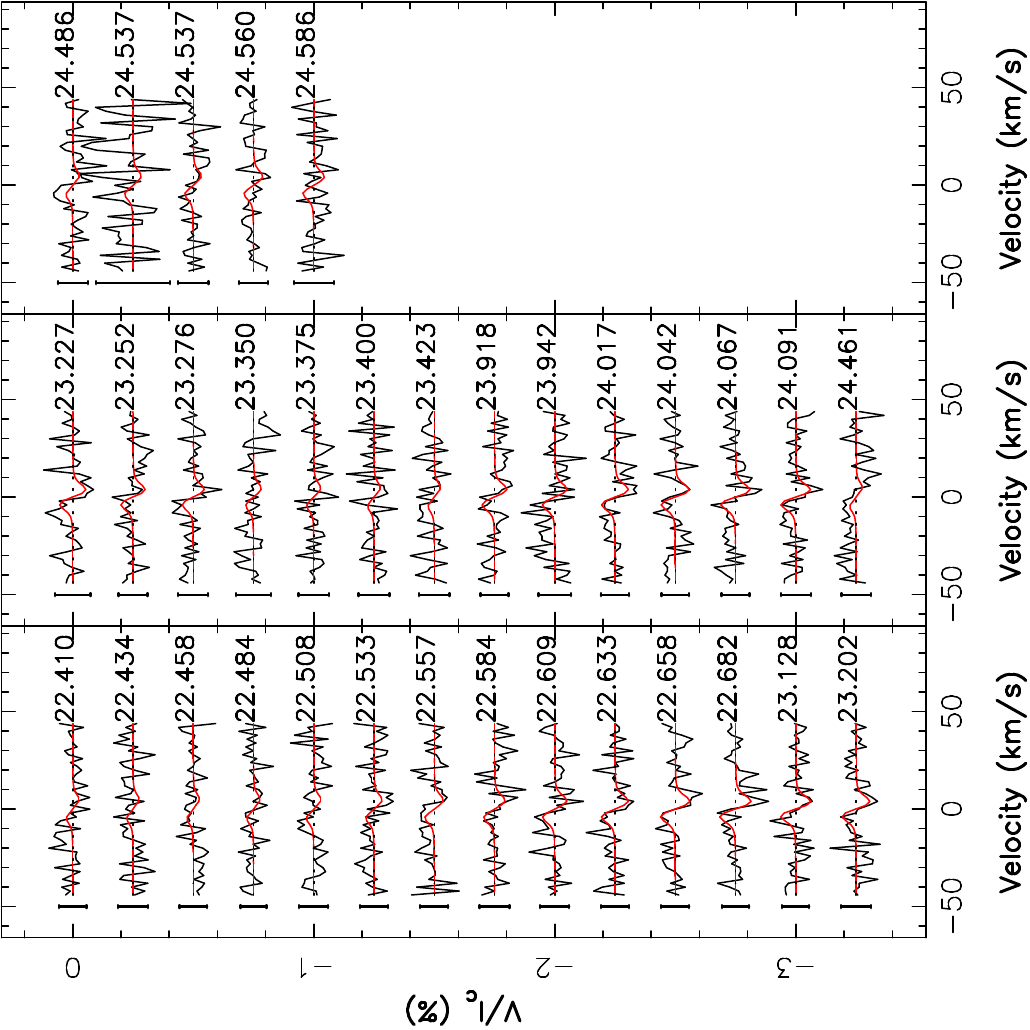}
    \caption{Same as Fig.~\ref{Fig:Gl905_StVFit} for Gl~617B.}
    \label{Fig:Gl617B_StVFit}
\end{figure}

\begin{figure}
\centering
	\includegraphics[height=\columnwidth, trim={0 0 0 0}, clip, angle = 270]{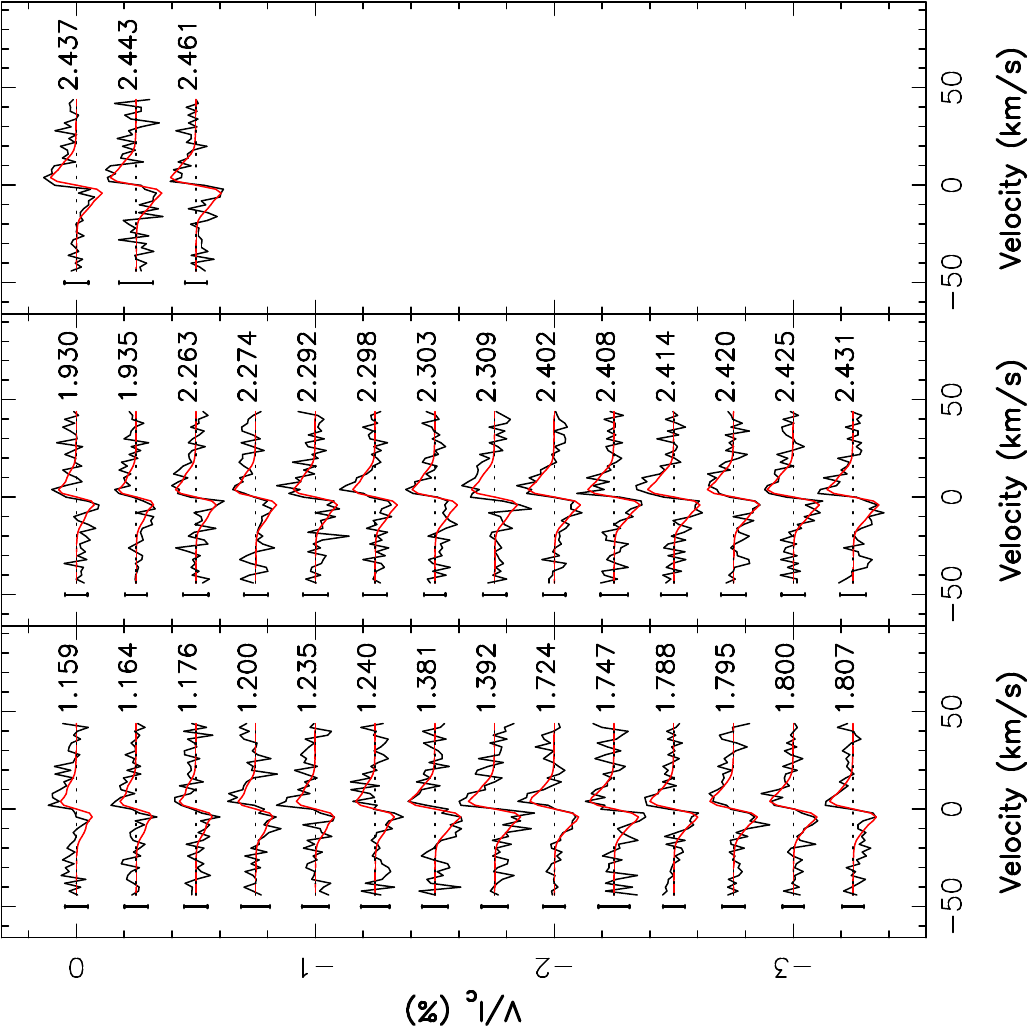}
\includegraphics[height=\columnwidth, trim={0 0 0 0}, clip, angle = 270]{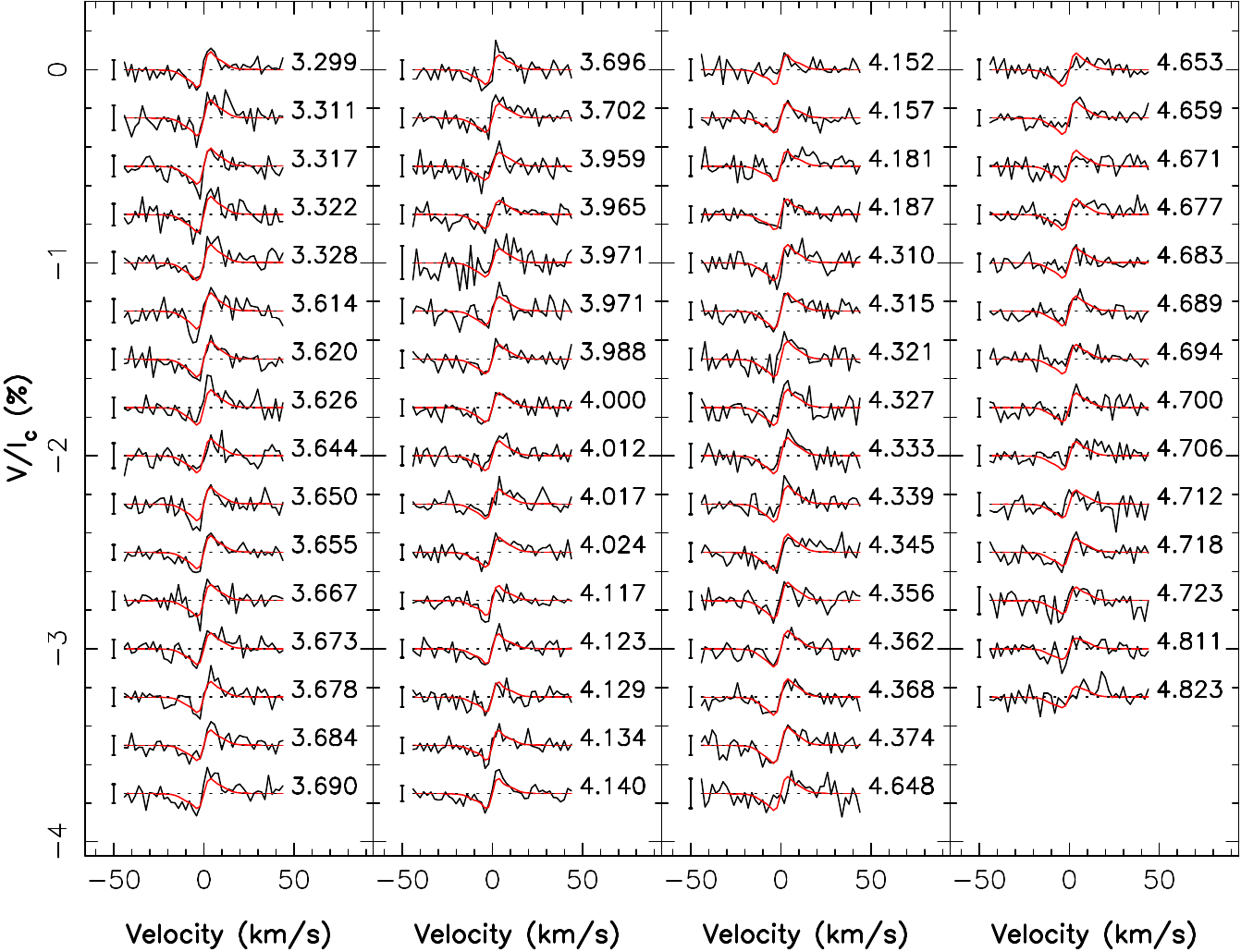}
\includegraphics[height=\columnwidth, trim={0 0 0 0}, clip, angle = 270]{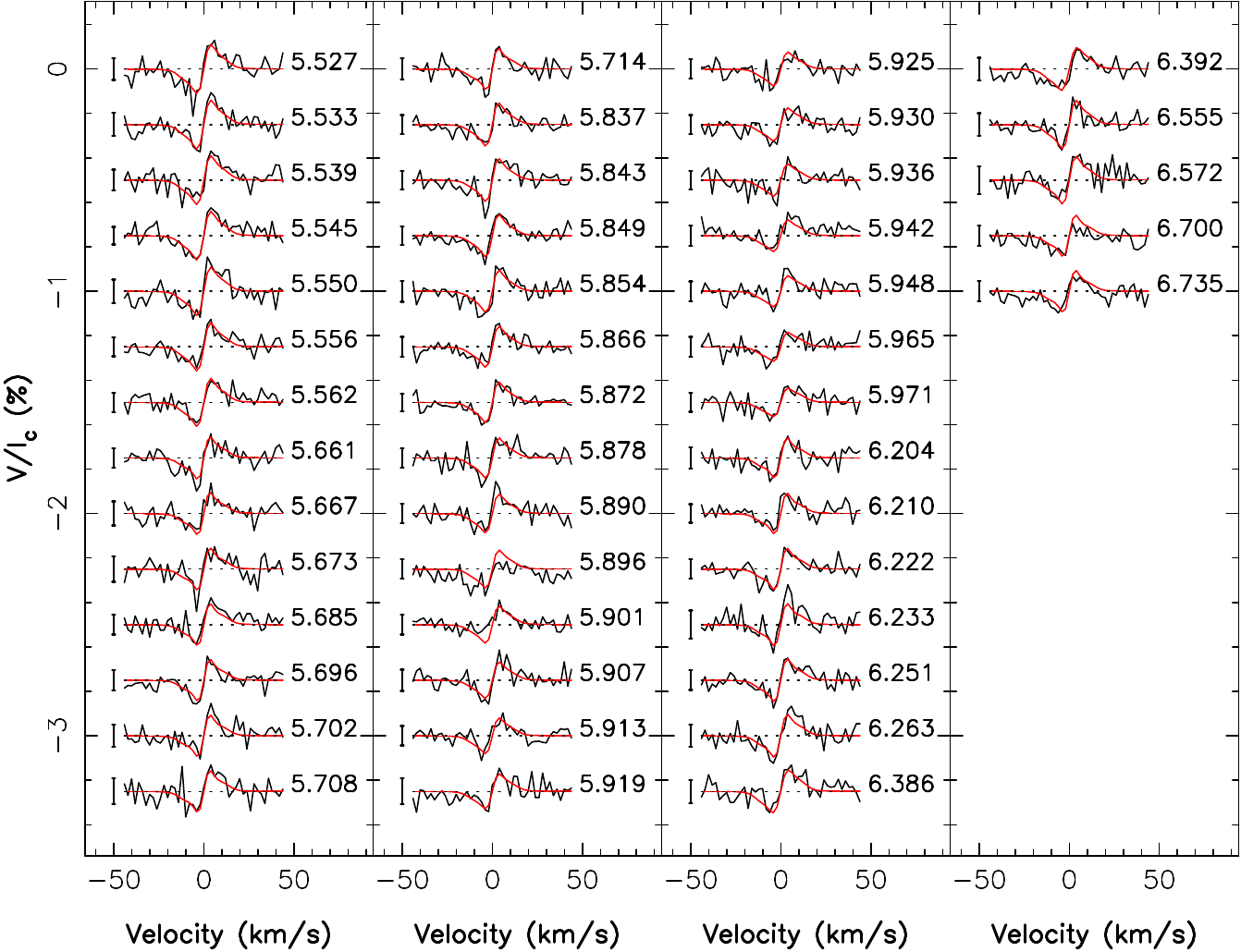}
    \caption{Same as Fig.~\ref{Fig:Gl905_StVFit} for Gl~408.}
    \label{Fig:Gl408_StVFit}
\end{figure}


\bsp	
\label{lastpage}
\end{document}